\documentclass[journal]{IEEEtran}

\ifCLASSINFOpdf
\else
  \usepackage[dvips]{graphicx}
\fi
\usepackage{url}
\usepackage[utf8]{inputenc}
\usepackage[T1]{fontenc}
\usepackage{url}
\usepackage{ifthen}
\usepackage{cite}
\usepackage[cmex10]{amsmath}
\usepackage{stfloats}
\usepackage{graphicx}
\usepackage{epstopdf}
\usepackage{amsfonts}
\usepackage{enumitem}
\usepackage{amssymb}
\usepackage{mathrsfs}
\usepackage{bm}
\usepackage{mdwmath}
\usepackage{mdwtab}
\usepackage{dsfont}
\usepackage{color}
\usepackage{verbatim}
\usepackage{array}
\usepackage{epstopdf}
\usepackage{multirow}
\usepackage{booktabs}
\usepackage{lineno}
\usepackage{amsmath}
\usepackage{makecell}
\usepackage{algorithmic}
\usepackage{tabularx}

\usepackage{amsthm}
\usepackage{tabularx}
\usepackage{graphicx}
\usepackage{wrapfig}
\usepackage{cite}
\usepackage{subcaption}
\usepackage{pifont}
\usepackage[font=footnotesize]{caption}

\usepackage{amsthm}
\usepackage{flushend}
\usepackage[ruled,vlined]{algorithm2e}

\begin{document}
\bstctlcite{IEEEexample:BSTcontrol}

\title{AI-driven Wireless Positioning: Fundamentals, Standards, State-of-the-art, and Challenges}

\author{Guangjin Pan, Yuan Gao,  \IEEEmembership{Member, IEEE}, Yilin Gao, Wenjun Yu, Zhiyong Zhong, \\  Xiaoyu Yang, Xinyu Guo, Shugong Xu, \IEEEmembership{Fellow, IEEE}
\thanks{This work was supported in part by  the National High Quality Program under Grant TC220H07D, in part by the National Key R\&D Program of China under Grant 2022YFB2902002, in part by the Innovation Program of Shanghai Municipal Science and Technology Commission under Grant 20511106603, and in part by Foshan Science and Technology Innovation Team Project under Grant FS0AAKJ919-4402-0060.}
\thanks{ Guangjin Pan was with the School of Communication and Information Engineering, Shanghai University, Shanghai, 200444, China. He is now with the Department of Electrical Engineering, Chalmers University of Technology, 41296 Gothenburg, Sweden (e-mail: guangjin.pan@chalmers.se).}

\thanks {Yuan Gao, Yilin Gao, Wenjun Yu, Zhiyong Zhong, Xiaoyu Yang, and Xinyu Guo are with the School of Communication and Information Engineering, Shanghai University, Shanghai, 200444, China (e-mail: \{gaoyuansie, gaoyilin, yuwenjun, zzy20010120, yangxiaoyu, guoxinyu \}@shu.edu.cn). }

\thanks {Shugong Xu is with the Department of Intelligent Science, Xi'an Jiaotong-Liverpool University, Suzhou 215123, China (e-mail: shugong.xu@xjtlu.edu.cn). Corresponding Author: Shugong Xu.}
}

\markboth{IEEE XXXX XXXXX, Vol. X, No. X, January 2025}
{Shell \MakeLowercase{\textit{et al.}}: Bare Demo of IEEEtran.cls for IEEE Journals}
\maketitle

\begin{abstract}
Wireless positioning technologies hold significant value for applications in autonomous driving, extended reality (XR), unmanned aerial vehicles (UAVs), and more. With the advancement of artificial intelligence (AI), leveraging AI to enhance positioning accuracy and robustness has emerged as a field full of potential. Driven by the requirements and functionalities defined in the 3rd Generation Partnership Project (3GPP) standards, AI/machine learning (ML)-based cellular positioning is becoming a key technology to overcome the limitations of traditional methods. This paper presents a comprehensive survey of AI-driven cellular positioning. We begin by reviewing the fundamentals of wireless positioning and AI models, analyzing their respective challenges and synergies. We provide a comprehensive review of the evolution of 3GPP positioning standards, with a focus on the integration of AI/ML in current and upcoming standard releases. Guided by the 3GPP-defined taxonomy, we categorize and summarize state-of-the-art (SOTA) research into two major classes: AI/ML-assisted positioning and direct AI/ML-based positioning. The former includes line-of-sight (LOS)/non-line-of-sight (NLOS) detection, time of arrival (TOA)/time difference of arrival (TDOA) estimation, and angle prediction; the latter encompasses fingerprinting, knowledge-assisted learning, and channel charting. Furthermore, we review representative public datasets and conduct performance evaluations of AI-based positioning algorithms using these datasets. Finally, we conclude by summarizing the challenges and opportunities of AI-driven wireless positioning.
\end{abstract}

\begin{IEEEkeywords}
Artificial intelligence, positioning technologies, 3GPP, cellular networks, 5G.
\end{IEEEkeywords}

\IEEEpeerreviewmaketitle

\section{Introduction}
\label{sec:Introduction}

\IEEEPARstart{W}{ith} the widespread deployment of 5G networks, wireless positioning technology has become a critical research area. Accurate positioning is indispensable for enabling the effective operation of systems and enhancing user experiences in applications such as intelligent transportation, emergency response, logistics tracking, and extended reality (XR). By providing precise location information, wireless positioning enables more efficient resource management, accurate service delivery, and enhanced security.

\begin{figure}[tbp]
    \centering
    \includegraphics[scale=0.4]{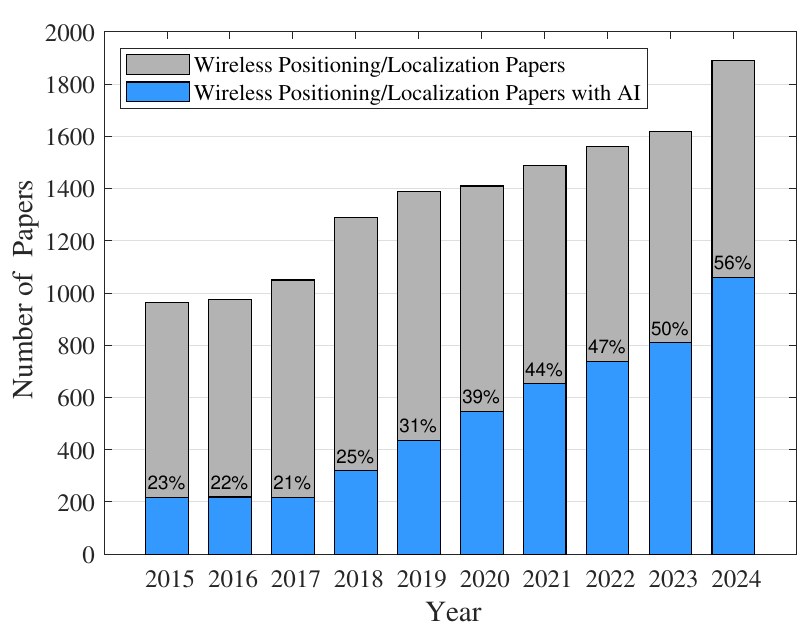}
    \caption{Comparison of the number of wireless positioning papers and AI-driven wireless positioning papers from 2015 to 2024.}
    \label{fig:wireless_ai_papers}
    \vspace{-4mm}
\end{figure}

Beyond communication, wireless positioning is an essential function of wireless networks. Its performance enhancement is critical for enabling a wide range of emerging applications and remains a valuable area of research. In outdoor environments, cellular positioning and global navigation satellite system (GNSS) technologies play pivotal roles, providing essential support for pedestrian navigation, autonomous driving, and unmanned aerial vehicle (UAV) localization \cite{Survey_del_peral-rosado_survey_2018}. These technologies not only enhance positioning accuracy but also improve user experience and safety. Indoors, advancements in technologies such as cellular networks, WiFi, Bluetooth, and ultra-wideband (UWB) have enabled precise localization within complex environments, which is crucial for applications like shopping mall navigation and emergency evacuation \cite{Survey_shit_survey_2019}. As part of the mobile communication infrastructure, cellular positioning is instrumental in both indoor and outdoor scenarios. However, traditional positioning methods, primarily based on geometric relationships like triangulation and trilateration, still face challenges in terms of accuracy, robustness, scalability, and adaptability to dynamic environments.

In recent years, the evolution of artificial intelligence (AI) has brought transformative changes to wireless positioning \cite{sze_efficient_2017}. AI technologies, particularly machine learning (ML) and deep learning, have significantly enhanced the accuracy and efficiency of positioning systems through their powerful data processing and pattern recognition capabilities \cite{Chaccour_Knowledge, XAI_Security}. AI algorithms can analyze complex wireless signal characteristics, identify environmental interference factors, and predict user locations with high precision. Furthermore, AI-driven systems can adapt to changing environments by learning from historical data, offering innovative solutions to the challenges of traditional techniques. Fig.~\ref{fig:wireless_ai_papers} illustrates the publication trends of wireless positioning-related papers over the past decade based on Google Scholar data.
It can be observed that since 2017, with the rapid advancement of AI technologies, AI-driven wireless positioning has gained increasing prominence within the broader field of wireless positioning research. By 2023 and beyond, more than half of the published papers in this area have incorporated AI technologies. As AI continues to advance, AI-driven wireless positioning is expected to play an increasingly critical role in the future.


\begin{table*}[t]
\centering
\caption{Comparison of Wireless Positioning Survey Papers (\ding{51}: Yes, \ding{55}: No, \ding{108}: Partial)}
\label{table:relatedwork}
\begin{tabular}{|p{0.6cm}|p{0.6cm}|p{7.8cm}|p{1.3cm}|p{1.3cm}|p{1.3cm}|p{1.4cm}|}
\hline
\textbf{Ref.} & \textbf{Year} & \textbf{Main Focus} & \textbf{AI/ML Related} & \textbf{3GPP Standards} & \textbf{Dataset} & \textbf{Cellular Positioning} \\ \hline
\cite{Survey_del_peral-rosado_survey_2009} & 2009 & TOA-based localization and NLOS mitigation techniques. & \ding{55} & \ding{55} & \ding{55} & \ding{55} \\ \hline
\cite{Survey_shastri_wi-fi_2016} & 2016 & WiFi fingerprint-based indoor positioning techniques. & \ding{55} & \ding{55} & \ding{55} & \ding{55} \\ \hline
\cite{Survey_khalajmehrabadi_modern_2017} & 2017 & Conventional and fingerprinting-based WLAN indoor localization. & \ding{108} & \ding{55} & \ding{55} & \ding{55} \\ \hline
\cite{Survey_Wei_Jammer_2017} & 2017 & Jammer localization techniques in multi-hop wireless networks. & \ding{55} & \ding{55} & \ding{55} & \ding{55} \\ \hline
\cite{Survey_del_peral-rosado_survey_2018} & 2018 & Evolution of cellular positioning technologies from 1G to 5G. & \ding{55} & \ding{51} & \ding{55} & \ding{51} \\ \hline
\cite{Survey_laoudias_survey_2018} & 2018 & Network localization, tracking, and navigation techniques. & \ding{108} & \ding{55} & \ding{55} & \ding{108} \\ \hline
\cite{Survey_jang_indoor_2019} & 2019 & Indoor localization approaches without offline fingerprint maps. & \ding{55} & \ding{55} & \ding{55} & \ding{55} \\ \hline
\cite{Survey_shit_survey_2019} & 2019 & Indoor localization techniques in WiFi, UWB, and RFID systems. & \ding{108} & \ding{55} & \ding{55} & \ding{55} \\ \hline
\cite{Survey_shit_ubiquitous_2019} & 2019 & Device-free localization for smart environment applications. & \ding{108} & \ding{55} & \ding{55} & \ding{55} \\ \hline
\cite{Survey_guo_survey_2020} & 2020 & Fusion-based localization systems for indoor positioning. & \ding{108} & \ding{55} & \ding{55} & \ding{108} \\ \hline
\cite{Survey_zhu_indoor_2020} & 2020 & ML-enhanced fingerprint-based indoor localization. & \ding{51} & \ding{55} & \ding{55} & \ding{55} \\ \hline
\cite{Survey_burghal_comprehensive_2020} & 2020 & ML technology for wireless positioning. & \ding{51} & \ding{55} & \ding{51} & \ding{108} \\ \hline
\cite{Survey_Chukhno_D2D_2022} & 2021 & D2D-based cooperative positioning techniques. & \ding{55} & \ding{51} & \ding{55} & \ding{51} \\ \hline
\cite{Survey_chen_tutorial_2022} & 2022 & THz localization techniques for future 6G networks. & \ding{108} & \ding{55} & \ding{55} & \ding{51} \\ \hline
\cite{Survey_shastri_review_2022} & 2022 & mmWave localization and sensing technologies. & \ding{108} & \ding{55} & \ding{55} & \ding{51} \\ \hline
\cite{Survey_Chen_Reconfigurable_2022} & 2022 & RIS-assisted localization methods for IoT applications. & \ding{55} & \ding{55} & \ding{55} & \ding{108} \\ \hline
\cite{Survey_trevlakis_localization_2023} & 2023 & High-accuracy localization technologies for 6G networks. & \ding{108} & \ding{51} & \ding{55} & \ding{51} \\ \hline
\cite{Survey_italiano_tutorial_2024} & 2024 & Comprehensive 5G positioning systems and standards. & \ding{108} & \ding{51} & \ding{55} & \ding{51} \\ \hline
\cite{Survey_sallouha_ground_2024} & 2024 & Localization techniques for integrated ground-air-space networks. & \ding{108} & \ding{55} & \ding{55} & \ding{108} \\ \hline
\cite{Survey_yang_positioning_2024} & 2024 & Comprehensive survey of indoor wireless positioning technologies. & \ding{108} & \ding{51} & \ding{55} & \ding{108} \\ \hline
\multicolumn{2}{|l|}{\textbf{\ \ Our Work}} & AI-driven positioning techniques for cellular networks, including fundamentals, 3GPP standards, SOTA, datasets, and challenges. & \ding{51} & \ding{51} & \ding{51} & \ding{51} \\ \hline
\end{tabular}
\end{table*}

\subsection{Related Work}
Wireless positioning is a widely researched topic, with extensive investigations focusing on a variety of techniques, scenarios, and applications. Below, we provide a detailed overview of existing surveys and explain how this paper, with its focus on AI-driven wireless positioning, distinguishes itself.

To address the challenges posed by wireless positioning, researchers have conducted numerous studies on the fundamental theories of positioning. Focusing on localization algorithms, the authors in \cite{Survey_del_peral-rosado_survey_2009} provide a comprehensive survey of time-of-arrival (TOA)-based localization algorithms. In \cite{Survey_shit_survey_2019}, the authors summarize indoor positioning algorithms and discuss systems based on WiFi, radio frequency identification (RFID), UWB, Bluetooth, and other technologies. The survey in \cite{Survey_guo_survey_2020} explores fusion-based indoor positioning techniques using data from cellular networks, WiFi, global positioning systems (GPSs), inertial navigation, cameras, and more. In \cite{Survey_shit_ubiquitous_2019}, the authors investigate device-free positioning algorithms. Regarding fingerprint-based positioning technologies, the authors in\cite{Survey_shastri_wi-fi_2016} and \cite{Survey_khalajmehrabadi_modern_2017} conduct detailed surveys, with \cite{Survey_jang_indoor_2019} specifically addressing methods that bypass offline fingerprint maps, and \cite{Survey_zhu_indoor_2020} delving deeper into intelligent algorithms for fingerprinting. Considering the impact of network capabilities on positioning technologies, the authors in \cite{Survey_Wei_Jammer_2017} survey techniques for jammer localization in multi-hop networks, providing critical insights into handling interference and security issues in positioning systems. In \cite{Survey_laoudias_survey_2018}, the authors explore advanced network localization and tracking technologies. Building on these works, the authors in \cite{Survey_yang_positioning_2024} review the latest research in positioning technologies within wireless networks. With advancements in technology, positioning capabilities in cellular networks have become increasingly important. In \cite{Survey_del_peral-rosado_survey_2018}, the authors investigate the standardization efforts for cellular positioning from 1G to 4G and summarize the key technologies for wireless positioning in 5G networks. Following this, the paper in \cite{Survey_italiano_tutorial_2024} provides an in-depth survey of positioning technologies in 5G networks, covering standardization, key elements, research trends, and performance analyses in real-world environments. Looking ahead to 6G networks, the authors in \cite{Survey_trevlakis_localization_2023} summarize the latest research in positioning technologies, including novel applications, supporting technologies, system models, critical performance indicators, and future research directions. Moreover, with advancements in communication technology, D2D networks \cite{Survey_Chukhno_D2D_2022}, millimeter-wave (mmWave) \cite{Survey_shastri_review_2022}, THz \cite{Survey_chen_tutorial_2022}, reconfigurable intelligent surface (RIS) \cite{Survey_Chen_Reconfigurable_2022}, and Ground-Air-Space Networks \cite{Survey_sallouha_ground_2024} are not only enhancing communication capabilities but also improving positioning performance.

The summary and comparison of existing survey papers are provided in Table.~\ref{table:relatedwork}. Some prior surveys have mentioned AI-related techniques for wireless positioning \cite{Survey_khalajmehrabadi_modern_2017,Survey_laoudias_survey_2018,Survey_shit_survey_2019,Survey_shit_ubiquitous_2019,Survey_guo_survey_2020,Survey_chen_tutorial_2022,Survey_shastri_review_2022,Survey_trevlakis_localization_2023,Survey_italiano_tutorial_2024,Survey_sallouha_ground_2024,Survey_yang_positioning_2024}. However, such discussions are typically brief and lack a systematic review of the technical landscape and evolution of AI/ML models in this domain. Among existing works, the survey papers in \cite{Survey_zhu_indoor_2020, Survey_burghal_comprehensive_2020} present relatively comprehensive reviews of AI/ML applications in wireless positioning. However, there are substantial differences between their works and ours. The survey in \cite{Survey_zhu_indoor_2020} primarily focuses on traditional ML techniques for positioning, while our work focuses more on advanced AI techniques (e.g., deep learning methods). The survey in \cite{Survey_burghal_comprehensive_2020} focuses on different AI technologies and discusses the roles of various AI technologies in positioning tasks. In contrast, our work focuses specifically on different types of positioning algorithms (such as AI/ML-assisted positioning and direct AI/ML positioning methods), and discusses how AI can enhance each of these categories. For wireless positioning standards, numerous studies have reviewed the progress of 3GPP standards \cite{Survey_del_peral-rosado_survey_2018,Survey_Chukhno_D2D_2022,Survey_trevlakis_localization_2023,Survey_italiano_tutorial_2024}. Nonetheless, the investigation into the integration of AI techniques within the 3GPP standardization framework remains insufficient. Our survey includes an in-depth analysis of AI-enabled positioning within the 3GPP framework. Regarding datasets, the survey in \cite{Survey_burghal_comprehensive_2020} broadly covers datasets for WiFi, Bluetooth, and other wireless technologies, with minimal attention given to cellular positioning datasets. Conversely, we place significant emphasis on datasets pertinent to cellular positioning. Similar to works such as \cite{Survey_del_peral-rosado_survey_2018,Survey_Chukhno_D2D_2022, Survey_chen_tutorial_2022, Survey_shastri_review_2022,Survey_trevlakis_localization_2023,Survey_italiano_tutorial_2024}, our survey focuses on the domain of cellular positioning. Considering the rapid advancements in AI for wireless positioning over the past five years (as shown in Fig.~\ref{fig:wireless_ai_papers}), we concentrate on reviewing AI-driven positioning solutions in cellular networks, aiming to provide targeted, up-to-date, and comprehensive insights into this fast-evolving field. Specifically, our goal is to comprehensively review cutting-edge AI-driven wireless positioning techniques for cellular networks, including their fundamentals, 3GPP standards, state-of-the-art (SOTA) algorithms, datasets, and open challenges.

\color{black}

\begin{figure*}[tb]
\includegraphics[scale=0.44]{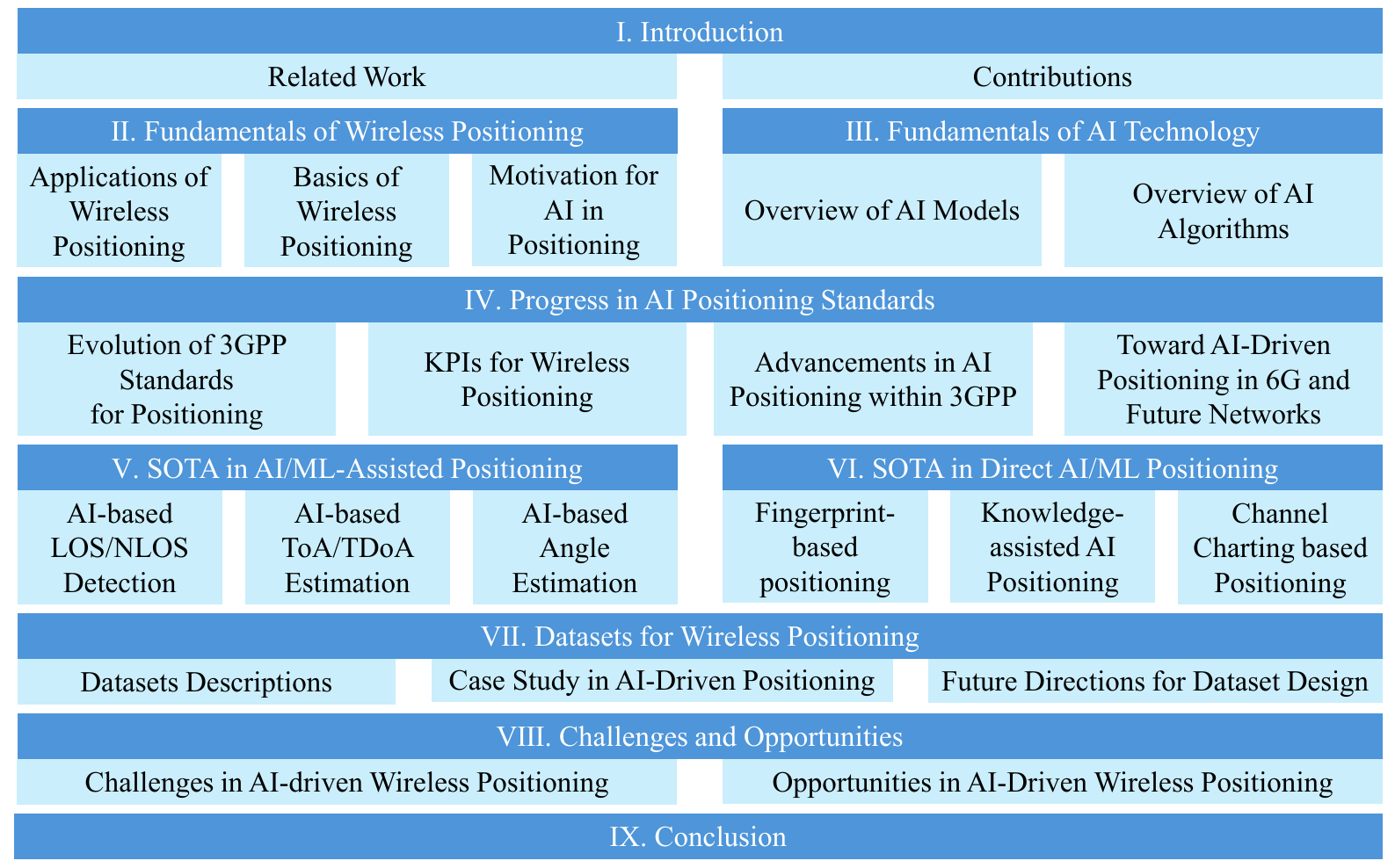}
\centering
\caption{The overall structure of this paper.}
\label{fig_sections}
\vspace{-4mm}
\end{figure*}

\subsection{Contributions}

As mentioned above, the rapid advancements in AI technology and the significant progress in AI-driven wireless positioning have motivated this comprehensive survey. The core contribution of our survey is to provide a focused and structured review of AI-driven cellular positioning, covering its fundamentals, standards, SOTA developments, datasets, and key challenges. The specific contributions of this paper are as follows:
\begin{itemize}
    \item \textbf{Foundation of AI and Wireless Positioning:} We present a comprehensive review of the foundational knowledge of both wireless positioning and AI technologies. For wireless positioning, we review key application scenarios, wireless channel models, and positioning algorithms. We further analyze the limitations of model-based positioning approaches and motivate the shift toward AI-driven solutions. On the AI side, we review representative neural network models and algorithms, and analyze their respective strengths, limitations, and suitability for different positioning tasks.

    \item \textbf{3GPP Standardization Progress:} We analyze the evolution of cellular positioning within the 3rd Generation Partnership Project (3GPP) framework, summarizing its advancements from early implementations to current 5G standards. We discuss the role of key performance indicators (KPIs) in wireless positioning and highlight the latest advancements in AI/ML-driven positioning solutions within the 3GPP standards. Moreover, we discuss potential future directions for AI-driven positioning in the upcoming 6G and beyond standards.

    \item \textbf{SOTA Research of AI-Driven Positioning\footnote{While our primary focus is on cellular-based positioning, we also selectively include non-cellular techniques in the SOTA review, to ensure completeness and to highlight transferable methodologies across different wireless technologies.}:} Based on 3GPP-defined frameworks, we summarize the SOTA in AI/ML-assisted positioning methods and direct AI/ML positioning methods. For AI/ML-assisted positioning methods, we focus on SOTA techniques for positioning parameter estimation, including AI-based line-of-sight (LOS)/non-line-of-sight (NLOS) scenarios detection, TOA/Time-difference-of-arrival (TDOA) estimation, and angle estimation algorithms. For direct AI/ML positioning methods, we classify and summarize the methods into fingerprint-based positioning, knowledge-assisted AI positioning, and channel charting based positioning, introducing the latest progress in each.

    \item \textbf{Datasets for Cellular Positioning:} We review publicly available datasets relevant to AI-based cellular positioning, summarizing their use cases, data characteristics, and limitations. In particular, we analyze two representative datasets (measurement-based MaMIMO and synthetic-based DeepMIMO) as case studies to evaluate AI models’ performance across different deployment scenarios. To facilitate reproducibility and accelerate research in this area, the corresponding implementation code has been released as open source on GitHub \footnote{The code is available at: https://github.com/guangjinpan/AI-Driven-Localization.}. We also identify current gaps in dataset availability and propose future directions for constructing more diverse and scalable datasets.

    \item \textbf{Challenges and Future Directions:} We discuss the major challenges in AI-driven wireless positioning, including data collection, accuracy in complex environments, and model generalization. We also present potential solutions to these challenges. Finally, we propose possible research directions and opportunities, providing guidance from both the perspectives of wireless technology development and AI technology advancements.
\end{itemize}

\color{black}

Therefore, the structure of this paper is illustrated in Fig. \ref{fig_sections}. In Sec. \ref{sec:Wireless}, we present the fundamentals of wireless positioning, while Sec. \ref{sec:AI} focuses on the fundamentals of AI technology. Sec. \ref{sec:3GPP} summarizes the progress in 3GPP AI/ML positioning standards. The SOTA advancements are reviewed in Sec. \ref{sec:AssistedPositioning} for AI/ML-assisted positioning methods and in Sec. \ref{sec:DirctPositioning} for direct AI/ML positioning methods. In Sec. \ref{sec:dataset}, we investigate publicly available datasets for wireless positioning, analyzing their characteristics and application scenarios. Sec. \ref{sec:Challenges} identifies the challenges and opportunities in AI-driven wireless positioning. Finally, Sec. \ref{sec:conclusion} concludes this paper.

\section{Fundamentals of Wireless Positioning} \label{sec:Wireless}


Wireless positioning is a key enabler for emerging applications. However, achieving high-precision positioning remains challenging due to the complex and dynamic nature of wireless propagation environments. In this section, we introduce the fundamentals of wireless positioning, focusing on real-world requirements and challenges and the motivation for AI integration. Specifically, we first review critical application scenarios that demand accurate positioning. Then, we present basic positioning principles, covering models of key measurements and positioning methods (including ranging-based, angle-based, fingerprint-based, and channel charting-based solutions). Finally, we deeply analyze the challenges faced by model-based approaches and discuss how AI-driven techniques can address them, providing a comparative analysis to motivate the integration of AI into positioning systems.

\subsection{Applications of Wireless Positioning}

Wireless positioning has emerged as a transformative technology, enabling a wide range of applications that improve efficiency, safety, and user experience across various domains. With the integration of AI and emerging wireless technologies such as RIS~\cite{10234214RIS}, Massive multiple-input multiple-output (MIMO)~\cite{zheng_flexible-position_2024}, THz communications \cite{Survey_chen_tutorial_2022}, non-terrestrial networks (NTN) \cite{dureppagari_ntn-based_2023}, near-field communications~\cite{chen_6g_2024}, and cell-free networks~\cite{10379122CELLFREE}, seamless and high-precision wireless positioning is poised to drive significant advancements across various industries. In this subsection, we introduce the importance of wireless positioning through several representative application scenarios.

\subsubsection{Intelligent Transportation}

Wireless AI positioning is pivotal in intelligent transportation systems, enabling precise vehicle localization that optimizes traffic flow, reduces congestion, and enhances road safety \cite{10251107_IntelligentTransportation, 10210349_IntelligentTransportation, 10554286_IntelligentTransportation}. Wireless positioning facilitates fleet management and route optimization for public transportation and ride-sharing services. Accurate location data also enhances road safety by enabling real-time hazard detection and proactive alerts to drivers \cite{10023991_IntelligentTransportation}.

\subsubsection{Autonomous Driving}

The success of autonomous driving depends heavily on high-precision positioning, which ensures vehicles can navigate safely and efficiently in complex environments. Wireless AI positioning offers the accuracy needed for lane-level navigation, seamless obstacle avoidance, and real-time decision-making in dynamic traffic scenarios \cite{10286277_AutonomousDriving, 10014536_AutonomousDriving, 9969877_AutonomousDriving}. Furthermore, positioning data is integral to vehicle-to-everything (V2X) communication, supporting synchronized interactions between vehicles and infrastructure to enhance traffic efficiency and safety.

\subsubsection{Extended Reality}
In XR applications, precise positioning of devices and users is crucial for creating seamless and immersive experiences. Whether in augmented reality (AR) navigation \cite{10007642XR}, virtual reality (VR) training simulations \cite{8851408XR}, or multiplayer gaming environments \cite{7907241XR}, wireless positioning ensures accurate tracking of movements and spatial relationships. This enhances the realism of interactive simulations and allows for seamless transitions between indoor and outdoor environments in XR-based applications.

\subsubsection{Indoor Tracking and Navigation}
Wireless positioning systems transform navigation within complex indoor environments \cite{10006721Indoor, 9094357Indoor, 6261513Indoor, nikonowicz_indoor_2024}, such as airports, shopping malls, and hospitals. These systems provide users with precise guidance to specific locations, improving efficiency and saving time in environments where traditional GPS signals are unreliable. For businesses, indoor navigation enhances operational workflows and provides opportunities for personalized user engagement, while improving overall customer experience.

\subsubsection{Public Safety}
Public safety applications greatly benefit from wireless AI positioning, particularly in emergency scenarios where speed and accuracy are critical \cite{6599064PublicSafety, moon_helps_2024}. By locating individuals in disaster-stricken or high-risk areas, such as collapsed buildings or burning structures, positioning systems streamline rescue operations and improve the success rates of life-saving interventions \cite{10034493PublicSafety}.

\subsubsection{Internet of Things}
Wireless positioning underpins the functionality of internet of things (IoT) ecosystems, enabling efficient management and monitoring of smart devices across various environments \cite{9703681IoT, Survey_Chen_Reconfigurable_2022}. In smart homes, factories, and agricultural settings, precise location data improves device interconnectivity and automation. By tracking equipment, inventory, and resources in real time, IoT-enabled positioning reduces operational inefficiencies and supports smarter decision-making.

\subsubsection{Security and Surveillance}
Security and surveillance systems utilize wireless AI positioning to monitor the real-time locations of personnel and assets in sensitive areas such as prisons, factories, and warehouses. By integrating real-time positional data, these systems enhance safety protocols and operational efficiency, ensuring the secure management of critical environments \cite{9543662SecuritySurveillance,10140186SecuritySurveillance}.

\subsubsection{Sports and Motion Analysis}
In sports and health domains, wireless positioning facilitates detailed motion analysis by tracking athletes’ positions and trajectories \cite{liu_survey_2023}. This data supports the optimization of training programs and enhances performance evaluations, providing athletes and coaches with actionable insights to refine techniques and strategies.

\subsubsection{Healthcare}
In healthcare settings, wireless positioning enables real-time tracking of patients, staff, and equipment, improving the delivery of medical services and overall operational efficiency \cite{van2016performance}. For instance, real-time location monitoring in critical care units, such as intensive care units or operating rooms, allows for rapid response to emergencies, potentially saving lives in time-sensitive situations.

\subsubsection{UAV positioning}

UAVs depend on wireless AI positioning for accurate navigation, collision avoidance, and stable operation in applications like surveillance, delivery, and disaster response \cite{akter_rfdoa-net_2021, peng2020uavpositioning, sandamini2023reviewuavpositioning}. In GPS-denied environments, such as indoors or dense urban areas, wireless positioning systems using technologies like UWB and 5G enable precise localization. AI-driven approaches further enhance positioning accuracy by integrating wireless signals with inertial and vision-based data, supporting tasks like agricultural spraying and infrastructure inspection. Despite challenges like interference and dynamic environments, advancements in multi-sensor fusion and cooperative positioning continue to improve UAV efficiency and safety.

\subsection{Basics of Wireless Positioning}

Wireless positioning estimates the location of a device by analyzing the propagation characteristics of wireless signals between transmitters and receivers. A fundamental concept is path loss, which describes the attenuation of signal power over distance \cite{rappaport2024wireless} and can be modeled as:
\begin{equation}
PL(d) = PL(d_0) + 10n \log_{10}\left(\frac{d}{d_0}\right),
\label{eq:path_loss}
\end{equation}
where $PL(d)$ represents the path loss as a function of distance $d$, $PL(d_0)$ is the path loss at a reference distance $d_0$, and $n$ is the path loss exponent, which depends on the environment. Based on path loss, received signal strength (RSS), such as received signal strength indicator (RSSI), and reference signal received power (RSRP), reflects this spatial relationship and thus forms the basis of range-based positioning. RSSI measures total received signal power, including noise and interference \cite{Xue2017rssi}, and is common in WiFi, Zigbee, and Bluetooth systems. RSRP, used in Long-Term Evolution (LTE)/5G \cite{3gpp2017rsrp}, isolates the power of the reference signal, offering better precision for cellular positioning. Using known transmission power and path loss models, distances can be inferred and fed into techniques like trilateration \cite{gezici2008survey}. However, RSSI/RSRP are sensitive to multipath, shadowing, and noise \cite{kotaru2015spotfi}, limiting their accuracy. To better model the wireless channel, channel state information (CSI) is used \cite{WuCSI2013CSI}. Unlike scalar RSSI/RSRP, CSI provides multidimensional amplitude and phase information for each subcarrier, enabling advanced signal modeling, especially for AI-driven positioning \cite{Wang2017CSIDeepLearning}.

\begin{figure}[tb]
\includegraphics[scale=0.31]{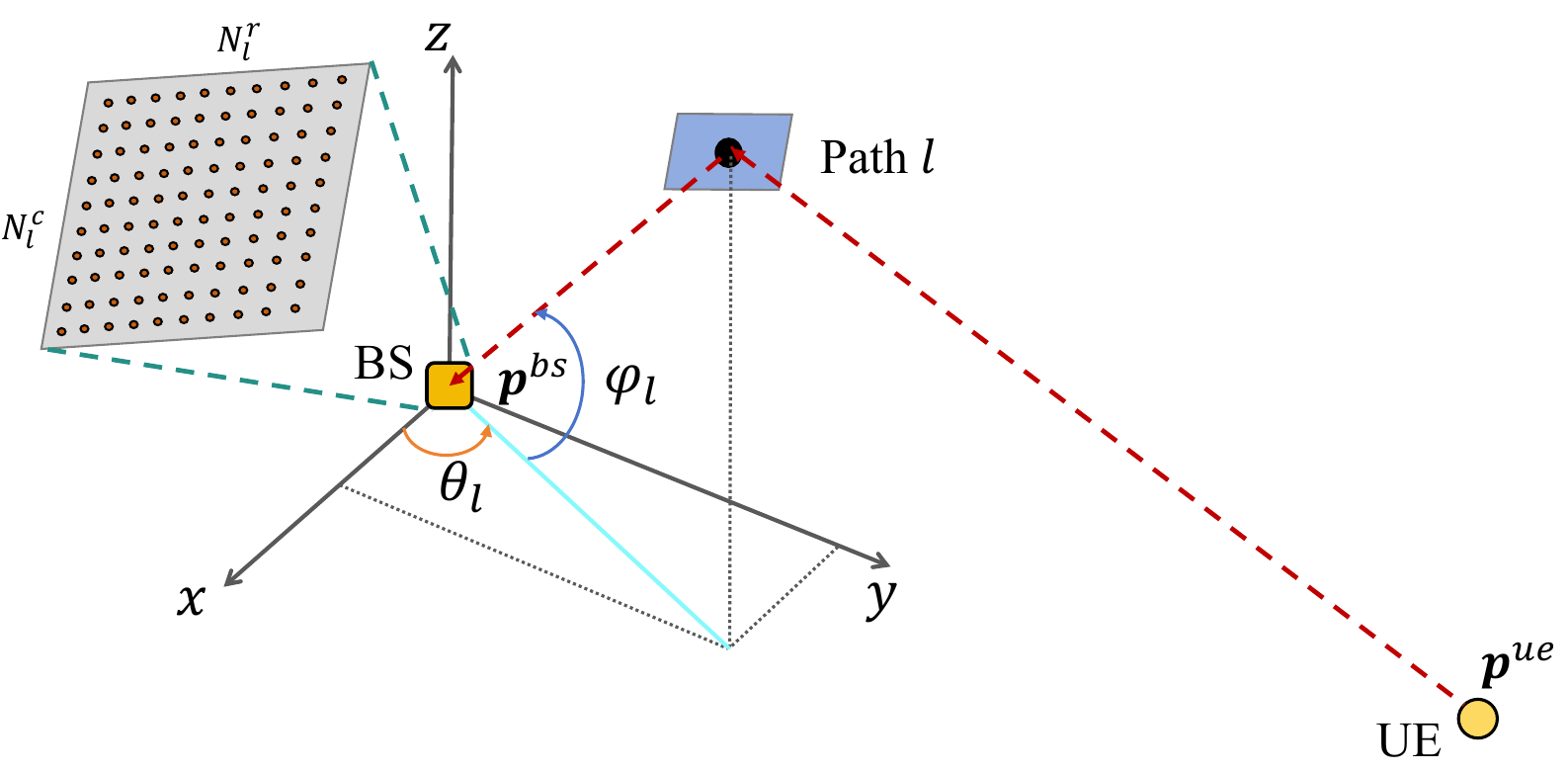}
\centering
\caption{MIMO-OFDM-based wireless positioning systems.}
\label{fig_csi}
\vspace{-4mm}
\end{figure}

As an illustrative example, we consider a MIMO-orthogonal frequency-division multiplexing (OFDM) system to model CSI in a wireless positioning system. As shown in Fig. \ref{fig_csi}, the BS is equipped with a uniform plane array, comprising $N_a$ antennas with $N_a^r$ in each row and $N_a^c$ in each column. The user equipment (UE) is equipped with a single omnidirectional antenna. The system contains $N_s$ subcarriers, each with a bandwidth of $B_s$. The locations of the BS and the UE can be represented as $\bm{p}^{ue}$ and $\bm{p}^{bs}$ respectively. The channel frequency response between the BS and the UE for subcarrier $s$ can be written as \cite{wu2021learning}
\begin{eqnarray}
\mathbf{H}_s=\sum_{l=1}^L\alpha_l\mathbf{a}(\theta_l,\varphi_l)e^{-j2\pi f_s\tau_l},
\end{eqnarray}
where $f_s$ is the subcarrier frequency, $\alpha_l$ and $\tau_l$ denote the complex gain and delay of path $l$, and $\mathbf{a}(\theta_l, \varphi_l)$ is the two-dimensional array response matrix. For the LOS path, $\tau_1$ and $(\theta_1, \varphi_1)$ represent the delay and angles-of-arrival (AOA), respectively.
Specifically, $\mathbf{a}(\theta_l, \varphi_l)$ can be expressed as
\begin{eqnarray}
    \mathbf{a}(\theta_l, \varphi_l) = \mathbf{a}_h(\theta_l, \varphi_l) \otimes \mathbf{a}_v(\varphi_l),
\end{eqnarray}
where $\mathbf{a}_h(\theta_l, \varphi_l)$ and $\mathbf{a}_v(\varphi_l)$ represent the elevation and azimuth components, respectively, and are defined as
\begin{eqnarray}
    \mathbf{a}_h(\theta_l, \varphi_l) & = & \left[1, e^{-j 2 \pi \frac{d^h}{\lambda} \sin \varphi_l \cos \theta_l}, \ldots, \right. \nonumber \\
    && \quad \left. e^{-j 2 \pi (M-1) \frac{d^h}{\lambda} \sin \varphi_l \cos \theta_l} \right]^\mathrm{T},  \\
    \mathbf{a}_v(\varphi_l) & = & \left[1, e^{-j 2 \pi \frac{d^v}{\lambda} \cos \varphi_l}, \ldots, \right. \nonumber \\
     && \quad \left. e^{-j 2 \pi (N-1) \frac{d^v}{\lambda} \cos \varphi_l} \right]^\mathrm{T}. 
\end{eqnarray}
Here, $\lambda$ is the wavelength, and $d^h$ and $d^v$ are the inter-antenna spacings in the horizontal and vertical directions, respectively, where $d^h = d^v = \frac{\lambda}{2}$. The full channel matrix in the frequency domain can then be expressed as $\mathbf{H} \triangleq [\mathbf{H}_1, \cdots, \mathbf{H}_s, \cdots, \mathbf{H}_{N_s}]$.

The channel frequency response (CFR) matrix $\mathbf{H}$ contains rich spatial and temporal information that directly relates to the physical characteristics of the propagation environment. Therefore, the research goal of wireless positioning is to determine a function $\mathcal{F}(\cdot)$ that can estimate an accurate position of the device based on the channel response $\mathbf{H}$. Therefore, the wireless positioning problem can be expressed as the following optimization problem:
\begin{align} 
\min_{\mathcal{F(\cdot)}} & \quad ||\hat{\bm{p}}^{ue} - \bm{p}^{ue}||_2^2, \\ 
\text{s.t.} & \quad \hat{\bm{p}}^{ue} = \mathcal{F}(\mathbf{H}), 
\end{align} 
where $\bm{p}^{ue}$ and $\hat{\bm{p}}^{ue}$ represent the true position and the estimated position, respectively. However, this is a challenging problem due to the complexity of the wireless propagation environment, which includes factors such as multipath propagation, noise, and NLOS conditions. Two main approaches exist:
\begin{itemize}
    \item Model-based: Model-based solutions extract geometric features (e.g., TOA, AOA) from CSI, then estimate the location via geometric relationships.
    \item Data-driven (AI-based): AI-based solutions use CSI directly as input to train models (e.g., deep learning) that learn the CSI-to-location mapping end-to-end.
\end{itemize}
In the following sections, we provide a detailed introduction to the principles of these methods, including their advantages, limitations, and typical application scenarios.


\begin{figure*}[h]
    \centering
    \captionsetup[subfigure]{justification=centering, font=small, skip=2pt} 
    \begin{subfigure}[t]{0.18\linewidth}
        \centering
        \includegraphics[scale=0.35]{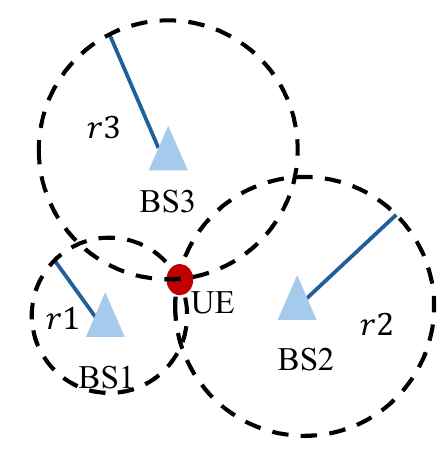}
        \caption{TOA-based positioning.}\label{fig:TOA}
    \end{subfigure}
    \begin{subfigure}[t]{0.18\linewidth}
        \centering
        \includegraphics[scale=0.35]{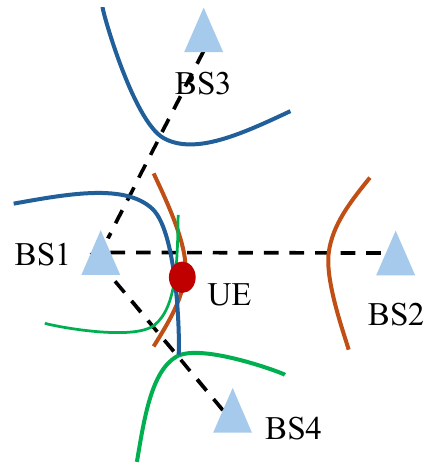}
        \caption{TDOA-based positioning.}\label{fig:TDOA}
    \end{subfigure}
        \begin{subfigure}[t]{0.18\linewidth}
        \centering
        \includegraphics[scale=0.34]{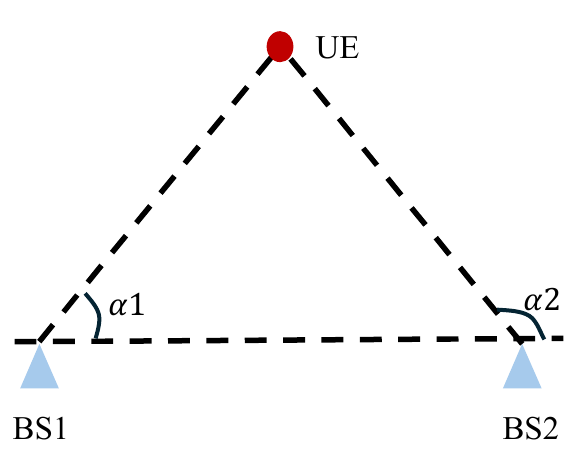}
        \caption{Angle-based positioning.}\label{fig:Angle}
    \end{subfigure}
    \begin{subfigure}[t]{0.4\linewidth}
        \centering
        \includegraphics[scale=0.32]{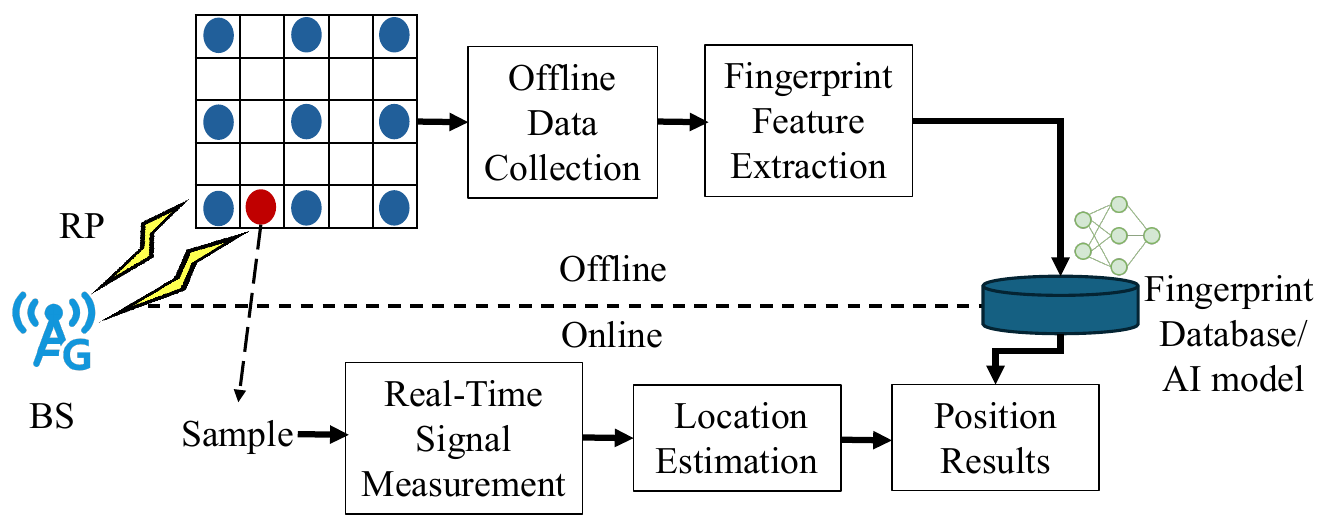}
        \caption{Fingerprint-based positioning.}\label{fig:fingerprint}
    \end{subfigure}
    \caption{Schematic diagram of positioning technology based on TOA, TDOA, angle and fingerprint.}
    \vspace{-3mm}
\end{figure*}

\subsubsection{Ranging-based Positioning}

High-precision positioning can be achieved by estimating the distances between a device and multiple BSs. With geometric relationships, accurate positioning is possible when distance estimations to at least three BSs are available. As shown in Fig. \ref{fig:TOA}, the intersection of circles representing these distances identifies the device's location. Adding more BSs and improving distance estimation accuracy further enhances positioning precision.

Among distance estimation methods, as previously mentioned, RSS-based techniques leverage the relationship between signal strength and distance \cite{Zanella_Best_2016, Choi_outdoor_2022, Coluccia_on_2018, Zanella_RSS_2014, shit_ai-enabled_2021}. However, their accuracy is limited by multipath propagation, interference, and environmental variability, making them unsuitable for high-precision applications.
\color{black} %
To address these limitations, TOA (also referred to as time-of-flight (TOF) \cite{Survey_italiano_tutorial_2024}) estimation offers a more accurate alternative \cite{Survey_italiano_tutorial_2024}. TOA measures the propagation delay of the LOS path, and the distance is calculated as $d = c \cdot t_{toa}$, where $t_{toa}$ is the signal propagation time, equal to the propagation delay of the LOS path $\tau_1$, and $c \approx 3 \times 10^8$ m/s is the speed of light. TOA can be estimated from CSI using signal processing algorithms \cite{8904430pilot, 9601204pilot, 9064586pilot, 7518638crlb}, expressed as:
\begin{equation}
t_{toa} = \mathcal{F}_{\text{toa}}(\hat{\mathbf{H}}).
\end{equation}
where $\mathcal{F}_{\text{toa}}(\cdot)$ is the estimation function and $\hat{\mathbf{H}}$ is the estimated channel. The estimation accuracy depends on signal bandwidth, signal-to-noise ratio (SNR), and sampling rate. 
According to the Cramer-Rao lower bound, the variance of the estimation error is inversely proportional to ${B^2 \cdot \text{SNR}}$ \cite{7518638crlb, 7501864crlb}. Advanced techniques, like multiple signal classification (MUSIC) \cite{MUSICToA} and estimation of signal parameters via rotational invariance techniques (ESPRIT) \cite{ESPRITToA}, can improve TOA accuracy, but their performance may degrade under multipath and dynamic environments. Moreover, TOA-based systems require tight clock synchronization between BS and UE, as clock offsets introduce systematic errors \cite{8636966toaSynchronization, 9465678toaSynchronization, RobusttoaSynchronization, choi_enhanced_2022} $t_{toa}^{\text{measured}} = t_{toa} + \Delta t_{clk} + e_{toa}$, where $\Delta t_{clk}$ and $e_{toa}$ represents the clock synchronization error and estimation error. To mitigate synchronization issues, round-trip time (RTT) estimation is often used, which eliminates the need for UE-BS synchronization by measuring total round-trip delay \cite{10472472RTT, 9151400RTT, 10493073RTT}. Since the RTT is calculated using the clock in the transmitter, it effectively cancels out the impact of clock misalignment.

Alternatively, when BSs are synchronized, TDOA offers another solution \cite{8964427TDOA, 7132722TDOA, 5664806TDOA}. By subtracting the TOA measurements from two BSs, TDOA inherently eliminates clock offset errors, thus relaxing the need for UE-side synchronization. Based on TDOA, the algorithms perform positioning via hyperbolic intersections, as shown in Fig.~\ref{fig:TDOA}. This method, also known as observed TDOA (OTDOA) in LTE/5G systems \cite{8030544OTDOA}, avoids the need for UE-side synchronization and benefits from increasing BS count and estimation accuracy.

\subsubsection{Angle-based positioning}

Angle-based positioning techniques estimate the location of a device by determining angular parameters, specifically the AOA \cite{7012090AOA, 8125734AOA, 9681696aoa} and angle-of-departure (AOD) \cite{10119056AOD, 10741287AOD, 10615899AOD}, using the spatial characteristics of signals received or transmitted by antenna arrays. AOA represents the incoming direction of a signal, typically estimated from uplink signals, while AOD refers to the signal's transmission direction, often derived from downlink. For 3D positioning, both elevation angle $\theta_1$ and azimuth angle $\phi_1$ of the LOS path are required, and can be extracted from CSI as:
\begin{equation}
[\theta_1, \phi_1] = \mathcal{F}_{\text{aoa}}(\hat{\mathbf{H}}).
\end{equation}
For 2D positioning, we only need to estimate $\theta_1$. The estimation accuracy is bounded by the Cramer-Rao lower bound (CRLB) \cite{4838824CRLB}:
\begin{equation} \text{Var}(\theta_1) \propto \frac{1}{ (2 \pi d_a / \lambda)^2  \text{SNR} \sin^2{\theta_1} N(N-1)(2N-1)}, \label{eq:crlb_aoa} 
\end{equation}
where $d_a$ is the inter-antenna spacing, $N$ is the number of antennas in the array. The estimation method of AOD is similar to that of AOA.

Similar to ranging-based methods, angle-based positioning infers location by intersecting spatial direction estimates from multiple BSs, as shown in Fig.~\ref{fig:Angle}. A simple method involves beam scanning, selecting the direction with maximum received power as the AOA. More advanced techniques, such as MUSIC \cite{9965430AOAmusic, 5618528AOAmusic} and ESPRIT \cite{1395953AOAesprit, 10295576AOAesprit}, provide higher angular resolution by exploiting signal subspace properties.
A key advantage of angle-based techniques is that they do not require time synchronization between BSs and the device, unlike TOA or TDOA. However, their performance depends heavily on antenna array resolution and the robustness of the estimation algorithm. While higher resolution improves accuracy, it also increases hardware and computational complexity.
\color{black}

\subsubsection{Fingerprint-based Positioning}

Fingerprint-based positioning establishes a mapping between wireless signal features and spatial locations, typically through an offline database or AI model training. A mapping function $\mathcal{F}_{\text{fp}}(\cdot)$ is learned from an offline dataset $D_{\text{offline}}=\{\bm{x}_i,\bm{p}_i\}$, where $\bm{x}_i$ denotes fingerprint features and $\bm{p}_i$ is the corresponding user location.

As shown in Fig.~\ref{fig:fingerprint}, this approach involves two phases: offline and online. In the offline phase, signal characteristics such as RSSI \cite{9718215RSSI, 9969867RSSI}, RSRP \cite{zhou2024indoor, 10556755RSRP, 8462098RSRP, 7997470RSRP, 10628101RSRP, 8555731RSRP}, CSI \cite{8673800FPToA, 9438641FPToA, 9259006FPToA, 10336870FPToA, 8187642FPToA}, or TOA \cite{9843909FPToA, 8939702FPToA} are collected at reference points (RPs) to build a fingerprint database or radio map. Features may be extracted from raw data or used directly, and either stored in a database or used to train an AI/ML model that approximates $\mathcal{F}_{\text{fp}}(\cdot)$. In the online phase, real-time measurements $\bm{x}$ from the UE are acquired. Position estimation is then performed by either matching against the fingerprint database or inputting $\bm{x}$ into the trained model, resulting in an estimated position $\hat{\bm{p}} = \mathcal{F}_{\text{fp}}(\bm{x})$.

Fingerprint-based positioning relies on learning from a historical database of wireless measurements but involves high deployment costs due to extensive data collection and maintenance. Adapting AI models to map signal features to locations demands significant computational resources and expertise. Additionally, dynamic wireless environments can degrade positioning accuracy over time due to data aging and environmental changes. This necessitates periodic updates to the fingerprint database or retraining of the AI model to maintain high accuracy levels.

\subsubsection{Channel Charting based Positioning}

Channel charting is a novel approach leveraging CSI to enable the pseudo-positioning of users by modeling relationships between different channels \cite{ferrand_wireless_2023}. It learns a mapping from CSI to a lower-dimensional channel chart, where distances in the latent space represent dissimilarity metrics between channels \cite{deng_multipoint_2018,medjkouh_unsupervised_2018,huang_improving_2019, yassine_optimizing_2024}. These distances indicate user proximity, where greater channel similarity implies closer placement on the channel chart \cite{ferrand_wireless_2023}. Besides positioning, channel charting techniques have been explored in various tasks, such as SNR prediction \cite{CC_SNR_Prediction}, pilot assignment and reuse \cite{CC_Pilot_allocation_1, CC_Pilot_allocation_2, CC_Pilot_Reuse}, beam tracking \cite{CC_Beam_Tracking}, and wireless resource optimization \cite{CC_Resource_Optimization}. In contrast to fingerprint-based methods, which depend on pre-collected fingerprints and external RPs to achieve absolute positioning, channel charting aims to uncover latent spatial relationships with minimal reliance on labeled data.

Channel charting, a form of manifold learning, encompasses classical methods such as multidimensional scaling (MDS), Sammon mapping \cite{studer2018channel}, and t-SNE \cite{aghajari_multi-point_2023}, alongside deep learning approaches like Siamese Neural Networks and triplet-based dimensionality reduction. In Siamese-based channel charting systems, dissimilarity metrics define relationships between CSI samples in the latent space. The channel charting learning process aims to ensure that latent space distances preserve the high-dimensional relationships captured by the dissimilarity metrics. A commonly used MDS-based loss function can be given by \cite{medjkouh_unsupervised_2018}
\begin{equation}
    \mathcal{L}_{MDS} = \sum_{i,j} w_{ij} \left( \| \mathbf{z}_i - \mathbf{z}_j \|_2 - d(\mathbf{H}_i, \mathbf{H}_j) \right)^2,
\end{equation}
where $\mathbf{z}_i$ and $\mathbf{z}_j$ are both the outputs of the neural network and the low-dimensional representations of the CSI samples $\mathbf{H}_i$ and $\mathbf{H}_j$. $d(\mathbf{H}_i, \mathbf{H}_j)$ is the dissimilarity metric in the original high-dimensional CSI space. $w_{ij}$ is a weight factor, often inversely proportional to $d(\mathbf{H}_i, \mathbf{H}_j)$, to prioritize preserving relationships for closer samples. This loss ensures that the distances in the latent space (\(\|\mathbf{z}_i - \mathbf{z}_j\|\)) align with the dissimilarity metrics derived from the high-dimensional CSI space, preserving spatial relationships. Two common measures of difference include:
\begin{enumerate}
    \item \textbf{Euclidean Distance} \cite{pihlajasalo_absolute_2020, agostini_not-too-deep_2022, agostini_federated_2022}:
    \begin{equation}
        d_E(\mathbf{H}_i, \mathbf{H}_j) = \|\mathbf{H}_i - \mathbf{H}_j\|_2,
    \end{equation}

    \item \textbf{Cosine Similarity}\cite{yassine_leveraging_2022, taner_channel_2023, stephan_angle-delay_2024}:
    \begin{equation}
        d_C(\mathbf{H}_i, \mathbf{H}_j) = 1 - \frac{\langle \mathbf{H}_i, \mathbf{H}_j \rangle}{\|\mathbf{H}_i\|_2 \cdot \|\mathbf{H}_j\|_2},
    \end{equation}
\end{enumerate}

Triplet-based dimensionality reduction is another approach \cite{yassine_leveraging_2022, ferrand_triplet-based_2021, taner_channel_2023, vindas_multi-site_2024}. In Triplet-based channel charting, relationships between CSI samples are learned through triplets \((\mathbf{H}_i, \mathbf{H}_j, \mathbf{H}_k)\). Each triplet consists of an anchor (\(\mathbf{H}_i\)), a positive sample (\(\mathbf{H}_j\)) that is similar to the anchor, and a negative sample (\(\mathbf{H}_k\)) that is dissimilar. The objective is to map these samples into a latent space:
\begin{equation}
\| \mathbf{z}_i - \mathbf{z}_j \|_2 < \| \mathbf{z}_i - \mathbf{z}_k \|_2,
\end{equation}
where $( \mathbf{z}_i, \mathbf{z}_j, \mathbf{z}_k )$ are the low-dimensional representations of \(\mathbf{H}_i, \mathbf{H}_j, \mathbf{H}_k\), respectively.

\begin{figure}[h]
    \centering
    \captionsetup[subfigure]{justification=centering, font=small, skip=2pt} 
    \begin{subfigure}[t]{\linewidth}
        \centering
        \includegraphics[scale=0.28]{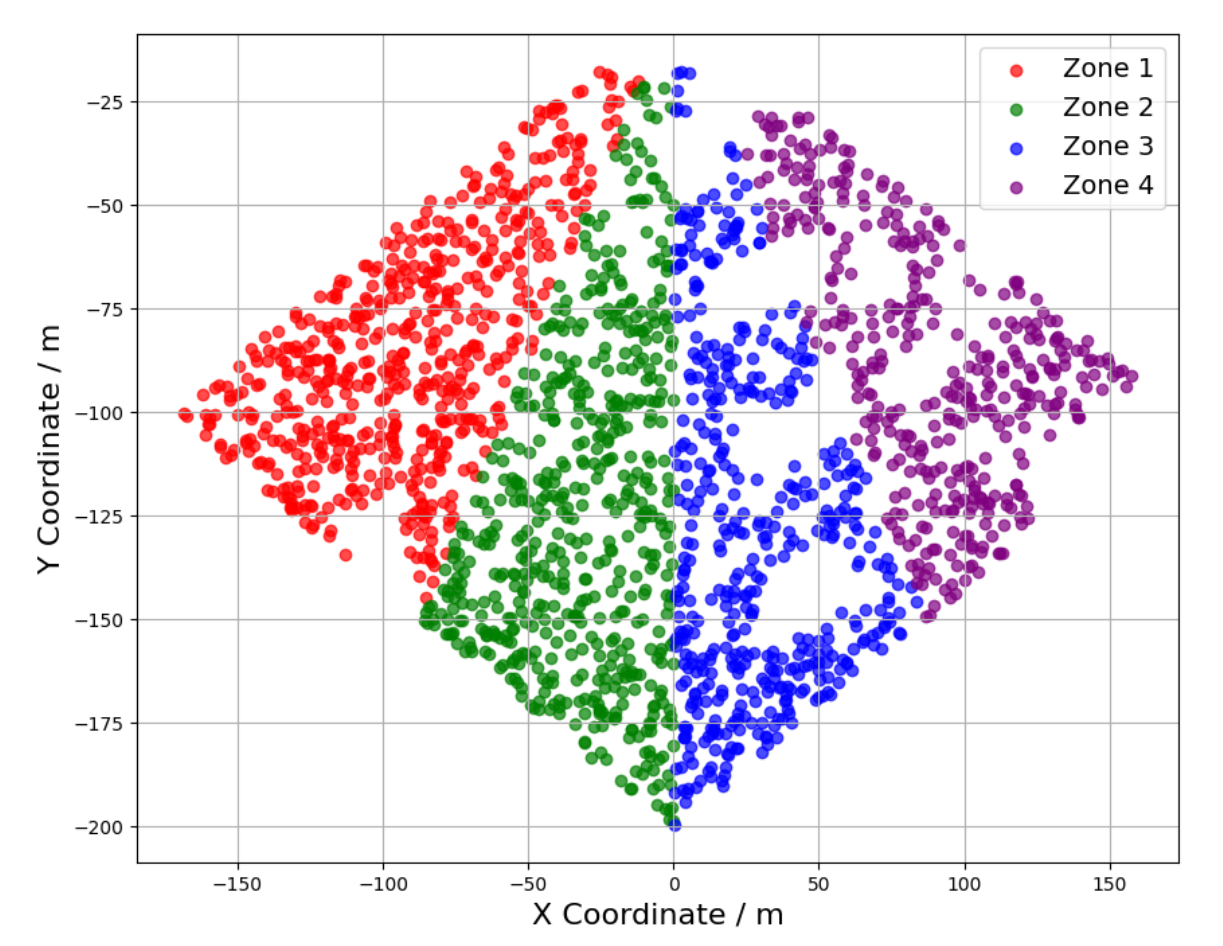}
        \caption{The actual physical space distribution of sampling points.}
    \end{subfigure}
    \begin{subfigure}[t]{\linewidth}
        \centering
        \includegraphics[scale=0.28]{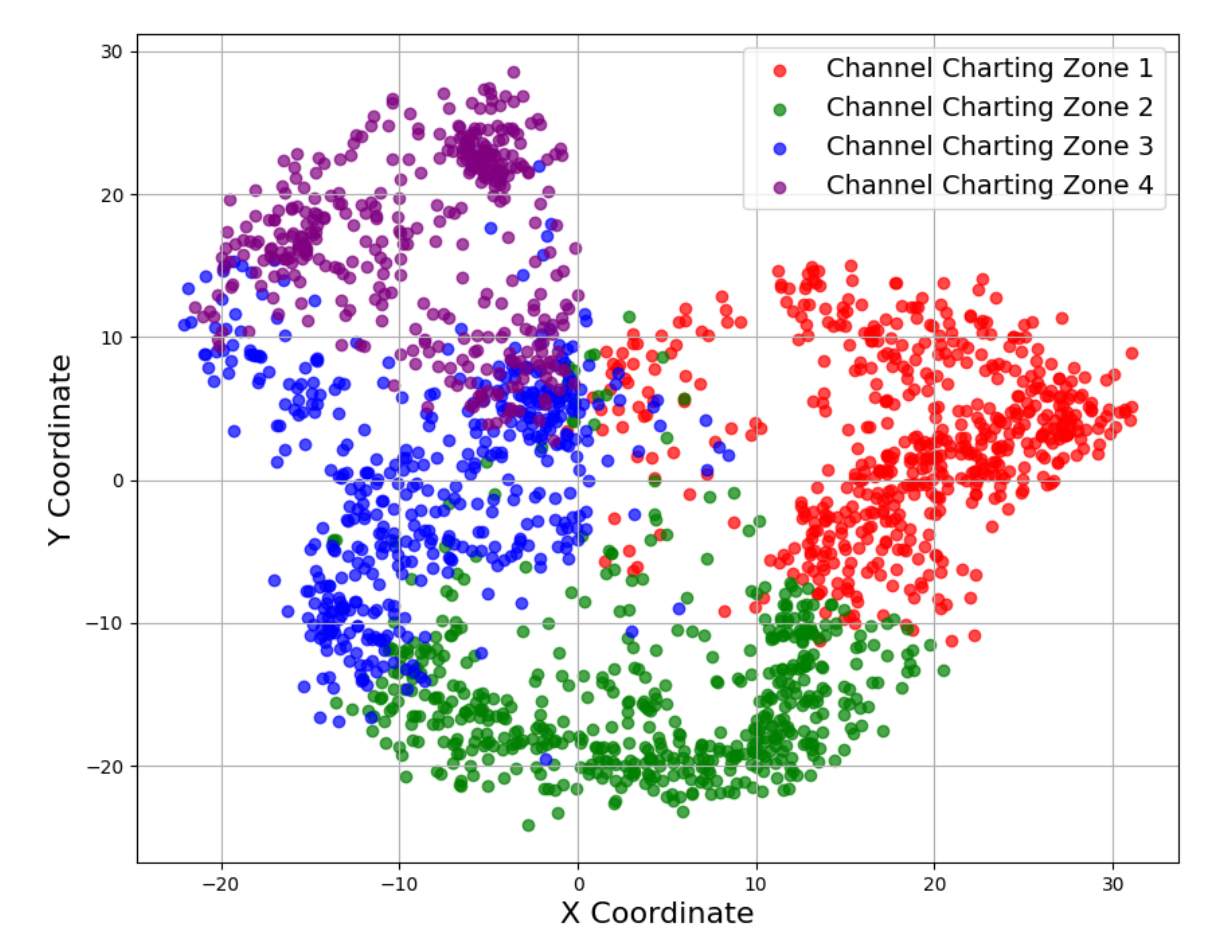}
        \caption{Latent space distribution of sampling points.}
    \end{subfigure}
    \caption{Schematic diagram of channel charting. The relative proximity in the latent space correlates with the physical proximity of the devices.}
    \label{fig_channel_charting}
    \vspace{-2mm}
\end{figure}

The corresponding triplet loss function can be written as \cite{ferrand_triplet-based_2021}
\begin{equation}
\mathcal{L}_{\text{triplet}} = \sum_{i,j,k} \max\left( 0, \| \mathbf{z}_i - \mathbf{z}_j \|_2^2 - \| \mathbf{z}_i - \mathbf{z}_k \|_2^2 + \delta \right),
\end{equation}
where $ \delta$ is a margin that enforces a minimum separation between positive and negative pairs. As shown in Fig.~\ref{fig_channel_charting}, the left figure illustrates the physical space distribution of multiple sampling channels. In contrast, the right figure demonstrates the distribution of these points in the latent space after applying the channel charting approach. The relative proximity in the latent space correlates with the physical proximity of the devices.

The primary advantage of channel charting-based positioning lies in its ability to operate without relying on a large number of reference points RPs, significantly reducing costs and complexity. Furthermore, channel charting continuously benefits from newly acquired channel information rather than being limited by static historical databases \cite{wang_how_2022}. However, channel charting based positioning techniques rely heavily on the quality of the low-dimensional mapping, and there is often no universal metric to evaluate the effectiveness of these mapping algorithms \cite{ferrand_wireless_2023}. Additionally, considering the highly nonlinear relationship between channel-based Dissimilarity metrics and physical distances, designing good metrics to improve positioning accuracy remains a significant challenge.

\subsection{Motivation for AI in Positioning}

\begin{table*}[ht]
\centering
\caption{Comparative Analysis of Model-Based and AI-Driven Positioning Techniques}
\label{tab:Model_vs_AI_Positioning}
\renewcommand{\arraystretch}{1.5} 
\begin{tabular}{|p{2.4cm}|p{2.9cm}|p{5.5cm}|p{5.5cm}|}
\hline
\textbf{Positioning Method} & \textbf{Key Procedure} & \textbf{Challenges in Model-Based Positioning} & \textbf{Advantages of AI-Driven Positioning} \\
\hline
\multirow{4}{*}{\makecell[l]{\vspace{1ex}Ranging/Angle- \\based Positioning \vspace{1ex}}} 
& Parameter Estimation & 
Model-based estimation methods are prone to large errors under NLOS, multipath and hardware impairments.  & 
AI technologies learn multipath and hardware impairment patterns from data to improve estimation accuracy. \\
\cline{2-4}
& Geometry-based Position Estimation& 
Geometric algorithms are sensitive to measurement errors. It also requires prior error statistics, limited adaptability in dynamic environments. & 
Adaptively learns uncertainties; dynamically adjusts BS contributions without explicit modeling. \\
\hline
\multirow{4}{*}{\makecell[l]{\vspace{1ex}Fingerprint-based \\ Positioning\vspace{1ex}}} 
& Database Construction & 
It relies on low-dimensional RSSI features with poor discriminative power, leading to weak generalization. It also has difficulty in handling fingerprint aging and environmental changes.
 & 
AI models can map high-dimensional fingerprints (e.g., CSI) to locations with fine-grained features and robustly adapt to fingerprint aging and environmental changes via fine-tuning or transfer learning. AI algorithms also support radio map augmentation based on generative modeling. \\
\cline{2-4}
& Fingerprint Matching & 
Traditional algorithms such as KNN are vulnerable to noise, environmental variability, and database sparsity; matching complexity increases with database size.& 
DL-based AI models learn complex non-linear mappings between fingerprints and locations to improve robustness and accuracy; inference speed remains constant as database size grows. \\
\hline
\multirow{2}{*}{\makecell[l]{  Channel Charting-\\ based Positioning\vspace{1ex}}} 
& Channel Embedding & 
Model-based channel charting struggles with the high-dimensional and non-linear nature of CSI data, making feature extraction difficult.
 & 
AI-driven channel charting employs self-supervised learning to learn embeddings that better preserve spatial relationships, resulting in smoother, more continuous, and more accurate channel charts.
 \\
\hline
\end{tabular}
\vspace{-2mm}
\end{table*}

Although model-based wireless positioning algorithms have been extensively studied for decades, solely relying on handcrafted models still faces significant limitations in complex real-world environments. Building upon the previous discussion of various positioning technologies, this subsection analyzes the inherent challenges encountered by model-based positioning methods and discusses how AI-driven approaches address these issues effectively. A comparative summary is provided in Table.~\ref{tab:Model_vs_AI_Positioning}.

\subsubsection{Ranging-based and Angle-based Positioning}
As mentioned above, in ranging-based and angle-based positioning methods, the localization process typically involves two key stages. First, the signal parameters, such as distance (TOA, TDOA) or angle (AOA, AOD), are estimated. Second, the user's location is determined based on the estimated parameters using a geometry-based positioning algorithm. We analyze the challenges in each of these stages as follows:
\begin{itemize}
    \item \textbf{Parameter Estimation:}  Accurate parameter estimation between BSs and UEs is critical for ranging-based and angle-based positioning. However, under NLOS conditions, the absence of a direct propagation path causes severe biases in the measured parameters, leading to substantial estimation errors \cite{Survey_del_peral-rosado_survey_2009}. Even in LOS scenarios, multipath propagation introduces interference between the direct path and other paths, leading to ambiguities in signal peak detection that severely degrade estimation accuracy \cite{wang2020multipath}. Additionally, hardware impairments at the transmitter or receiver can further degrade estimation accuracy \cite{ozturk2022impact}. AI-based techniques can learn to recognize multipath features and hardware impairment patterns from data, thereby enabling more accurate and robust parameter estimation even in challenging environments \cite{bayraktar_ris-aided_2024, ozturk2022impact}.

    \item \textbf{Geometry-based Position Estimation:} 
    Based on the estimated parameters, model-based geometric algorithms, such as triangulation, are used to infer the user's position. However, inaccuracies in distance or angle measurements can significantly amplify positioning errors. While some model-based methods attempt to mitigate these errors using techniques such as weighted least squares \cite{Survey_del_peral-rosado_survey_2009}, they heavily rely on prior knowledge of error statistics, which is difficult to obtain and maintain in dynamic real-world environments. AI-driven positioning approaches can adaptively learn the uncertainty in parameter estimation, such as identifying NLOS links or weighting the reliability of BSs based on measured channel features, thereby achieving more robust and accurate localization.
\end{itemize}

\subsubsection{Fingerprint-based Positioning}
Fingerprint-based positioning first collects wireless fingerprints at known RPs and constructs an offline fingerprint database. During the online positioning phase, the system matches the current signal measurements against the database to determine the UE's location. In the following, we analyze the limitations of model-based approaches in the database construction and fingerprint matching steps, and describe how AI technologies address these challenges:
\begin{itemize}
    \item \textbf{Database Construction:}  
    Under model-based algorithms, fingerprinting relies on low-dimensional wireless measurements (typically RSSI) at known RPs and interpolating or fitting a radio map over the target area. Due to the limited dimensionality and weak descriptive power of RSSI features, the constructed radio maps often suffer from poor generalization and low robustness \cite{lee_voronoi_2012}. Furthermore, model-based approaches have difficulty in handling problems of fingerprint aging and environmental changes in wireless propagation environments \cite{Survey_jang_indoor_2019}.  
    In comparison, AI models can directly learn the mapping between high-dimensional fingerprints (e.g., CSI matrices) and locations, eliminating the need to construct and interpolate complex fingerprint databases. This enables the algorithm to extract more fine-grained spatial features from high-dimensional CSI measurements to assist positioning \cite{sun2019fingerprint,sun2018single,wang2016csi}. Building on this capability, AI methods also exhibit strong resilience to fingerprint aging and environmental dynamics through fine-tuning or transfer learning, significantly reducing the need for re-acquisition of fingerprints when the deployment environment changes \cite{LOS_sun_channel_2023,xiang_self-calibrating_2021,yang_updating_2021}. Moreover, AI models can assist in constructing enhanced radio maps. Specifically, generative AI models \cite{zhang_rme-gan_2023,hu20233d} can capture multi-dimensional relationships across time, space, and frequency domains, allowing for channel extrapolation to augment the radio map, thereby significantly improving positioning performance across wider spatial and spectral ranges \cite{zhang_rme-gan_2023, yang_updating_2021,hu20233d}.

    \item \textbf{Fingerprint Matching:}  
    Traditional fingerprint matching typically employs simple algorithms such as k-nearest neighbors (KNN) or weighted KNN, which compute similarities between observed and stored fingerprints \cite{wang2020novel,10556755RSRP}. These methods are vulnerable to noise, environmental variability, and database sparsity, leading to degraded positioning accuracy. Moreover, as the size of the offline database grows, the computational complexity of traditional matching methods increases significantly. DL-based AI models can replace conventional matching by learning complex, non-linear mappings between fingerprint features and spatial locations, where the inference speed remains constant regardless of the size of the offline database. Neural network models can robustly extract high-level features from noisy fingerprints, achieving superior localization performance and robustness.
\end{itemize}

\subsubsection{Channel Charting-based Positioning}
Model-based channel charting struggles with the high-dimensional and non-linear nature of CSI data, making feature extraction difficult. In contrast, AI-driven channel charting addresses these limitations by employing self-supervised learning techniques (e.g., Siamese networks \cite{medjkouh_unsupervised_2018}, triplet loss \cite{ferrand_triplet-based_2021}) to learn embeddings that better preserve true spatial relationships among devices, resulting in smoother, more continuous, and more accurate channel charts even in complex propagation environments.

\subsection{Lessons Learned}

Wireless positioning holds tremendous potential across a wide range of application domains. Although existing model-based positioning algorithms have made significant contributions, achieving high-precision localization in real-world environments remains challenging due to the inherent complexity, dynamics, and unpredictability of wireless propagation. Ranging-based, angle-based, fingerprint-based, and channel charting-based positioning techniques have each advanced the field from different perspectives and have laid a solid foundation for AI-driven wireless positioning. However, these techniques also face critical limitations when deployed in complex and dynamic environments. To address the various challenges inherent in positioning tasks, AI models offer a new paradigm by learning complex high-dimensional mappings directly from wireless measurements. AI-driven positioning technologies have demonstrated significant potential in enhancing localization accuracy, robustness, and scalability. This motivates the integration of AI into wireless positioning systems to meet the stringent requirements of next-generation applications.

\color{black}

\section{Fundamentals of AI Technology} \label{sec:AI}
Advanced AI models and algorithms play a critical role in enhancing the accuracy and efficiency of AI-driven wireless positioning systems. In this section, we will provide a review of classic AI models and algorithms.

\subsection{Overview of AI Models}

In the realm of AI, neural network architectures play a pivotal role in enabling machines to learn from data and make predictions or classifications. This subsection introduces four mainstream neural network models commonly applied in AI-driven wireless positioning systems: fully-connected neural networks (FCNNs), convolutional neural networks (CNNs), long short-term memory networks (LSTMs), and Transformers. As these architectures have been extensively covered in existing literature \cite{Survey_burghal_comprehensive_2020,jiao2024advanced}, we do not elaborate on their internal mechanisms again in this survey. Instead, this subsection focuses on analyzing the computational complexity of each model and discusses its respective applicability to different wireless positioning scenarios. A detailed comparative summary is presented in Table.~\ref{tab:AI_model_comparison}.

\begin{table*}[t]
\centering
\renewcommand\arraystretch{1.4} 
\caption{Comparison of Different AI Models for Wireless Positioning}
\label{tab:AI_model_comparison}
\begin{tabular}{|m{1.3cm}|m{2.80cm}|m{3.6cm}|m{4.0 cm}|m{4.1cm}|}
\hline
\textbf{Model} & \textbf{Per-Layer Complexity} & \textbf{Advantages} & \textbf{Disadvantages} & \textbf{Suitable Positioning Tasks} \\ \hline

FCNN & $O(d^{\text{out}}_{\text{FCNN}} d^{\text{in}}_{\text{FCNN}})$ & Simple architecture; efficient for low-dimensional features. & Large number of parameters; prone to overfitting; limited capability in feature extraction. & Position estimation based on small-scale features (e.g., RSSI, RSRP); mapping from estimated parameters to coordinates. \\ \hline

CNN & 
$\begin{aligned}
O(&H_{\text{out}} W_{\text{out}} C_{\text{in}}\\
 & C_{\text{out}} H_{\text{Kernel}} W_{\text{Kernel}})
\end{aligned}$
& Strong at local feature extraction; enables efficient parameter sharing & Limited capability in modeling global dependencies. & Process high-dimensional wireless signals (e.g., CSI); build radio maps for fingerprint-based localization systems. \\ \hline

LSTM & $\begin{aligned} O( & T  \! \times  \! ({d^{\text{in}}_{\text{LSTM}}} + {d^{\text{out}}_{\text{LSTM}}}) \\ & \times {d^{\text{out}}_{\text{LSTM}}}) \end{aligned}$ & Capture long-term dependencies and dynamically adapts to sequential inputs. & Difficult to parallelize; high training and inference complexity. & Handle time-domain CIRs; model dynamic CSI; track device trajectories. \\ \hline

Transformer & $O(T^2 d_{\text{Trans}})$ & Strong at modeling global dependencies; effective in integrating multi-source sequences. & High computational and memory costs for long sequences; requires large datasets for effective training. & Suitable for large-scale CSI modeling, dynamic environment positioning, and multi-modal data fusion. \\ \hline

\end{tabular}
\end{table*}

\color{black}

\subsubsection{FCNNs} 

FCNNs are foundational architectures where each neuron in a layer is connected to all neurons in the next layer~\cite{cite:mlp-1, cite:mlp-2, sze_efficient_2017}. Denote $d^{\text{in}}_{\text{FCNN}}$ and $d^{\text{out}}_{\text{FCNN}}$ as the number of input and output neurons, respectively. The computational complexity of an FCNN layer is $O(d^{\text{out}}_{\text{FCNN}} \times d^{\text{in}}_{\text{FCNN}})$. FCNNs are efficient for small-scale feature extraction and perform well in simple regression and classification tasks. However, their large number of parameters and limited feature extraction capability often lead to overfitting, particularly when training data is insufficient or the underlying data structure is highly complex. Therefore, neural networks based on FCNNs are typically employed for position regression using small-scale localization features (e.g., RSSI, RSRP) \cite{9969867RSSI, adege2018applying}, or to approximate mappings from estimated signal parameters to position estimates \cite{pan2020deep}.

\subsubsection{CNNs} 

CNNs are feedforward architectures based on convolutional operations, widely used for spatial feature extraction~\cite{cite:cnn-1, cite:cnn-2}.  
Given an input feature map $\bm{x}_{\text{CNN}} \in \mathbb{R}^{C_{\text{in}} \times H_{\text{in}} \times W_{\text{in}}}$, a convolutional layer computes an output feature map $\bm{y}_{\text{CNN}} \in \mathbb{R}^{C_{\text{out}} \times H_{\text{out}} \times W_{\text{out}}}$ using $C_{\text{out}}$ convolutional kernels. $C_{\text{in}}, H_{\text{in}}, W_{\text{in}}$ denote the number of input channels, the height, and the width of the input feature map, respectively, and $C_{\text{out}}$, $H_{\text{out}}$, $W_{\text{out}}$ represent the corresponding dimensions of the output feature map generated by $C_{\text{out}}$ convolutional filters. 
Each kernel has a size of $C_{\text{in}} \times H_{\text{Kernel}} \times W_{\text{Kernel}}$, and the computational complexity per layer is $O(H_{\text{out}} W_{\text{out}} C_{\text{in}} C_{\text{out}} H_{\text{Kernel}} W_{\text{Kernel}})$.
CNNs significantly reduce the number of parameters compared to FCNNs by exploiting local connectivity and weight sharing, making them highly efficient for high-dimensional data. However, their local feature extraction mechanism limits their ability to capture long-range dependencies. In wireless positioning tasks, CNNs can effectively process high-dimensional wireless signals, such as CSI matrices \cite{mylonakis_novel_2024,NB-2020-Pan,LOS_si_lightweight_2023,li2023automatic}. For fingerprint-based localization, CNNs can also facilitate radio map construction for localization tasks \cite{RadioUNet2021Levie}.

\subsubsection{LSTMs} 
LSTM networks~\cite{cite:lstm-1, cite:lstm-2} are a class of recurrent neural networks (RNNs). They are designed to model sequential data by introducing gated mechanisms to regulate information flow. Given an input sequence $\{\bm{x}_t\}_{t=1}^T$ with $\bm{x}_t \in \mathbb{R}^{d^{\text{in}}_{\text{LSTM}}}$, an LSTM layer produces hidden states $\{\bm{y}_t\}_{t=1}^T$ with $\bm{y}_t \in \mathbb{R}^{d^{\text{out}}_{\text{LSTM}}}$, where $d^{\text{in}}_{\text{LSTM}}$ and $d^{\text{out}}_{\text{LSTM}}$ denote the input and hidden dimensions, respectively, and $T$ is the sequence length. The computational complexity per layer is $O(T ({d^{\text{in}}_{\text{LSTM}}} + {d^{\text{out}}_{\text{LSTM}}}){d^{\text{out}}_{\text{LSTM}}})$. An LSTM layer employs a forget gate, input gate, and output gate to selectively discard irrelevant past information, incorporate new inputs, and generate updated hidden states. These mechanisms enable LSTMs to capture long-term dependencies while dynamically adapting to new inputs. However, due to their sequential nature, LSTMs are inherently difficult to parallelize across time steps, resulting in high overall training and inference complexity compared to feedforward architectures. Beyond LSTM, alternative models such as gated recurrent units (GRUs)~\cite{cite:gru-dey2017gate}, bidirectional RNNs (BiRNNs)~\cite{cite:biLSTM-graves2013hybrid}, and temporal convolutional networks (TCNs)~\cite{cite:TCNlea2017temporal} have been proposed for sequence modeling. GRU offers efficiency with fewer parameters, BiRNN captures bidirectional dependencies, and TCN achieves scalable performance by replacing recurrence with causal convolutions. Sequence-based neural networks are widely applied in wireless positioning tasks, particularly for processing time-domain channel impulse responses (CIRs), modeling dynamic channel \cite{chen_deep_2023}, and tracking device trajectories \cite{wu_diffractive_2022}.

\subsubsection{Transformer} 

Transformers are neural network architectures based on attention mechanisms, and have revolutionized AI across multiple domains~\cite{cite:transformer-1, cite:transformer-2, Jonathan_Estimating_2024}. Given an input sequence $\{\bm{x}_i\}_{i=1}^T$ with $\bm{x}_i \in \mathbb{R}^{d_{\text{Trans}}}$, a Transformer layer produces output representations $\{\bm{z}_i\}_{i=1}^T$ with $\bm{z}_i \in \mathbb{R}^{d_{\text{LSTM}}}$, where $T$ is the sequence length and $d_{\text{LSTM}}$ is the feature dimension. The computational complexity of the self-attention operation per layer is $O(T^2 d_{\text{LSTM}})$, which imposes significant resource demands, particularly for long sequences~\cite{cite:transformer-4-lu2021soft}. The core component is the self-attention mechanism, which assigns dynamic weights to different positions in the input sequence, enabling the model to capture global dependencies~\cite{cite:transformer-2}. Furthermore, Transformers require large datasets for effective training, which may limit their efficiency when applied to small-scale datasets or simpler tasks. In wireless positioning tasks, Transformers can capture global relationships from channel data, leading to improved localization performance \cite{salihu_self-supervised_2024, xu2024swin,cho2024transformer}. They can also handle variable-length signal sequences and integrate multi-source information \cite{Sensing2024Cui}, making them well-suited for diverse wireless network configurations.

\subsection{Overview of AI Algorithms}

\begin{table*}[htbp]
\centering
\caption{Comparison of Different Learning Techniques for Wireless Positioning}
\label{tab:learning_comparison}
\renewcommand{\arraystretch}{1.4}
\begin{tabular}{|p{2.8cm}|p{4.4cm}|p{4.2cm}|p{4.8cm}|}
\hline
\textbf{Learning Technique} & \textbf{Advantages} & \textbf{Limitations} & \textbf{Suitable Positioning Tasks} \\
\hline
{Transfer Learning} &
Reduce labeling effort; facilitates domain adaptation. &
Sensitive to domain shifts; prone to overfitting with limited target data. &
Adaptation of pretrained positioning models from simulations to real-world environments or across different scenarios. \\
\hline
{Meta Learning} &
Enable fast adaptation from few samples; promotes task generalization. &
Computationally intensive; performance depends on task formulation. &
Few-shot indoor positioning in unseen environments or new deployment locations. \\
\hline
{Continual Learning} &
Enable long-term adaptation; eliminates the need for full retraining. &
Prone to catastrophic forgetting; increased model complexity. &
Online adaptation of positioning models in dynamic environments, e.g., layout changes or varying user densities. \\
\hline
GANs &
Facilitate data augmentation; addresses class imbalance. &
Unstable training; susceptible to mode collapse. &
Generation of synthetic wireless signal data to enhance model training under limited labeled data conditions. \\
\hline
\end{tabular}
\end{table*}

In AI-driven wireless positioning, let $\bm{\theta}$ denote the parameters of the neural network. Given a training dataset $\mathcal{D}_{\text{train}} = \{(\bm{x}_i, {\bm{p}}^{\text{ue}}_i)\}$, where $\bm{x}_i$ is the input wireless signal and $\bm{p}^{\text{ue}}_i$ is the position of the UE, the objective is to minimize the positioning error across the dataset. Accordingly, a general positioning loss function can be formulated as
\begin{align}
    \min_{\bm{\theta}} \; \mathbb{E}_{\{(\bm{x}_i, {\bm{p}}^{\text{ue}}_i)\} \sim \mathcal{D}_{\text{train}}} \ell(\mathcal{F}(\bm{x}_i; \bm{\theta}), {\bm{p}}^{\text{ue}}_i)
    \label{equ:train}
\end{align}
where $\ell(\cdot, \cdot)$ denotes a positioning loss function, e.g., Euclidean distance between the predicted and true positions. $\mathcal{F}(\cdot,\bm{\theta})$ represents the AI model for position estimation with parameters $\bm{\theta}$. To learn optimal $\bm{\theta}$ that enhance the positioning performance of AI models in highly dynamic wireless environments, a variety of learning paradigms, such as transfer learning, meta learning, continual learning, and generative adversarial networks (GANs), have been extensively studied. In this subsection, we introduce these AI algorithms commonly used in wireless positioning to highlight their roles and advantages in improving localization performance. A comparison of different learning techniques for wireless positioning is presented in Table.~\ref{tab:learning_comparison}.

\subsubsection{Transfer Learning} 
Transfer learning leverages knowledge from a source domain to improve learning in a related target domain, especially when labeled data in the target domain is scarce \cite{cite:transfer-1, cite:transfer-2}. In wireless positioning, it enables models trained on simulated or related environments to generalize to new scenarios, reducing data collection costs. However, its effectiveness depends on the similarity between source and target domains. Large domain shifts may hinder performance, and small target datasets can cause overfitting during fine-tuning. The fine-tuning process can be formulated as
\begin{align}
    \min_{\bm{\theta}_{\text{target}}} &\quad \mathbb{E}_{\{(\bm{x}_i, {\bm{p}}^{\text{ue}}_i)\} \sim \mathcal{D}_{\text{target}}} \ell(\mathcal{F}(\bm{x}_i; \bm{\theta}_{\text{target}}(\bm{\theta}_{\text{source}})), {\bm{p}}^{\text{ue}}_i),
\end{align}
where $\bm{\theta}_{\text{source}}$ denotes the parameters learned from the source domain, $\bm{\theta}_{\text{target}}$ denotes the parameters for the target domain. $ \bm{\theta}_{\text{target}}(\bm{\theta}_{\text{source}}))$ means $\bm{\theta}_{\text{target}}$ is initialized using $\bm{\theta}_{\text{source}}$. During fine-tuning, $\bm{\theta}_{\text{target}}$ is further updated to minimize the loss on the target task using the target domain data $\mathcal{D}_{\text{target}}$.

\subsubsection{Meta Learning}
Meta learning, or “learning to learn,” trains models across a distribution of tasks so that they can quickly adapt to new tasks using only a few labeled samples \cite{metasurvey}. In wireless positioning, it enables rapid adaptation to new environments, users, or deployment conditions, making it suitable for dynamic scenarios \cite{gao2023metaloc,pu_bayesian_2024}. Its strengths lie in few-shot adaptability, but its performance is sensitive to task design and may suffer from high computational cost. A typical formulation is
\begin{align}
    \!  \!  \! \min_{\bm{\theta}_{\text{meta}}} & \
    \mathbb{E}_{\mathcal{T} \! \sim  \! p(\mathcal{T})}
    \mathbb{E}_{\{(\bm{x}_j, \bm{p}_j^{\mathrm{ue}})\} \sim \mathcal{D}_\mathcal{T}^{\text{query}}}
     \! \ell\big(\mathcal{F}(\bm{x}_j, \bm{\theta}_\mathcal{T}^{*}(\bm{\theta}_{\text{meta}})),\, \bm{p}_j^{\mathrm{ue}}\big) \\
    \text{s.t.} 
    & \ \bm{\theta}_\mathcal{T}^{*}  \! =  \! \arg  \! \min_{\bm{\theta}_\mathcal{T}}  \! \mathbb{E}_{\{(\bm{x}_i, \bm{p}_i^{\mathrm{ue}})\} \sim \mathcal{D}_\mathcal{T}^{\text{support}}}
 \ell\big(\mathcal{F}(\bm{x}_i, \bm{\theta}_\mathcal{T}(\bm{\theta}_{\text{meta}})),\,  \! \bm{p}_i^{\mathrm{ue}}\big), \nonumber
\end{align}
where $\bm{\theta}_\mathcal{T}$ and $\bm{\theta}_{\text{meta}}$ are the task-specific parameters and meta-parameters, respectively. $\mathcal{D}_\mathcal{T}^{\text{support}}$ and $\mathcal{D}_\mathcal{T}^{\text{query}}$ denote the support set and query set for task $\mathcal{T}$, respectively. Each task $\mathcal{T}$ is sampled from a task distribution $p(\mathcal{T})$. The task-adapted parameters $\bm{\theta}_T^*(\bm{\theta}_{\text{meta}})$ are obtained via a few steps of gradient descent on the support set, initialized from the meta-parameters $\bm{\theta}_{\text{meta}}$.

\subsubsection{Continual Learning} 
Continual learning \cite{cite:Continuous-3, cite:Continuous-4}, also known as lifelong learning \cite{cite:Continuous-5} or incremental learning \cite{cite:Continuous-1, cite:Continuous-2}, enables models to incrementally learn from a sequence of tasks or data distributions without forgetting previously acquired knowledge \cite{cite:Continuous-1, cite:Continuous-3, cite:Continuous-5}. Unlike transfer learning, its goal is not just adaptation but long-term knowledge accumulation. In wireless positioning, continual learning allows models to adapt to evolving network conditions, environments, or hardware upgrades without retraining from scratch. It supports continual updates in dynamic systems, but may suffer from catastrophic forgetting or increased computational cost due to the need for memory or regularization mechanisms. Formally, assume that a sequence of datasets $\{ \mathcal{D}^t_{\text{cont}} = \{(\bm{x}_i, \bm{p}^{\text{ue}}_i)\} \}_{t=1}^T$ becomes available over time, where $T$ denotes the total number of continual learning steps (or tasks). Let $\mathcal{D}^t_{\text{cont}}$ represent the dataset used at the $t$-th update. The model learns parameters $\bm{\theta}_t$ by minimizing a continual learning loss while preserving knowledge from previous updates:
\begin{align}
\min_{\bm{\theta}_t} & \quad \mathbb{E}_{(\bm{x}_i, \bm{p}_i^{\text{ue}}) \sim \mathcal{D}^t_{\text{cont}}}
 \ell(\mathcal{F}(\bm{x}_i; \bm{\theta}_t), \bm{p}_i^{\text{ue}}) \nonumber \\ 
& \quad + \lambda \cdot \Omega(\bm{\theta}_t, \bm{\theta}_{1:t-1})
\end{align}
where $\Omega(\cdot)$ is a regularization term that mitigates catastrophic forgetting by constraining parameter updates based on previously learned tasks, and $\lambda$ is a hyperparameter balancing stability and plasticity.

\subsubsection{Generative Adversarial Networks} 

GANs consist of a generator and a discriminator trained in opposition: the generator creates synthetic data, while the discriminator distinguishes between real and fake samples \cite{cite:gan-1, cite:gan-2}. This adversarial process enables GANs to learn to generate realistic wireless signal data from position inputs. In wireless positioning, GANs are effective for data augmentation, especially when labeled data is scarce or hard to collect. They can generate realistic synthetic data under diverse environments, improving model generalization and robustness. GANs also help address class imbalance and enhance data diversity. However, their training is often unstable, requiring careful tuning, and the generated data may suffer from mode collapse or lack of physical fidelity in some cases. Let $\mathcal{D}_{\text{GAN}} = \{(\bm{x}_i, \bm{p}_i^{\text{ue}})\}$ denote the real data used to train the GAN, where $\bm{x}_i$ is the input wireless signal and $\bm{p}_i^{\text{ue}}$ is the corresponding user position. The generator $G(\tilde{\bm{p}}^{\text{ue}}, \bm{\theta}_G)$ synthesizes wireless signal features $\tilde{\bm{x}}$ conditioned on a sampled position $\tilde{\bm{p}}^{\text{ue}} \sim p(\bm{p}^{\text{ue}})$, while the discriminator $D(\bm{x}, \bm{\theta}_D)$ aims to distinguish real samples from generated ones. $\bm{\theta}_G$ and $\bm{\theta}_D$ denote the parameters of the generator and discriminator, respectively. The training objective of a GAN can be formulated as
\begin{align}
\min_{\bm{\theta}_G} \max_{\bm{\theta}_D} & \quad 
\mathbb{E}_{(\bm{x}_i, \bm{p}_i^{\text{ue}}) \sim \mathcal{D}_{\text{GAN}}} 
\left[\log D(\bm{x}_i, \bm{\theta}_D)\right] \nonumber \\
& \quad +
\mathbb{E}_{\tilde{\bm{p}}^{\text{ue}} \sim p(\bm{p}^{\text{ue}})} 
\left[\log\left(1 - D(\tilde{\bm{x}}, \bm{\theta}_D)\right)\right] \nonumber \\
\text{s.t.} & \quad \tilde{\bm{x}} = G(\tilde{\bm{p}}^{\text{ue}}, \bm{\theta}_G)
\end{align}

\color{black}

\subsubsection{Other Advanced AI Techniques}

Beyond the AI algorithms discussed, advanced techniques such as knowledge distillation \cite{cite:KnowledgeDistillation}, ensemble learning \cite{cite:EnsembleLearning}, and self-supervised learning \cite{cite:Self-Supervised} offer additional avenues to enhance wireless positioning systems. For example, knowledge distillation involves transferring knowledge from a large, complex model (teacher) to a smaller, more efficient model (student), improving inference speed and scalability while maintaining high accuracy. By combining predictions from multiple models, ensemble learning improves robustness and accuracy, particularly in heterogeneous environments. Self-supervised learning leverages unlabeled data by generating pseudo-labels through pretext tasks, reducing the reliance on labeled datasets and enabling models to learn effective feature representations.

As AI technologies continue to evolve, these advanced techniques and future innovations will play a critical role in overcoming challenges and achieving higher accuracy, efficiency, and robustness in wireless positioning systems. Integrating these methodologies will remain an ongoing focus in research and practical deployments.

\subsection{Lessons Learned}
Based on the comparison of mainstream AI models and algorithms in wireless positioning, and their relevance to specific positioning tasks, it becomes evident that the choice of models and the limitations of algorithms must be closely aligned with practical system configurations and application scenarios. For AI models, FCNNs are suitable for simple, low-dimensional positioning tasks, but often result in a large number of parameters. CNNs efficiently extract local features from high-dimensional spatial data such as CSI matrices, but struggle to model global dependencies. Sequential models like LSTMs effectively capture temporal dynamics in CIR data and device trajectories, though their inherently sequential structure limits computational efficiency. Transformers excel at capturing global dependencies and integrating multi-source information, but they impose heavy demands on both data volume and computational resources. Regarding AI algorithms, each learning paradigm exhibits distinct strengths and limitations. Transfer learning is effective in leveraging knowledge from related domains when labeled data is scarce, but its success heavily depends on the similarity between source and target domains. Meta learning provides fast adaptability to new tasks, though its practicality is constrained by high computational complexity and sensitivity to task design. Continual learning supports long-term adaptability in dynamic environments, but strategies are required to mitigate catastrophic forgetting. GANs significantly enhance dataset diversity and address class imbalance through synthetic data generation, but face challenges such as training instability. In addition, other advanced techniques such as self-supervised learning offer promising ways to reduce the reliance on labeled data, while knowledge distillation further improves deployment efficiency by compressing large models into lightweight alternatives. Ultimately, achieving robust and efficient AI-driven wireless positioning demands a carefully tailored integration of models and algorithms, precisely aligned with the specific requirements of each application scenario.

\color{black}

\section{Progress in AI Positioning Standards} \label{sec:3GPP}

In this section, we provide an overview of 3GPP positioning standards, from 1G to advanced AI/ML-supported techniques in 5G. We also discuss positioning KPIs, deployment scenarios, and future directions toward AI-native positioning in 6G.

\subsection{Evolution of 3GPP Standards for Positioning}

1G networks used analog communication and lacked standardized positioning. Positioning was primarily based on signal strength (e.g., RSS) for BS selection and handovers. Some proprietary methods like TDOA and AOA in advanced mobile phone system (AMPS) signals, supported emergency and ITS applications \cite{Survey_del_peral-rosado_survey_2018, reed1998overview1G}, but accuracy was limited.

2G introduced digital communication and standardized positioning in global system for mobile communications (GSM) \cite{Survey_del_peral-rosado_survey_2018}, including Cell-ID, Time Advance, enhanced observed time difference (E-OTD), and assisted GPS (A-GPS) \cite{3gpp-22.071-Locationservices}. These improvements enabled services like E911 and laid the foundation for future enhancements.

3G technologies such as universal mobile telecommunications system (UMTS) and CDMA2000 introduced OTDOA, UTDOA, A-GPS \cite{3gpp-25.305-UTRAN}, and radio frequency pattern matching (RFPM) \cite{reed1998overview1G}. CDMA2000 also used advanced forward link trilateration (AFLT) with A-GPS for better synchronization. These methods reduced positioning errors to 25–200 meters, addressing broader application needs.

In LTE Release 9, positioning became formally standardized. Key methods included \cite{3gpp-36.305-E-UTRAN_R9}: 
\begin{itemize}
    \item \textbf{Enhanced Cell-ID (E-CID):} Improved accuracy by utilizing RSRP, and transmission time differences between mobile terminals and BSs.
    \item \textbf{OTDOA:} Introduced positioning reference signal (PRS), transmitted in low-interference subframes to enable terminals to measure time differences and compute reference signal time difference (RSTD).
    \item \textbf{Assisted GNSS (A-GNSS):} Supported GPS and other satellite navigation systems, with network assistance to provide differential corrections in signal-restricted environments.
\end{itemize}
To support cellular positioning, PRS signals were transmitted in dedicated subframes to minimize interference and enhance OTDOA performance. Furthermore, LTE positioning protocols, such as LTE positioning protocol (LPP), were defined to facilitate efficient communication between mobile terminals and positioning servers. Then, LTE-advanced (LTE-A) further enhanced positioning via UTDOA, improved OTDOA, RFPM \cite{3gpp-36.809-RFPM}, and advanced PRS performance. Additional techniques included D2D positioning, MIMO vertical estimation, and integration of Bluetooth, WLAN, and barometer data \cite{3gpp-36.855-positioningenhancements}.

With the evolution of 5G technologies, positioning capabilities in 3GPP standards have advanced significantly to meet the growing demands of diverse applications. In Release 16, the groundwork for 5G positioning was laid by defining positioning use cases and service requirements \cite{3gpp-22.872, 3gpp-22.261}. These efforts were aimed at addressing the initial needs for positioning in 5G networks. Subsequent studies focused on enabling new radio (NR) positioning techniques across both FR1 (sub-6 GHz) and FR2 (mmWave) bands. Key aspects included the specification of NR Downlink and Uplink reference signals to support techniques such as Downlink TDOA (DL-TDOA), Downlink AOD (DL-AOD), Uplink TDOA (UL-TDOA), Uplink AOA (UL-AOA), Multi-cell RTT, and E-CID. Additionally, the NR positioning protocol A (NRPPa) was introduced \cite{3gpp-38.455}, providing a standardized communication framework for positioning in 5G networks. To meet the demands of emerging 5G applications and vertical industries requiring higher accuracy, lower latency, and improved reliability, 3GPP launched the "Study on NR Positioning Enhancements" in Release 17 \cite{3gpp-38.857}. Key advancements included exploring positioning in in-coverage, partial coverage, and out-of-coverage scenarios to support diverse use cases, including sidelink-based positioning for V2X applications \cite{3gpp-38.845}. Release 18 introduced further improvements \cite{3gpp-38.859}, such as PRS and sounding reference signal (SRS) bandwidth aggregation, NR carrier phase measurements, low-power high-accuracy support, and positioning for RedCap UEs. Sidelink positioning for V2X was also expanded.

For AI/ML aspect, starting with Release 18, 3GPP initiated a study on AI/ML positioning \cite{3gpp-38.843}. This study explores the potential of AI/ML to improve positioning accuracy, details the general AI/ML framework, use cases for AI/ML positioning, and evaluation metrics and common KPIs for AI/ML positioning. These developments aim to complement traditional positioning methods by leveraging data-driven models to process large datasets and provide more accurate results.

\begin{table*}[t]
    \caption{5G Positioning performance requirements}
    \label{table:KPI}
    \hspace{1.5cm}
    \begin{tabularx}{\textwidth}{|c|ccccccc|}
    \hline
 
\multirow{3}{*}{\makecell{Positioning \\Service Level}} & \multicolumn{7}{c|}{Positioning requirements} \\ \cline{2-8} 
& \multicolumn{2}{c|}{Absolute Positioning} & \multicolumn{2}{c|}{Relative Positioning} & \multicolumn{1}{c|}{\multirow{2}{*}{\begin{tabular}[c]{@{}l@{}}\ \ Service \\ Availability\end{tabular}}} & \multicolumn{1}{c|}{\multirow{2}{*}{Latency}} & \multirow{2}{*}{Mobility} \\ \cline{2-5}

& \multicolumn{1}{c|}{Horizontal} & \multicolumn{1}{c|}{Vertical} & \multicolumn{1}{c|}{Horizontal} & \multicolumn{1}{c|}{Vertical} & \multicolumn{1}{c|}{} & \multicolumn{1}{c|}{} &  \\ \hline
 
\multirow{7}{*} 

Level 1 & \multicolumn{1}{c|}{$10$ m} & \multicolumn{1}{c|}{\multirow{2}{*}{$3$ m}} & \multicolumn{1}{c|}{\multirow{4}{*}{N/A}} & \multicolumn{1}{c|}{\multirow{4}{*}{N/A}} & \multicolumn{1}{c|}{$95$ \%} & \multicolumn{1}{c|}{\multirow{5}{*}{1 s}} & 
\multicolumn{1}{c|}{\multirow{4}{*}{\begin{tabular}[c]{@{}l@{}}$\leq 30$ km/h (indoor)\\ $\leq 250$ km/h (rural and urban)\\$\leq 500$ km/h (trains) \\ \end{tabular}}} 
 \\ \cline{1-2} \cline{6-6}
 
Level 2 & \multicolumn{1}{c|}{$3$ m} & \multicolumn{1}{c|}{}  & \multicolumn{1}{c|}{}& \multicolumn{1}{c|}{} & \multicolumn{1}{c|}{\multirow{3}{*}{$99$ \%}} & \multicolumn{1}{c|}{} & \multicolumn{1}{c|}{} 
 \\ \cline{1-3}
Level 3 & \multicolumn{1}{c|}{$1$ m} & \multicolumn{1}{c|}{\multirow{2}{*}{$2$ m}} & \multicolumn{1}{c|}{} & \multicolumn{1}{c|}{} & \multicolumn{1}{c|}{} & \multicolumn{1}{c|}{} & \multicolumn{1}{c|}{} 
 \\ \cline{1-2} 
Level 5 & \multicolumn{1}{c|}{$0.3$ m} & \multicolumn{1}{c|}{}  & \multicolumn{1}{c|}{} & \multicolumn{1}{c|}{} & \multicolumn{1}{c|}{} & \multicolumn{1}{c|}{} & \multicolumn{1}{c|}{} 
 \\ \cline{1-5} \cline{8-8} 

Level 7 & \multicolumn{1}{c|}{N/A} & \multicolumn{1}{c|}{N/A} & \multicolumn{1}{c|}{$0.2$ m} & \multicolumn{1}{c|}{$0.2$ m} & \multicolumn{1}{c|}{}& \multicolumn{1}{c|}{} & $\leq 30$ km/h (indoor and outdoor) \\ \cline{1-8} 

Level 4 & \multicolumn{1}{c|}{$1$ m} & \multicolumn{1}{c|}{$2$ m} & \multicolumn{1}{c|}{\multirow{2}{*}{N/A}} & \multicolumn{1}{c|}{\multirow{2}{*}{N/A}} & \multicolumn{1}{c|}{\multirow{2}{*}{$99.9$ \%}} & 
\multicolumn{1}{c|}{$15$ms} &$\leq 30$ km/h indoor  \\ \cline{1-3} \cline{7-8} 
Level 6 & \multicolumn{1}{c|}{$0.3$ m} & \multicolumn{1}{c|}{$2$ m} & \multicolumn{1}{c|}{} & \multicolumn{1}{c|}{} & \multicolumn{1}{c|}{} & \multicolumn{1}{c|}{$10$ms} & $\leq 30$ km/h indoor \\ \hline
\end{tabularx}
\end{table*}

\subsection{KPIs for Wireless Positioning}


Positioning performance is assessed through a set of critical metrics that ensure the accuracy and reliability of location services. The 5G system is designed to deliver positioning services that meet the performance requirements outlined in Table. \ref{table:KPI}. These requirements are inclusive of all UE types, including specialized UEs such as V2X and machine-type communication (MTC) devices.

In 3GPP \cite{3gpp-22.261}, 5G positioning use cases are divided into 5G positioning service area and 5G enhanced positioning service area. The former includes indoor and outdoor (rural and urban) scenarios. Indoor scenarios include locations inside buildings such as offices, hospitals, and industrial facilities. Outdoor scenarios support the positioning of vehicles and trains at speeds up to 250 km/h and 500 km/h. And the 5G enhanced positioning service area further supports dense urban areas (up to 60 km/h), vehicles in tunnels, and railway positioning. In addition to horizontal and vertical positioning accuracy, 3GPP Release 19 also introduces positioning service availability and positioning service latency to support mission-critical services. A total of 7 positioning service levels are defined, including 6 levels of absolute positioning accuracy requirements and 1 relative positioning accuracy requirement. In the absolute positioning accuracy requirements, the horizontal accuracy ranges from 10 m to 0.3 m, and the vertical accuracy ranges from 3 m to 2 m. The positioning service availability ranges from 95\% to 99.9\%. The scenarios of positioning service levels 1, 2, 3, and 5 require a 1s level latency, while positioning service levels 4 and 6 require a ms-level latency, which is only required in indoor scenarios of 5G enhanced positioning services, such as collaboration and collision avoidance of mobile robots and factory scenarios. For the relative positioning scenario of positioning service level 7, it is applicable to the positioning of two UEs within 10 m or the distance between a UE and a 5G positioning node within 10 m.

When evaluating AI/ML-based positioning, in addition to the common KPIs mentioned above, the performance of the AI model needs to be considered. For example, various overheads include open-air overhead, auxiliary information overhead, model delivery and transmission, etc. The complexity of the model's inference needs to be considered, which includes the complexity of pre-processing and post-processing, the computational complexity in tera operations per second (TOPS), floating-point operations per second (FLOPS), and multiplication-accumulation operations (MACs), and the potential difference between actual complexity and evaluation complexity due to platform dependencies and optimization solutions. In addition, regardless of the underlying algorithm, model complexity should be reported in terms of the number of real-valued model parameters and operations. Finally, for model monitoring, performance indicators that need to be considered include the accuracy and relevance of monitoring indicators, related overhead, computational and memory complexity, and latency, which reflects the timeliness of monitoring results and response operations.

\subsection{Advancements in AI Positioning within 3GPP}

To address the ongoing challenges in 5G positioning technologies, particularly in complex environments with multipath effects and NLOS scenarios, 3GPP Release 18 introduced an AI/ML study initiative \cite{3gpp-38.843}. This initiative explores the potential applications of AI/ML in positioning to improve accuracy and reliability under challenging conditions. In this section, we present the 3GPP standards for AI/ML-driven positioning from three perspectives: lifecycle management (LCM) framework, model deployment, and model inputs and outputs.

\subsubsection{Lifecycle Management for AI/ML models} 

\begin{figure}[tb]
\includegraphics[scale=0.28]{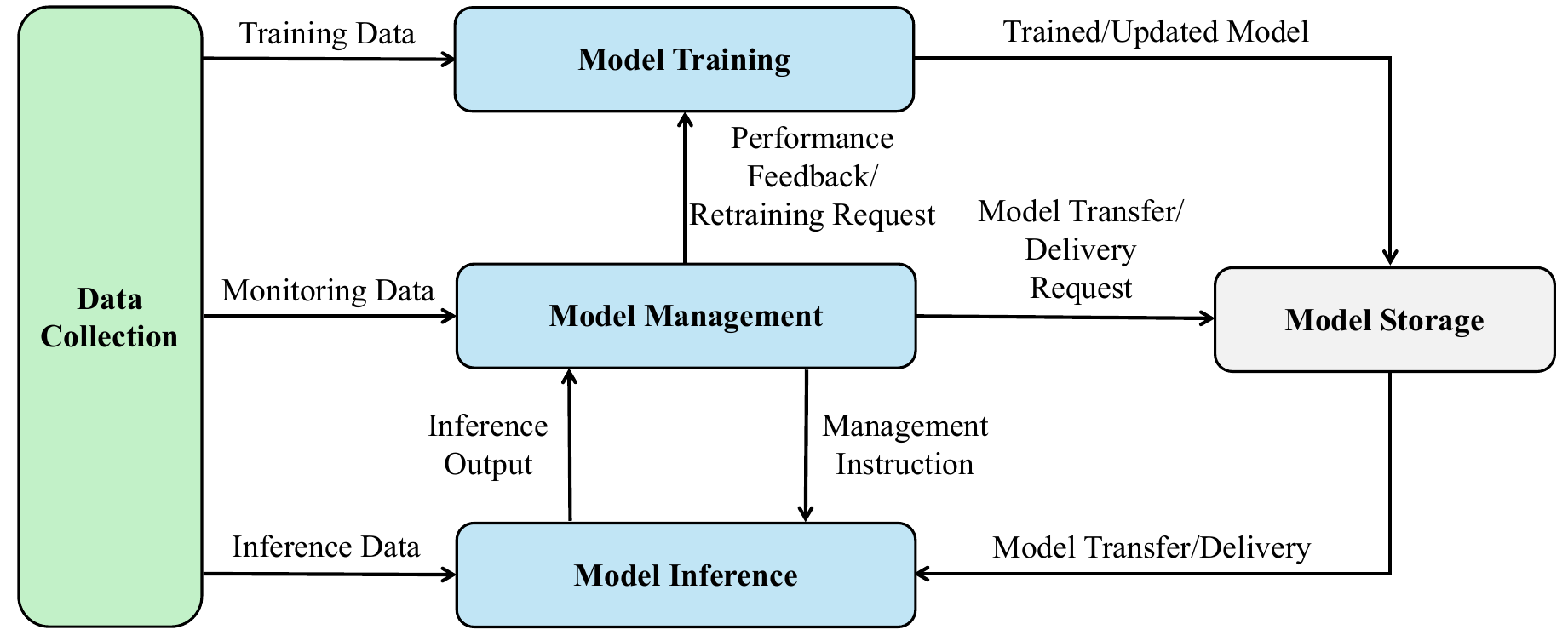}
\centering
\caption{Schematic diagram of AI/ML model LCM.}
\label{fig_3GPPAI}
\vspace{-2mm}
\end{figure}

Taking into account the needs of AI positioning algorithms, such as data collection, model training, updating, inference, and transmission, 3GPP Release 18 defines a comprehensive LCM framework for AI/ML positioning modules. As shown in Fig. \ref{fig_3GPPAI}, the LCM framework encompasses several key stages:
\begin{itemize}
    \item \textbf{Data Collection:} 
    The data collection function provides the input data required for model training, management, and inference. This data includes training data, monitoring data, and inference data, all essential for AI/ML model operation. In the context of positioning, this module enables gNodeB (gNBs) to acquire rich and diverse data (e.g., CSI, RSRP) based on real-time network configuration and environmental conditions. These data are subsequently used for model training or inference. When more training data is needed, this module can also employ generative AI techniques such as GANs to augment datasets through synthetic data generation.
    
    \item \textbf{Model Training:} 
    The model training function executes the training, validation, and testing of AI/ML models. It may also generate performance metrics for model evaluation. Additionally, this function handles data preparation, such as preprocessing, cleaning, formatting, and transformation, using the training data provided by the data collection function. In positioning systems, this module can support techniques like transfer learning and meta learning to enable cross-domain model adaptation. Moreover, it is responsible for model compression and optimization, such as pruning or quantization, to reduce complexity while maintaining accuracy, enabling efficient deployment on resource-constrained devices.

    \item \textbf{Model Management:} 
    The model management function oversees the operation and monitoring of AI/ML models, including decisions related to model selection, activation, switching, and rollback. It ensures the correctness of inference operations based on data received from data collection and inference functions. This function plays a central role in managing multiple model versions customized for different environments and user KPI requirements. On one hand, it selects suitable models based on channel characteristics collected by the Data Collection module and the user-specific KPIs such as accuracy or latency. On the other hand, when degradation is detected (e.g., due to model aging), it initiates retraining or model replacement. Overall, this module coordinates interactions across the lifecycle pipeline to guarantee that positioning services meet real-time performance targets.

    \item \textbf{Model Inference:} 
    The inference function applies the AI/ML model to input data provided by the data collection function to produce outputs. This function also performs data preparation, such as preprocessing, when required, to ensure accurate inference results. This module executes the core inference process for positioning tasks, producing either location coordinates or intermediate outputs such as position-related signal parameters.

    \item \textbf{Model Storage:} 
    The model storage function stores trained or updated models, which can later be used for inference. This function also acts as a key point for model transfer/delivery processes and other protocol terminations. This component defines the structure of available models and manages their storage after training. It also tracks the complexity and accuracy of each stored model to support optimal model selection during runtime.

\end{itemize}
These modules collectively address mechanisms for data collection, model training, updating, and sharing in wireless positioning, facilitating the effective deployment of AI/ML algorithms in wireless networks. Specifically, the flexibility of the LCM framework provides the foundation for advanced AI capabilities such as transfer learning, online model adaptation, and network-aware configuration. These functions are expected to be vital for enabling intelligent and context-aware positioning services in 5G-Advanced and future 6G networks.

\subsubsection{AI/ML Models Deployment}  

\begin{figure*}[tb]
\includegraphics[scale=0.44]{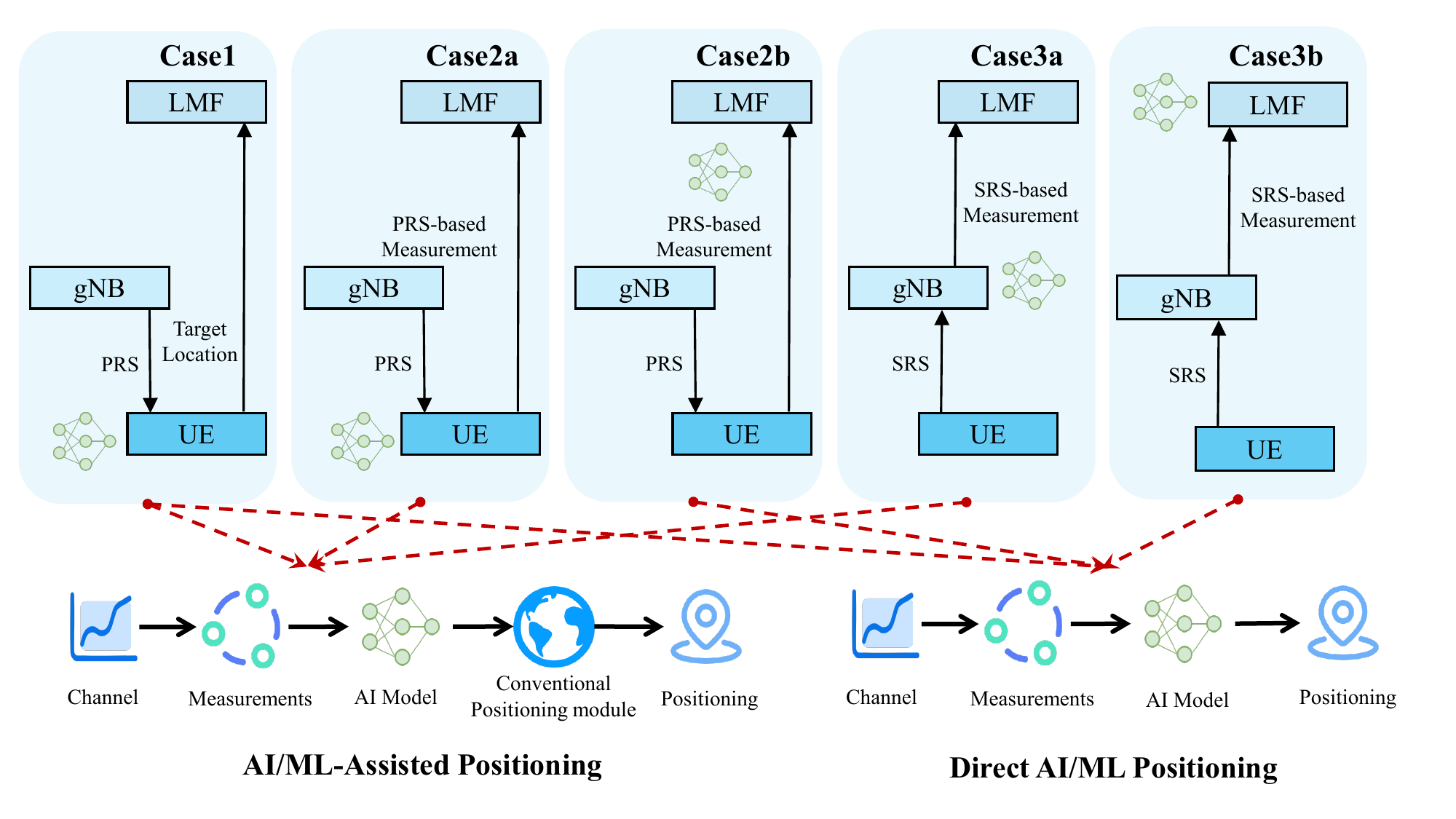}
\centering
\caption{Schematic diagram of AI/ML positioning cases and categories.}
\label{3GPPAIcase}
\end{figure*}

\begin{table*}[t]
\centering
\caption{Comparison of AI/ML-Based Positioning Deployment Scenarios}
\renewcommand{\arraystretch}{1.2} 
\setlength{\tabcolsep}{8pt} 
\label{tab:deployment_comparison}
\begin{tabular}{|
>{\centering\arraybackslash}p{3.0cm}|  
>{\centering\arraybackslash}p{2.5cm}|  
>{\centering\arraybackslash}p{2.5cm}|  
>{\centering\arraybackslash}p{2.5cm}|  
>{\centering\arraybackslash}p{2cm}|  
>{\centering\arraybackslash}p{2cm}|
}
\hline
\textbf{} & \textbf{Case 1} & \textbf{Case 2a} & \textbf{Case 2b} & \textbf{Case 3a} & \textbf{Case 3b} \\
\hline
{Computation Location} & UE & UE & LMF & gNB & LMF \\
\hline
{Communication Overhead} & Low & Medium & High & Low & Low \\
\hline
{Signal Characteristics} & Downlink interference possible & Downlink interference possible & Downlink interference possible & Lower transmit power than PRS & Lower transmit power than PRS \\
\hline
{Scalability} & Single-BS localization & Single-/Multi-BS localization & Single-/Multi-BS localization & Single-BS localization & Single-/Multi-BS localization \\
\hline
\end{tabular}
\end{table*}

Based on the role of AI in wireless positioning, AI-driven positioning techniques are categorized into two main types: \textbf{Direct AI/ML Positioning} and \textbf{AI/ML-Assisted Positioning}. 

In AI/ML-assisted positioning, the AI/ML models do not directly output the UE location through end-to-end learning. Instead, they enhance the positioning process by providing improved measurements or probabilistic estimates. For example, AI/ML models can output probabilities for LOS or NLOS conditions, refine ranging estimates (such as (TOA or OTDOA), or enhance angular measurements (e.g., AOA or AOD). These outputs are used to improve the accuracy of traditional positioning techniques by addressing uncertainties and inaccuracies in the measurement process.

In contrast, direct AI/ML positioning utilizes AI/ML models to determine the UE's location directly. These models take raw wireless channel observations as inputs and estimate the UE's position without relying on intermediate measurement steps. A prominent example of this approach is fingerprint-based positioning, where inputs such as CIR or power delay profile (PDP) are fed into an AI/ML model to directly estimate the UE's location.

Based on the deployment location of AI/ML models within network entities and the distinction between direct and assisted positioning methods, 5 deployment cases have been identified. As shown in Fig. \ref{3GPPAIcase}, the AI/ML models can be deployed on the UE, gNB, or the location management function (LMF), enabling flexibility in their application across different network architectures and positioning. In Fig. \ref{3GPPAIcase} and Table. \ref{tab:deployment_comparison}, we provide a detailed description of each deployment scenario below.
\begin{enumerate}
    \item \textbf{Case 1:} In this case, the AI/ML model is deployed locally on the UE. The model can support both direct AI/ML positioning and AI/ML-assisted positioning using downlink PRS. This deployment incurs low uplink communication overhead but requires significant computational resources on the UE side. It is less scalable for multi-BS scenarios and is thus better suited for single-BS localization. Moreover, the use of PRS may lead to downlink interference.

    \item \textbf{Case 2a:} In this case, the AI/ML model is deployed on the UE. The UE utilizes the AI/ML model to perform measurements on the downlink PRS and transmits the measurement results to the LMF for positioning. This scenario only supports AI/ML-assisted positioning. It offers moderate uplink communication overhead and good scalability for both single- and multi-BS localization. However, it still imposes high computational demands on the UE and may suffer from downlink PRS interference.

    \item \textbf{Case 2b:} In this case, the AI/ML model is deployed on the LMF. The UE performs measurements on the downlink PRS and transmits the results to the LMF, where the AI/ML model determines the UE's location. This setup supports direct AI/ML positioning and offers high scalability for both single- and multi-BS localization, while requiring small computational resources on the UE. The cost is the increased uplink communication overhead due to the raw signal transmission and the potential PRS interference in the downlink.

    \item \textbf{Case 3a:} In this case, the AI/ML model is deployed on the gNB. The gNB uses the AI/ML model to perform measurements on the uplink SRS and transmits the results to the LMF for positioning. This scenario only supports AI/ML-assisted positioning. It reduces the computational requirements on the UE side, but may result in limited participation from neighboring gNBs since SRS power is lower than PRS. In addition, since the AI model is deployed on the gNB, it is mainly applicable to single-BS scenarios.

    \item \textbf{Case 3b:} In this case, the AI/ML model is deployed on the LMF. The gNB performs measurements on the uplink  SRS and transmits the results to the LMF, where the AI/ML model determines the UE's position. This scenario only supports direct AI/ML positioning. This approach minimizes the computational load on the UE and supports both single- and multi-BS positioning, providing good scalability. However, the relatively low power of the SRS may limit the number of participating gNBs, potentially affecting performance in sparse deployments.
\end{enumerate}

\subsubsection{Model Inputs and Outputs}

For model training, training data can be generated by various network entities, including the UE, positioning reference unit (PRU), gNB, or LMF. For LMF-side model inference (Cases 2b and 3b), the input data is generated by the UE or gNB and terminates at the LMF. For gNB-side model inference (Case 3a), the input data are directly available within the gNB, reducing latency and ensuring efficient processing.

In cellular positioning, various measurements serve as critical inputs for AI/ML models to achieve accurate localization. These inputs are primarily derived from reference signals such as DL PRS and UL SRS, which are reused from existing 3GPP specifications but can be enhanced with AI-specific configurations. Key input types include high-dimensional information like delay profiles (DP), PDP, CIR, and CIR phase data. Additionally, timing-related measurements such as TOA, RSTD, RTT, and angle-related metrics extracted from CSI can be leveraged as model inputs. Power metrics like DL PRS-RSRP and UL SRS-RSRP also contribute to improving positioning accuracy.

The outputs of AI/ML models depend on the type of positioning. For AI/ML-assisted positioning, models output refined measurements or probabilistic estimates, such as TOA, OTDOA, LOS/NLOS probabilities, or angular measurements (AOA, AOD). For direct AI/ML positioning, models output the estimated UE location.

\subsection{Toward AI-Driven Positioning in 6G and Future Networks}

Although the formal standardization process for 6G has not yet begun, it is widely anticipated that AI will play a central role in positioning services within AI-native network architectures. AI will no longer serve merely as an auxiliary tool, but will become a core component of next-generation network design. Based on this vision, we also discuss potential standardization directions for AI-based positioning technologies in future 6G networks in this section.

\subsubsection{Native AI/ML Integration in Air Interface and Protocol Design}

Unlike current 5G systems that rely on predefined mechanisms, such as PRS patterns, 6G is expected to adopt AI-native system designs, in which AI is deeply integrated not only into position estimation but also into broader aspects of air interface and protocol operation. Several standardization opportunities in this area include:
\begin{itemize}
    \item \textbf{AI-Driven Pilot and Signaling Design:} AI can be used to adaptively design reference signals and signaling strategies based on environmental conditions. For instance, dynamic adjustment of pilot density, beam configuration, or frequency band allocation can improve localization robustness in challenging environments such as THz communication \cite{Survey_chen_tutorial_2022}, RIS-assisted networks \cite{Survey_Chen_Reconfigurable_2022}, and ultra-dense deployments.

    \item \textbf{Joint Multi-Task AI Model Architectures:} Future systems may adopt unified, multi-task AI models capable of jointly performing functions such as channel estimation, prediction, feedback compression, and positioning within a shared learning framework \cite{salihu_self-supervised_2024, pan2025large}. This design can reduce model redundancy, improve computational efficiency, and enable task-aware inference pipelines.

    \item \textbf{AI-Based Protocol Adaptation:} AI models can dynamically sense environmental variations (e.g., blockage, channel fluctuation, user mobility) and adapt key protocol parameters, such as scheduling periods, feedback granularity, and retransmission strategies, to optimize positioning accuracy and service continuity.
\end{itemize}

\subsubsection{Multi-Source and Multi-Modal Positioning}
6G systems are expected to natively support multi-source and multi-modal positioning by integrating signals from diverse infrastructures (such as cell-free networks \cite{10379122CELLFREE}, NTN \cite{dureppagari_ntn-based_2023}, and RIS\cite{Survey_Chen_Reconfigurable_2022}) alongside heterogeneous sensory data. By combining radio signals from different access technologies with inputs from vision, LiDAR, inertial sensors, and map-based information, future systems could enable high-precision positioning and sensing applications. To realize these capabilities, the following aspects are expected to be addressed in future standardization efforts: 
\begin{itemize}
    \item \textbf{Unified Data Interfaces for Sensor and Network Fusion:} Define standardized data formats and APIs for fusing radio and non-radio data sources across network elements and devices. This includes time synchronization, metadata structuring, and channel/measurement abstraction layers.

    \item \textbf{Scene-Aware Information Exchange Mechanisms:} Establish standardized protocols for the exchange of environment-specific information (e.g., 3D maps and building layouts) among access points, edge servers, and user equipment. These mechanisms aim to enable collaborative localization and enhance situational awareness across heterogeneous networks. In addition, 6G can also consider the integration and synchronization of digital twin systems \cite{shi2025digital}, which can provide dynamic, high-fidelity representations of physical environments to further improve positioning accuracy, consistency, and adaptability in complex scenarios

    \item \textbf{Cross-Domain Privacy-Preserving Learning Mechanisms:} Develop standard-compliant methods to enable model training and update across different stakeholders (e.g., network operators, application vendors, and device manufacturers) without compromising data privacy or ownership.

\end{itemize}

\subsubsection{AI-Specific Performance Metrics and QoS Requirements}

Although existing 3GPP standards provide comprehensive KPIs for positioning accuracy, latency, these metrics are not tailored to AI-based localization algorithms and frameworks. As AI-driven positioning becomes a core enabler in 6G networks, it is essential to establish performance evaluation criteria that capture the unique characteristics, resource demands, and operational behavior of different AI models and deployment strategies. Future standardization should explicitly define the fundamental parameters of AI-driven positioning systems, including:
\begin{itemize}
\item \textbf{Network Resource Consumption:} Quantify both communication overhead (e.g., channel and feature transmission) and computational resource usage (e.g., FLOPs, memory, inference time).
\item \textbf{Data Transmission Volume:} Measure the size of AI model inputs and outputs, including uplink/downlink data volume involved in distributed inference or collaborative learning \cite{sun2025energy}.
\item \textbf{Model Update Dynamics:} Define metrics for training frequency, online adaptation cost, and update latency in dynamic or multi-user environments.
\end{itemize}
Building on these foundational parameters, AI-driven positioning systems should also be evaluated using a new class of AI-aware QoS indicators, including but not limited to:

\begin{itemize}
\item \textbf{Latency:} It includes as end-to-end latency, transmission delay, and inference latency. Specifically, end-to-end latency refers to the total time from signal reception to position output, encompassing data preprocessing, feature extraction, model inference, and result post-processing. 
\item \textbf{Positioning Accuracy and Model Aging Rate:} In addition to static positioning accuracy under standard conditions, it is important to evaluate the rate at which model performance degrades over time or across environments. This reflects the frequency at which models need to be retrained or updated to maintain acceptable accuracy.
\item \textbf{Energy Efficiency:} The energy consumption of the positioning process includes the computational energy in the inference process and the communication energy for data exchange.
\end{itemize}

\subsection{Lessons Learned}
The evolution of 3GPP standards reflects the growing maturity and significance of wireless positioning in cellular networks. Building upon traditional techniques, R18 marks a key milestone by introducing AI/ML frameworks into standardized positioning, enabling improved accuracy and robustness in complex scenarios such as NLOS conditions and dynamic environments. However, the integration of AI also brings new technical and standardization challenges, including the need for specialized performance metrics and comprehensive lifecycle management. The diversity of deployment options (across UE, gNB, and LMF), reveals inherent trade-offs between communication overhead, computational demands, and scalability. Furthermore, as positioning in 6G becomes tightly coupled with sensing and semantic understanding of the environment, standardization must also address challenges including AI-native network design, cross-modal data fusion, and privacy-preserving collaborative learning. This section highlights that the success of AI-driven positioning in future networks will rely not only on advances in algorithms but also on system-level orchestration and standardization frameworks tailored to the unique characteristics of AI models.

\color{black}

\section{SOTA in AI/ML-Assisted Positioning} \label{sec:AssistedPositioning}

\begin{table*}[t]
\centering
\caption{Summary of LOS/NLOS Classification Works}
\label{table:LOSClassification}
\begin{tabular}{|
>{\centering\arraybackslash}p{1.8cm}|  
>{\raggedright\arraybackslash}p{3.2cm}|  
>{\raggedright\arraybackslash}p{5cm}|  
>{\raggedright\arraybackslash}p{6cm}|    
}
\hline
\textbf{Backbone} & \textbf{Algorithm} & \textbf{Input Feature} & \textbf{Key Insight} \\
\hline

\multirow{3}{*}{Traditional ML }  
& GMM \cite{LOS_fan_non-line--sight_2019}, GBDT \cite{LOS_huang_nlos_2019}, k-NN, SVM, GMM, K-means \cite{LOS_zhang_wireless_2020} 
& Signal energy, delay statistics \cite{LOS_fan_non-line--sight_2019}, CIR statistics \cite{LOS_huang_nlos_2019, LOS_zhang_wireless_2020} 
& Handle simple features, rely on manual extraction from wireless channels.

\\
\hline

\multirow{2}{*}{FCNN} 
& Supervised learning \cite{LOS_sang_identification_2020, LOS_huang_machine_2020} 
& CIR statistics \cite{LOS_sang_identification_2020, LOS_huang_machine_2020} 
& Provide stronger feature extraction than traditional ML, but still rely on manually extracted features. \\
\hline

\multirow{10}{*}{CNN} 
& Supervised learning \cite{LOS_si_lightweight_2023, LOS_zhu_simple_2023, LOS_zheng_channel_2020, LOS_jiang_uwb_2020, LOS_cui_nlos_2021, LOS_wang_multi-classification_2022, LOS_zhang_novel_2022, LOS_deng_uwb_2023} &  CIR \cite{LOS_si_lightweight_2023, LOS_zhu_simple_2023, LOS_jiang_uwb_2020, LOS_deng_uwb_2023}, PAS images \cite{LOS_zheng_channel_2020}, Morlet-transformed CIR \cite{LOS_cui_nlos_2021}, Wavelet-packet images \cite{LOS_wang_multi-classification_2022}, CSI eigenmatrix/vector \cite{LOS_zhang_novel_2022}, and GAF \cite{LOS_deng_uwb_2023}  & Handle high-dimension complex channel features, reduces manual feature engineering via deep feature extraction.\\
\cline{2-4}
& Transfer learning \cite{LOS_sun_channel_2023, LOS_nkrow_transfer_2024} & CIR \cite{LOS_zhu_simple_2023}, Stockwell Spectrograms \cite{LOS_sun_channel_2023} &  Enable cross-scenario generalization and adaptation. \\
\cline{2-4}
& GAN \cite{LOS_nkrow_transfer_2024} & CIR \cite{LOS_nkrow_transfer_2024} &Enhance model robustness by generating diverse training data. \\
\cline{2-4}
& Autoencoder \cite{LOS_tedeschini_latent_2023, LOS_tedeschini_cooperative_2023}  & ADCPM \cite{LOS_tedeschini_latent_2023, LOS_tedeschini_cooperative_2023} & Learn latent representations of channel features to improve classification performance. \\
\hline

\multirow{4}{*}{Temporal Model}
& Supervised learning \cite{LOS_liu_uwb_2022, LOS_kim_uwb_2023, LOS_lv_uwb_2024}& CIR \cite{LOS_liu_uwb_2022, LOS_kim_uwb_2023}, and CGWT \cite{LOS_lv_uwb_2024} & Capture temporal dependencies in CIR sequences using temporal models.\\
\cline{2-4}
& Transfer learning \cite{LOS_li_uwb_2024} & CIR \cite{LOS_li_uwb_2024} & Further enable cross-scenario generalization and adaptation.\\
\hline

\multirow{2}{*}{Transformer}
& Supervised learning \cite{LOS_zhou_deep_2024, yang2025fuzzy} & CIR \cite{yang2025fuzzy}, CIR + statistics \cite{LOS_zhou_deep_2024} & Leverage self-attention mechanisms to fuse spatial and statistical features.\\
\hline

\end{tabular}
\end{table*}

As discussed above, AI/ML-assisted positioning leverages AI technologies for estimating positioning-related parameters, which are subsequently used to enhance the performance of traditional positioning algorithms. This approach is necessary because, while conventional estimation techniques are effective under ideal conditions, they often face significant challenges in real-world environments, including multipath propagation, signal distortion, and hardware imperfections. The advent of ML and DL offers new possibilities for addressing these challenges by employing data-driven methods to model the inherent complexity and nonlinearity of wireless environments. ML and DL techniques not only improve the robustness of parameter estimation but also enhance real-time performance in dynamic scenarios. Based on these advancements, we conduct a detailed survey of AI-based positioning parameter estimation algorithms, with a specific focus on AI-based LOS/NLOS detection, TOA/TDOA estimation, and angle estimation algorithms.

\subsection{AI-based LOS/NLOS Detection}

The wireless environment significantly impacts positioning accuracy. LOS scenarios, with unobstructed transmitter-receiver paths, yield more accurate distance and angle measurements, while NLOS scenarios, involving reflections and obstructions, degrade reliability due to multipath effects and delays. Thus, identifying LOS and NLOS conditions is critical for ensuring positioning accuracy. We surveyed existing works in this area and summarized them in Table. \ref{table:LOSClassification}.

Prior to the advancements in deep learning, significant work had already been done in this area, employing ML algorithms such as support vector machines (SVM) \cite{LOS_sang_identification_2020}, random forest \cite{LOS_sang_identification_2020}, gradient boosting decision tree (GBDT) \cite{LOS_huang_nlos_2019}, Gaussian mixture models (GMM) \cite{LOS_fan_non-line--sight_2019}, and k-means \cite{LOS_zhang_wireless_2020} for LOS/NLOS identification. With the development of AI technology, FCNN networks have also been widely used in this task. In \cite{LOS_sang_identification_2020} and \cite{LOS_huang_machine_2020}, the authors compare three ML classifiers, i.e. SVM, random forest, and FCNN, to identify LOS and NLOS.

Subsequently, more advanced models and algorithms have been widely studied for LOS/NLOS identification. These works leverage deep learning backbones such as CNN \cite{LOS_si_lightweight_2023, LOS_zhu_simple_2023, LOS_zheng_channel_2020, LOS_jiang_uwb_2020, LOS_cui_nlos_2021, LOS_wang_multi-classification_2022, LOS_zhang_novel_2022, LOS_deng_uwb_2023, LOS_sun_channel_2023, LOS_nkrow_transfer_2024, LOS_tedeschini_latent_2023, LOS_tedeschini_cooperative_2023}, temporal models (including LSTM \cite{LOS_li_uwb_2024, LOS_kim_uwb_2023}, GRU \cite{LOS_liu_uwb_2022}, and TCN \cite{ LOS_lv_uwb_2024}), and Transformer-based architectures \cite{LOS_zhou_deep_2024, yang2025fuzzy} to enhance classification performance. To enrich the input feature space and improve model discrimination capability, a variety of representations have been proposed, such as power angular spectrum (PAS) \cite{LOS_zheng_channel_2020}, wavelet-based transformations \cite{LOS_cui_nlos_2021, LOS_wang_multi-classification_2022}, eigen features \cite{LOS_zhang_novel_2022}, Gramian angular fields (GAF) \cite{LOS_deng_uwb_2023}, Stockwell spectrograms \cite{LOS_sun_channel_2023}, angle-delay cannel power matrix (ADCPM) \cite{LOS_tedeschini_latent_2023, LOS_tedeschini_cooperative_2023}, and complex gaussian wavelet transform (CGWT) \cite{LOS_lv_uwb_2024}. Furthermore, advanced learning techniques have been introduced to improve performance. In \cite{LOS_sun_channel_2023, LOS_nkrow_transfer_2024} and \cite{LOS_li_uwb_2024}, transfer learning is employed to enhance cross-scenario generalization and reduce the overhead of retraining models for new environments. In \cite{LOS_nkrow_transfer_2024}, GAN is employed to generate representative CIR samples for the target domain, effectively reducing the need for extensive data collection and manual labeling. Autoencoders \cite{LOS_tedeschini_latent_2023, LOS_tedeschini_cooperative_2023} provide a powerful means for unsupervised feature extraction and dimensionality reduction. They learn compact latent representations that effectively distinguish LOS from NLOS paths. In parallel, to reduce computational complexity and enable real-time deployment, several studies have proposed lightweight network architectures \cite{LOS_si_lightweight_2023, LOS_zhu_simple_2023, LOS_zhang_novel_2022} without sacrificing much accuracy.
\color{black}

Furthermore, various works have leveraged LOS/NLOS identification to improve positioning algorithms. For instance, the authors in \cite{LOS_kim_uwb_2023} utilize LSTM to estimate approximate NLOS errors, correcting the positioning estimation results to enhance positioning accuracy. In UWB scenarios, the paper \cite{LOS_deng_uwb_2023} uses CNN for NLOS identification, greatly enhancing ranging accuracy and reducing positioning errors based on identification results. In 5G channels, the authors in \cite{LOS_tedeschini_cooperative_2023} use ADCPM for NLOS identification and select different positioning methods based on the identified LOS/NLOS results to enhance positioning accuracy.

\subsection{AI-based TOA/TDOA Estimation}

As mentioned above, TOA and TDOA are fundamental techniques in wireless positioning, providing the basis for accurate positioning by measuring the time it takes for a signal to propagate between transmitters and receivers. However, traditional TOA/TDOA estimation methods face significant challenges in real-world environments, including the presence of multipath propagation, NLOS conditions, and hardware imperfections. Recent advancements in AI can effectively enhance TOA/TDOA estimation by leveraging data-driven learning techniques.

\begin{table*}[t]
\centering
\caption{Summary of TOA/TDOA Estimation Works}
\label{table:TOAestimation}
\begin{tabular}{|
>{\centering\arraybackslash}p{1.8cm}|  
>{\raggedright\arraybackslash}p{3.2cm}|  
>{\raggedright\arraybackslash}p{5cm}|  
>{\raggedright\arraybackslash}p{6cm}|    
}
\hline
\textbf{Backbone} & \textbf{Algorithm} & \textbf{Input Feature} & \textbf{Key Insight} \\
\hline

\multirow{2}{*}{Traditional ML} 
& SVM \cite{yin_entropy-based_2015, wang_semi-supervised_2021,liu_machine_2023}, PCA \cite{savic_kernel_2016}, KNN \cite{liu_machine_2023} 
& Channel statistics (signal amplitude, energy, delay spread, etc.) \cite{yin_entropy-based_2015, wang_semi-supervised_2021,savic_kernel_2016,liu_machine_2023}
& Handle simple features, relies on manual extraction from wireless channels.

\\
\hline

\multirow{2}{*}{FCNN}
& Supervised learning \cite{dvorecki_machine_2019, kirmaz_toa_2023} 
& CSI \cite{dvorecki_machine_2019} and CIR \cite{kirmaz_toa_2023} 
& Provide stronger feature extraction than traditional ML, but still relies on manually extracted features. \\
\hline

\multirow{4}{*}{CNN} 
& Supervised learning \cite{bialer_deep_2018, sun_deep_2019, hsiao_super-resolution_2021, luo_toa_2019, NB-2020-Pan, niitsoo_deep_2019, feigl_robust_2021,abbasi_novel_2021, wei_deep-learning-based_2023} & CIR \cite{ hsiao_super-resolution_2021, niitsoo_deep_2019, feigl_robust_2021,abbasi_novel_2021, wei_deep-learning-based_2023}, channel cross-correlation \cite{bialer_deep_2018, luo_toa_2019, NB-2020-Pan} and IQ sample \cite{sun_deep_2019}  & Handle high-dimension complex channel features, reduces manual feature engineering via deep feature extraction.\\
\cline{2-4}
& VAE \cite{li_variational_2023} & CIR \cite{li_variational_2023}&  Enable cross-scenario generalization and adaptation. \\
\hline

\multirow{2}{*}{Temporal Model}
& Supervised learning \cite{lynch_localisation_2020}& CIR \cite{lynch_localisation_2020 } & Capture temporal dependencies in CIR sequences using temporal models.\\
\hline

\multirow{2}{*}{Transformer}
& Supervised learning \cite{Jonathan_Multipath_2023, Jonathan_Estimating_2024} & CIR \cite{Jonathan_Multipath_2023, Jonathan_Estimating_2024} & Leverage self-attention mechanisms to fuse spatial and statistical features.\\
\hline

\end{tabular}
\end{table*}

Early studies applied traditional ML algorithms to TOA estimation, focusing on mitigating errors in distance measurements \cite{yin_entropy-based_2015,savic_kernel_2016, wang_semi-supervised_2021}. For example, the authors in \cite{yin_entropy-based_2015, wang_semi-supervised_2021} propose an SVM approach for error mitigation for ranging estimation. In \cite{savic_kernel_2016}, the authors use kernel principal component analysis (PCA) for TOA estimation. In \cite{liu_machine_2023}, the authors combine ML algorithms with Kalman Filters for TOA estimation using 5G downlink signals. While these methods provide initial improvements, their inability to effectively interpret high-dimensional channel features limits their robustness and adaptability in complex environments.

To overcome the limitations of traditional ML, deep learning techniques have been employed to enhance TOA/TDOA estimation. These methods leverage the ability of neural networks to model complex, nonlinear relationships in data, achieving superior accuracy \cite{dvorecki_machine_2019, kirmaz_toa_2023, bialer_deep_2018, hsiao_super-resolution_2021, sun_deep_2019, luo_toa_2019, niitsoo_deep_2019, feigl_robust_2021, abbasi_novel_2021,lynch_localisation_2020, wei_deep-learning-based_2023,li_variational_2023, Jonathan_Multipath_2023, Jonathan_Estimating_2024}. These works leverage neural network backbones such as FCNNs \cite{dvorecki_machine_2019, kirmaz_toa_2023}, CNNs \cite{bialer_deep_2018, sun_deep_2019, hsiao_super-resolution_2021, luo_toa_2019, NB-2020-Pan, niitsoo_deep_2019, feigl_robust_2021,abbasi_novel_2021, wei_deep-learning-based_2023, li_variational_2023}, temporal models \cite{lynch_localisation_2020}, and Transformers \cite{Jonathan_Multipath_2023, Jonathan_Estimating_2024}. The related works are summarized in Table.~\ref{table:TOAestimation}. For traditional ML-based methods, input features are typically derived from channel statistics such as signal amplitude, energy, and delay spread. In contrast, deep learning approaches can directly learn from raw or minimally processed channel information, including CIR, CSI, cross-correlation signals, and even raw in-phase and quadrature (IQ) samples.  With the development of model architectures, neural networks are increasingly capable of modeling and extracting high-dimensional representations from these raw inputs through end-to-end training. This enables more robust feature learning and improves the generalization of TOA/TDOA estimation models in complex wireless environments. For example, in \cite{hsiao_super-resolution_2021}, the authors use neural networks to generate high-resolution CIR for accurate TOA estimation. In the 5G system, the authors propose a CNN-based algorithm for TOA estimation \cite{sun_deep_2019}. In the IoT system, the authors in \cite{NB-2020-Pan} propose a CNN-based algorithm by generating fine-grained features from full-band and resource-block-based reference signals, leveraging spectrogram-like cross-correlation feature maps to directly project time-frequency domain variations into TOA results. In addition, some studies \cite{jang_convolutional_2021, wang_2d-cnn-based_2021, pan_doa_2023} consider joint angle and TOA estimation to achieve higher accuracy to achieve higher accuracy estimation.

In addition to the fundamental approach of end-to-end training with neural networks, recent research also leverages advanced AI techniques \cite{li_variational_2023} to address challenges such as data scarcity and robustness in TOA/TDOA estimation. In \cite{li_variational_2023}, the authors propose inter-instance variational autoencoders (VAEs), which use variational inference with latent variables to simultaneously estimate distance and identify environmental conditions. Furthermore, to solve the difficulty of obtaining datasets, in \cite{choi_enhanced_2022}, the paper combines neural networks with the Fine Timing Measurement (FTM) protocol to enhance RTT ranging. This work introduces an unsupervised learning framework that uses naturally accumulated sensor data to reduce data collection overhead. The authors in \cite{choi_sensor-aided_2022} utilize sensor data gathered during regular application use to reconstruct device trajectories for designing an unsupervised learning technique for TOA estimation.

\subsection{AI-based Angle Estimation}

\begin{table*}[t]
\centering
\caption{Summary of Angle Estimation Works}
\label{table:AOAestimation}
\begin{tabular}{|
>{\centering\arraybackslash}p{1.8cm}|  
>{\raggedright\arraybackslash}p{3.2cm}|  
>{\raggedright\arraybackslash}p{5.5cm}|  
>{\raggedright\arraybackslash}p{5.5cm}|    
}
\hline
\textbf{Backbone} & \textbf{Algorithm} & \textbf{Input Feature} & \textbf{Key Insight} \\
\hline

\multirow{3}{*}{Traditional ML} 
& SVM \cite{yang_machine-learning-based_2021}
& PDP and path loss \cite{yang_machine-learning-based_2021}
& High-dimensional channel of the antenna array makes it difficult to deploy traditional ML algorithms.
\\
\hline

\multirow{5}{*}{FCNN} 
& Supervised learning \cite{huang_deep_2018, xiang_novel_2019, chen_deep_2020,barthelme_doa_2021, khan_angle--arrival_2019} 
& Received signal \cite{huang_deep_2018} and covariance matrix \cite{xiang_novel_2019, chen_deep_2020, barthelme_doa_2021}, and spectrum vector from MUSIC \cite{khan_angle--arrival_2019}.
& Leverage the nonlinear fitting capabilities of CNNs to improve performance. \\
\cline{2-4}
& Autoencoder \cite{liu_direction--arrival_2018} & covariance matrix \cite{liu_direction--arrival_2018}& Learn latent representations of channel features to improve performance. \\
\hline

\multirow{5}{*}{CNN} 
& Supervised learning \cite{zhu_two-dimensional_2020,xiang_improved_2020,papageorgiou_deep_2021,akter_rfdoa-net_2021,chen_sdoa-net_2024,liu_model-driven_2024, cao_complex_2020, naoumi_complex_2024,wang_deep_2021, xu_deep_2023, mylonakis_novel_2024, kassir_improving_2024, zheng_deep_2024} & CIR \cite{wang_deep_2021,xu_deep_2023}, CSI \cite{naoumi_complex_2024}, covariance matrix \cite{zhu_two-dimensional_2020, papageorgiou_deep_2021, xiang_improved_2020, liu_model-driven_2024, chen_sdoa-net_2024, cao_complex_2020}, correlation matrix \cite{mylonakis_novel_2024, kassir_improving_2024}, and received signal \cite{akter_rfdoa-net_2021, zheng_deep_2024}  & Handle high-dimension complex channel features for angle estimation.\\
\cline{2-4}
& Transfer learning \cite{tian_vehicle_2022, guo_transfer_2022, pan_doa_2023 } & covariance matrix \cite{tian_vehicle_2022}, correlation matrix \cite{guo_transfer_2022}, and received signal \cite{pan_doa_2023}&  Enable cross-scenario generalization and adaptation. \\
\hline

\multirow{2}{*}{Temporal Model} 
& Supervised learning \cite{xiang_improved_2021}& Covariance matrix \cite{xiang_improved_2021} & Capture temporal dependencies of the channel using temporal models.\\
\hline

\multirow{2}{*}{Transformer}
& Supervised learning \cite{liu_transformer-based_2022, Wu_TransAoA_2025} & Received signal \cite{liu_transformer-based_2022, Wu_TransAoA_2025} & Leverage self-attention mechanisms to capture long-term dependencies of channels.\\
\hline

\end{tabular}
\end{table*}

The related works for angle estimation are summarized in Table.~\ref{table:AOAestimation}. In the era of traditional ML, algorithms such as SVM \cite{yang_machine-learning-based_2021} have been employed for angle estimation. For example,
in \cite{yang_machine-learning-based_2021}, the paper presents an SVM-based AOA estimation method for vehicular communications using power PDP and path loss received by each element of the antenna
array. However, angle estimation inherently relies on antenna arrays, resulting in high-dimensional input features that are often difficult to process effectively using traditional ML algorithms. This complexity limits the performance and scalability of traditional methods in practical scenarios.

Compared to traditional ML techniques, neural networks have been employed due to their superior fitting capabilities, enabling more accurate angle estimation. Some studies demonstrate the potential of neural networks for angle estimation based on FCNN \cite{huang_deep_2018, xiang_novel_2019, chen_deep_2020,barthelme_doa_2021}. For instance, in \cite{huang_deep_2018}, a deep learning-based framework integrates FCNNs for efficient channel and DOA estimation in massive MIMO systems. To achieve super-resolution AOA estimation, the authors in \cite{khan_angle--arrival_2019} firstly obtain the MUSIC spectrum and then regard it as the input of the FCNN, which increases the AOA estimation performance. In \cite{barthelme_doa_2021}, the paper proposes a neural network-based method to reconstruct full-array covariance matrices from subarray samples, enabling improved DOA estimation with MUSIC. With the development of neural network models, the accuracy of angle estimation has been further improved based on CNN \cite{zhu_two-dimensional_2020,xiang_improved_2020,papageorgiou_deep_2021,akter_rfdoa-net_2021,chen_sdoa-net_2024,liu_model-driven_2024, cao_complex_2020, naoumi_complex_2024,wang_deep_2021, xu_deep_2023, mylonakis_novel_2024, kassir_improving_2024, zheng_deep_2024}, LSTM \cite{xiang_improved_2021}, Transformer \cite{liu_transformer-based_2022, Wu_TransAoA_2025}. For example, in \cite{zhu_two-dimensional_2020}, the paper introduces a deep ensemble learning approach for 2D DOA estimation, combining multiple independently trained CNNs to map spatial covariance matrices to azimuth and elevation angles to achieve angle estimation. In \cite{xiang_improved_2020}, the authors demonstrate the importance of phase features to improve DOA estimation accuracy, and propose a neural network framework for DOA estimation. Considering low-SNR conditions, the authors in \cite{papageorgiou_deep_2021} present a CNN-based method for robust DOA estimation, framing the problem as a multi-label classification task using sample covariance matrices. Considering that the channels in wireless systems are complex, the authors in \cite{cao_complex_2020} introduce a complex-valued deep learning framework, and use virtual covariance matrices to handle spherical wave effects and complex signal features for near-field DOA estimation. In \cite{naoumi_complex_2024}, the paper employs a complex neural network-based deep learning approach and a parameterized algorithm for joint AOA and AOD estimation, enhancing computational efficiency via channel matrix preprocessing and coarse timing estimation. In addition, the authors in \cite{xiang_improved_2021} propose an LSTM-based DOA estimation method that enhances phase features to improve accuracy and robustness to array imperfections. Additionally, in \cite{liu_transformer-based_2022}, the authors propose a Transformer-based signal denoising network with temporal attention to enhance AOA estimation accuracy in indoor NLOS environments.
\color{black}

Furthermore, advancements in AI technologies have further enhanced the performance of angle estimation algorithms. Autoencoders, known for their powerful feature extraction capabilities, have been increasingly applied in this domain. For example, in \cite{liu_direction--arrival_2018}, the paper introduces a DNN framework combining autoencoders and parallel classifiers to achieve robust DOA estimation. This framework effectively addresses array imperfections and demonstrates strong generalization to unseen scenarios. In addition, transfer learning can improve the robustness of the DOA algorithm through domain transfer. The authors in \cite{guo_transfer_2022} propose a transfer learning method using a ResNet for AOA estimation in massive MIMO systems, leveraging shared features across channel models to reduce data requirements and avoid training separate networks for each channel. Similarly, in \cite{pan_doa_2023}, the authors propose a deep transfer learning-based DOA and TOA joint estimation algorithm using a multi-task network with shared-private structure. In the vehicle positioning system \cite{tian_vehicle_2022}, the authors perform 2D DOA estimation for incoherently distributed sources with massive MIMO, employing transfer learning and attention mechanisms for improved accuracy, robustness, and efficiency.


\subsection{Lessons Learned}
The transition from traditional ML to deep learning has significantly improved positioning accuracy, particularly in complex wireless environments characterized by multipath propagation, NLOS conditions, and dynamic interference. Deep learning demonstrates strong capabilities in capturing high-dimensional channel features and modeling nonlinear relationships, thereby reducing the need for manual feature engineering. To further improve performance, the choice of input features is very important. While early approaches rely heavily on manually extracted channel statistics, more recent studies have shown the benefits of using raw or minimally processed inputs such as CIR, CSI, and received signal matrices, which preserve richer spatial and temporal information for learning. However, robustness across domains remains a significant challenge. Although many deep models perform well in specific scenarios, their performance often degrades in unseen environments. To address this issue, techniques such as transfer learning, domain adaptation, and data augmentation have emerged as promising solutions for improving generalization and reducing reliance on large labeled datasets.
\color{black}

\section{SOTA in Direct AI/ML Positioning} \label{sec:DirctPositioning}

In this section, we discuss the SOTA in direct AI/ML positioning, categorizing techniques into Fingerprint-based Positioning, Knowledge-Assisted AI Positioning, and Channel Charting-Based Positioning based on modeling approaches and reliance on auxiliary information.
\begin{itemize}
    \item{Fingerprint-based Positioning:} This method directly applies AI/ML to map wireless signals (e.g., RSS or CSI) to user positions using pre-collected signal fingerprints. It achieves high accuracy positioning with comprehensive fingerprint databases but requires extensive data collection and frequent updates, making it resource-intensive and less adaptable to dynamic environments.

    \item{Knowledge-assisted AI positioning:} This approach integrates wireless domain and geometric domain knowledge into AI models to enhance fingerprint-based positioning. By embedding prior knowledge, it improves learning efficiency, positioning accuracy, and generalization, reducing reliance on exhaustive datasets and performing well in complex, dynamic environments.

    \item{Channel charting based positioning:} Channel charting based positioning uses learning methods to model relationships between signal sampling points, creating a pseudo-coordinate system from raw CSI data without relying on absolute positions or detailed environment knowledge. It is scalable and reduces dependency on labeled datasets, making it suitable for dynamic and large-scale scenarios.

\end{itemize}

\subsection{Fingerprint-based positioning}

As discussed earlier, fingerprint-based positioning involves three critical processes: fingerprint collection and feature extraction, fingerprint database construction, and fingerprint database updating. Below, we provide a detailed introduction to these three components.

\subsubsection{Fingerprint Collection and Feature Extraction}

\begin{table}[t]
\centering
\caption{Overview of Signal Types and Representations Used as Fingerprints in Localization}
\label{tab:fingerprint_signals}
\renewcommand{\arraystretch}{1.2}
\begin{tabular}{|p{4.5cm}|p{3.0cm}|}
\hline
\textbf{Signal / Representation} & \textbf{Representative Works} \\
\hline
RSS / RSRP / RSRQ & 
\cite{wang2020novel, chen2019learning, prasad2018machine, shao_indoor_2018, tarekegn_dfops_2021, klus_neural_2021, wang_5g1m_2024, li_outdoor_2020} \\

\hline
TOA / AOA Measurements & 
\cite{9843909FPToA, 8939702FPToA} \\

\hline
CFR / Spatial-Frequency Domain CSI & 
\cite{berruet2018delfin, foliadis_multi-environment_2023, chu_exploiting_2024} \\

\hline
CIR / Angle-Delay Domain CSI & 
\cite{hejazi2021dyloc, qiu_cooperative_2020, xu_cross-region_2024, xu2024swin, sun2018single, wu2021learning, gong2023deep} \\

\hline
Filtered / Preprocessed CSI & 
\cite{ruan_ipos-5g_2023, he2024transfer} \\

\hline
Fingerprint Fusion (e.g., RSS, CSI) & 
\cite{10612810fingerprint, 2019Zhang-LTE} \\

\hline
Sensor Fusion (e.g., WiFi, IMU, GNSS) & 
\cite{belmonte-hernandez_swiblux_2019, tarekegn_dfops_2021, klus_neural_2021} \\

\hline
\end{tabular}
\end{table}

In the fingerprint collection and feature extraction stage, wireless signals are collected offline at RPs, and relevant features are extracted to serve as fingerprints. The simplest forms of fingerprints include RSS \cite{wang2020novel,chen2019learning,prasad2018machine,shao_indoor_2018,tarekegn_dfops_2021,klus_neural_2021}, RSRP \cite{wang_5g1m_2024}, RSRQ \cite{li_outdoor_2020,wang_5g1m_2024}, and TOA/AOA measurements \cite{9843909FPToA, 8939702FPToA, zhang2021aoa}. However, these types of fingerprints are often insufficiently detailed and lack the reliability required for positioning performance due to factors such as device errors and measurement noise.

CSI, as a high-dimensional representation of wireless channels, provides rich spatial and temporal information for positioning. Various forms of CSI-based fingerprints have been explored in the literature. Some studies directly use the CFR as input, with either amplitude–phase or real–imaginary components encoded as two separate feature maps \cite{berruet2018delfin, foliadis_multi-environment_2023, chu_exploiting_2024}. To better capture high-resolution multipath characteristics, the discrete Fourier transform (DFT) is often applied to obtain CIR or angle-delay domain features such as the ADCPM \cite{xu2024swin, hejazi2021dyloc, qiu_cooperative_2020, xu_cross-region_2024, sun2018single}. Building upon this, technologies such as sparsity-enhanced ADCPM and refined beam domain channel matrices have been proposed to improve multipath discrimination and robustness \cite{wu2021learning, gong2023deep}. Given the presence of noise and hardware imperfections in practical systems, several works also explore preprocessing techniques for CSI to enhance the learnability and stability of fingerprints \cite{ruan_ipos-5g_2023, he2024transfer}. Moreover, recent studies demonstrate that fusing different types of wireless fingerprints (such as CSI and RSS) \cite{10612810fingerprint, 2019Zhang-LTE}, or integrating auxiliary sensors (like IMUs and GNSS) \cite{belmonte-hernandez_swiblux_2019, tarekegn_dfops_2021, klus_neural_2021} can further improve positioning accuracy and robustness, particularly in complex environments. Table.~\ref{tab:fingerprint_signals} summarizes representative fingerprint types used in wireless positioning.

\color{black}

\subsubsection{Fingerprint Database Construction}

\begin{table}[t]
\centering
\caption{Overview of Fingerprint Database Construction Methods}
\label{tab:fingerprint_database_methods}
\renewcommand{\arraystretch}{1.2}
\begin{tabular}{|p{3.5cm}|p{4.2cm}|}
\hline
\textbf{Method Category} & \textbf{Representative Works} \\
\hline
{Model-based interpolation method for radio map construction} & Linear interpolation \cite{bravenec_influence_2023}, Kriging interpolation \cite{sato_space-frequency-interpolated_2021, zuo_multi-phase_2018,sato_kriging-based_2017}, Gaussian process regression \cite{zhen_radio_2022}, Voronoi tessellation based interpolation\cite{lee_voronoi_2012} and multicomponent optimization and sparse recovery \cite{khalajmehrabadi_structured_2017}. \\
\hline

{AI-assisted interpolation method for radio map construction} & FCNN \cite{sato_performance_2019}, CNN \cite{chaves-villota_deeprem_2023}, LSTM \cite{roger_deep-learning-based_2024}, Transformer \cite{Chen2024Radio}, GAN \cite{li2019sparsely,zhang_rme-gan_2023}, and autoencoder \cite{teganya_deep_2022} \\
\hline

{AI-driven positioning based on fingerprint mapping}  & FCNN \cite{li_outdoor_2020}, CNN \cite{shao_indoor_2018,berruet2018delfin,foliadis2023multi,wu2021learning}, LSTM \cite{tarekegn_dfops_2021}, RNN \cite{hsieh_towards_2018,bai_dl-rnn_2020}, and Transformers \cite{han_crowdbert_2024, xu2024swin}. \\
\hline

\end{tabular}
\end{table}

An overview of fingerprint database construction methods is provided in Table. \ref{tab:fingerprint_database_methods}. Traditional fingerprint positioning relies on constructing a fingerprint database and determining user locations by comparing real-time measurements with the database. Interpolation methods are widely used across time, frequency, and spatial domains to reconstruct complete radio maps for positioning \cite{sato_space-frequency-interpolated_2021, zuo_multi-phase_2018}. Model-based interpolation techniques, such as linear interpolation \cite{bravenec_influence_2023}, Kriging interpolation \cite{sato_space-frequency-interpolated_2021, zuo_multi-phase_2018,sato_kriging-based_2017}, gaussian process regression \cite{zhen_radio_2022}, Voronoi tessellation-based interpolation\cite{lee_voronoi_2012}, as well as multicomponent optimization and sparse recovery \cite{khalajmehrabadi_structured_2017} have been extensively utilized for this purpose. All of these techniques attempt to construct reliable and adaptable radio maps from the perspectives of accuracy, complexity, and data efficiency. In parallel, AI-based interpolation methods have attracted increasing attention. These methods leverage a variety of learning models, including FCNNs \cite{sato_performance_2019}, CNNs \cite{chaves-villota_deeprem_2023}, LSTMs \cite{roger_deep-learning-based_2024}, and Transformers \cite{Chen2024Radio}. In addition, GANs \cite{li2019sparsely,zhang_rme-gan_2023} are employed to synthesize fingerprint data for unmeasured positions through adversarial learning, while autoencoders \cite{teganya_deep_2022} compress sparse radio maps into low-dimensional latent spaces and reconstruct interpolated results through decoding. These AI-driven approaches offer greater modeling flexibility and are better suited to capturing complex, nonlinear spatial variations in wireless signal propagation.

However, the fingerprint positioning approach based on the radio environment map depends heavily on the density of RPs and the efficiency of the query algorithm. It faces challenges such as high deployment costs and prolonged query times. With the advancement of AI technologies, existing studies increasingly replace fingerprint databases with AI models. AI algorithms learn the mapping between wireless fingerprints and locations from datasets, significantly improving positioning accuracy. This enhancement in accuracy largely depends on the design of the AI models, leading to the development of effective models using architectures like FCNN \cite{li_outdoor_2020}, CNN \cite{shao_indoor_2018,berruet2018delfin,foliadis2023multi,wu2021learning}, LSTM \cite{tarekegn_dfops_2021}, RNN \cite{hsieh_towards_2018,bai_dl-rnn_2020}, and Transformers \cite{han_crowdbert_2024, xu2024swin}. For example, in \cite{quezada-gaibor_surimi_2022}, the paper proposes a hybrid indoor positioning architecture combining CNN, LSTM, and GAN models to enhance training data and improve accuracy. DeepWiPos uses an LSTM-based framework with attention modules to fuse RSS and fingerprint spatial gradients, addressing RSS instability and spatial ambiguity \cite{yang_deepwipos_2023}. The paper in \cite{foliadis2023multi} investigates Residual Block-based CNN algorithms, using uplink beamformed CSI fingerprints with multiple spatial dimensions at both the BS and UE.

\begin{table*}[t]
\centering
\caption{Summary of AI-driven Fingerprint Update Works}
\label{table:databaseupdate}
\begin{tabular}{|
>{\centering\arraybackslash}p{1.5cm}| 
>{\centering\arraybackslash}p{0.7cm}|  
>{\centering\arraybackslash}p{2.8cm}|  
>{\centering\arraybackslash}p{0.9cm}|  
>{\centering\arraybackslash}p{1.5cm}|  
>{\raggedright\arraybackslash}p{7.1cm}| 
}
\hline
\textbf{Update Type} & \textbf{Ref.} & \textbf{Algorithm} & \textbf{Label Needed} & \textbf{Time/Space Shift} & \textbf{Key Insight} \\
\hline

\multirow{8}{*}{{\makecell{\\Radio Map \\ Update}}} 
& \cite{belmonte-hernandez_recurrent_2020} & GAN & Yes & Time & Use conditional GAN to generate time-aware RSSI maps for updating fingerprints. \\
\cline{2-6}
& \cite{zhang2023learning} & Few-shot Learning + GAN & Yes & Time & Augment fingerprints via GAN and infers location labels by spectral clustering. \\
\cline{2-6}
& \cite{zhu_intelligent_2022} & Incremental Learning & No & Time & Use incremental learning adapt to time-varying CSI without retraining CNN. \\
\cline{2-6}
& \cite{li2021transloc} & Model-based Transfer Learning & Yes & Time & Align heterogeneous RSS/CSI domains via feature mapping and weighted instance transfer without retraining. \\
\cline{2-6}
& \cite{yang_updating_2021} & Model-based Transfer Learning & Yes & Time & Use enhanced transfer component analysis to align domains and filters changed APs for radio map updating. \\
\hline

\multirow{12}{*}{\makecell{\\Neural \\ Network\\ Update}} 
& \cite{li2023multidomain} & Transfer Learning & Yes & Time + Space & Handle label sparsity and domain shifts via Bayesian domain transfer for cross-domain positioning. \\
\cline{2-6}
& \cite{li2024transfer} & Transfer Learning + Knowledge Distillation & Yes & Time + Space & Combine knowledge distillation with fine-tuning on real data to adapt from synthetic to real scenarios. \\
\cline{2-6}
& \cite{li2025d} & Transfer Learning & Yes & Time + Space & Use dual-domain adaptation and weight regularization to adapt CSI-based model across dynamic conditions. \\
\cline{2-6}
& \cite{xu_cross-region_2024} & Meta-Learning  & Yes & Space & Use cross-region fusion and meta-Learning to enable model adaptation. \\
\cline{2-6}
& \cite{gao2023metaloc} & Meta-Learning & Yes & Space & Learn environment-agnostic meta-parameters for rapid adaptation with few labeled samples. \\
\cline{2-6}
& \cite{etiabi_femloc_2024} & Federated Meta-Learning & Yes & Time + Space & Learn a meta-model via federated training to rapidly adapt to new environments. \\
\cline{2-6}
& \cite{pu_bayesian_2024} & Bayesian Meta-Learning & Yes & Time & Combine Bayesian meta-learning to enable fast adaptation of localization. \\
\hline
\end{tabular}
\end{table*}
\color{black}

\subsubsection{Fingerprint Database Updating}
Fingerprint database and AI model aging also pose critical challenges in wireless fingerprint-based positioning. Environmental changes, such as infrastructure updates, furniture rearrangements, or time variations, can cause the database and models to become outdated, reducing positioning accuracy. Addressing this issue requires efficient methods for maintaining and updating the database or AI models. To achieve this goal, crowdsourcing methods provide an effective solution by opportunistically collecting labeled fingerprints during routine device usage. Crowdsourced fingerprint updates leverage opportunistic real-world data, such as high-confidence labels obtained near base stations or in areas with strong GPS signals \cite{xiang_self-calibrating_2021, yu_effective_2019, xiang_crowdsourcing-based_2022}, as well as user movement trajectories obtained from pedestrian dead reckoning (PDR) \cite{he_indoor_2017, wang_optimal_2018, liu_adaptive_2019, wu_automatic_2018, huang_online_2019,9259006FPToA, yuan_adaptive_2024}. With the help of crowdsourced data, we have the opportunity to update the fingerprint to maintain positioning accuracy.

Advanced AI techniques have been widely applied to tackle the challenges of radio map aging and domain adaptation in fingerprint-based localization. Fingerprint databases built on RSS or CSI measurements may degrade over time or vary across environments. To address this, several works employ radio map updates using GANs to synthesize new fingerprints~\cite{belmonte-hernandez_recurrent_2020, zhang2023learning} or transfer learning to adapt old fingerprints to new domains~\cite{li2021transloc, yang_updating_2021, li2023multidomain}. In~\cite{zhu_intelligent_2022}, incremental learning has also been used to continuously adapt fingerprints without retraining. In contrast, when deep models replace explicit fingerprint databases, model adaptation becomes essential. Methods based on transfer learning~\cite{li2024transfer, li2025d}, meta-learning~\cite{xu_cross-region_2024, gao2023metaloc, pu_bayesian_2024}, and federated learning~\cite{etiabi_femloc_2024} aim to fine-tune neural networks with minimal labeled data. The comparison of related work is shown in Table. \ref{table:databaseupdate}.
\color{black}

\subsection{Knowledge-assisted AI Positioning}

Model-based positioning methods, such as trilateration, triangulation, and multilateration, utilize geometric principles and channel propagation models to estimate position based on signal properties. While these geometry-based methods are effective under some conditions, they often face significant challenges in complex or dynamic environments, particularly in scenarios affected by multipath effects and NLOS conditions. In contrast, fingerprint-based positioning uses AI to directly learn the mapping between channel (or features) and the position, achieving end-to-end positioning. However, as noted above, fingerprint positioning systems can experience performance degradation due to changes in the wireless environment. These systems also face significant difficulties in transferring across different environments or deployment scenarios, which limits their robustness and scalability.

To overcome these challenges, knowledge-assisted AI positioning integrates domain knowledge, such as geometry and channel models, into AI-driven approaches. By embedding physical insights into learning models, this hybrid method improves positioning accuracy and generalization, even in variable environments, while reducing the dependency on large labeled datasets. This subsection introduces two key branches of this approach: wireless-knowledge-assisted positioning and geometric-knowledge-assisted positioning.

\subsubsection{Wireless-knowledge-assisted Positioning}

\begin{table}[t]
\centering
\caption{Categorization of Wireless-Knowledge-Assisted Positioning Methods}
\label{tab:knowledge_assisted_positioning}
\begin{tabular}{|
>{\centering\arraybackslash}p{2.0cm}|
>{\centering\arraybackslash}p{3.5cm}|
>{\centering\arraybackslash}p{1.8cm}| }
\hline
\textbf{Category} & \textbf{Methods} & \textbf{References} \\
\hline
\multirow{2}{*}{{\makecell{ Knowledge \\ from wireless \\ system \\ characteristics}}} 
& Knowledge derived from network configurations & \cite{chen_deep_2023,Hu2023HoloFed} \\
\cline{2-3}
& Knowledge derived from different channel-related tasks  &  \cite{bayraktar_ris-aided_2024, guo_learning-based_2024} \\
\hline
\multirow{2}{*}{{\makecell{ Knowledge \\ from unlabeled \\  wireless data}}} 
& Generative-based SSL pretraining  & \cite{wang_signal-guided_2024, han_crowdbert_2024, pan2025large} \\
\cline{2-3}
& Contrastive-based SSL pretraining  &  \cite{pan2025large, deng2022supervised, gong_semisupervised_2024, zhang2025lessons, salihu_self-supervised_2024} \\
\hline
{Knowledge-driven mitigation for  deployment imperfections} 
& Deployment imperfections such as hardware impairments, calibration errors, and clock offsets.
& \cite{rivetti_spatial_2023, bayraktar_ris-aided_2024} \\
\hline
\end{tabular}
\end{table}

Wireless-knowledge-assisted positioning aims to enhance AI-based localization by embedding domain knowledge of wireless propagation into learning models. To systematically summarize existing approaches, Table.~\ref{tab:knowledge_assisted_positioning} categorizes representative methods based on the type of wireless knowledge utilized and its functional role in the positioning process. One effective strategy involves designing customized localization algorithms by leveraging the system characteristics of wireless networks. On one hand, knowledge derived from specific network configurations can be exploited to improve positioning accuracy \cite{chen_deep_2023, Hu2023HoloFed}. For instance, in FDMA systems, multi-frequency fusion has been proposed to harness subcarrier diversity for more robust localization \cite{chen_deep_2023}. On the other hand, system-level knowledge can be utilized through joint design with other communication tasks, such as channel estimation \cite{bayraktar_ris-aided_2024} and CSI feedback \cite{guo_learning-based_2024}, which share underlying characteristics with localization. For example, \cite{bayraktar_ris-aided_2024} jointly optimizes channel estimation and localization in RIS-aided mmWave systems. Similarly, \cite{guo_learning-based_2024} proposes an integrated learning framework in which CSI feedback and localization mutually benefit from each other.

Another way to incorporate wireless knowledge is by learning channel representations from large volumes of unlabeled data using self-supervised learning (SSL). SSL enables models to capture the intrinsic semantic structure of the wireless channel, such as spatial geometry, multipath propagation patterns, and user mobility characteristics, without relying on costly location labels. SSL approaches can be broadly categorized into two paradigms: generative-based methods (e.g., masked modeling) \cite{wang_signal-guided_2024, han_crowdbert_2024, pan2025large}, and contrastive-based methods \cite{pan2025large, deng2022supervised, gong_semisupervised_2024, zhang2025lessons, salihu_self-supervised_2024}. Once pre-trained using these strategies, the learned channel representations can be fine-tuned with a small amount of labeled data to achieve significantly improved localization performance \cite{salihu_self-supervised_2024, pan2025large}. In particular, the authors in \cite{salihu_self-supervised_2024} propose a Large Wireless Localization Model (LWLM) that combines generative-based and contrastive-based SSL objectives. By pretraining on large-scale CSI data, the model achieves performance gains across various localization tasks, demonstrating the effectiveness in capturing channel semantics relevant to positioning.

Beyond improving localization accuracy, wireless domain knowledge is also leveraged to address practical challenges in real-world wireless systems, such as hardware impairments, calibration errors, and clock offsets, all of which can significantly degrade positioning performance during deployment \cite{rivetti_spatial_2023, bayraktar_ris-aided_2024}. Considering hardware impairments, the authors in \cite{rivetti_spatial_2023} explore end-to-end positioning using an autoencoder architecture, effectively mitigating hardware-induced errors. In \cite{bayraktar_ris-aided_2024}, the authors address challenges specific to RIS-aided mmWave systems, such as clock offsets between transmitters and receivers, impairments at the transmit and receive antenna arrays, and coupling effects among RIS elements. 

\color{black}

\subsubsection{Geometric-knowledge-assisted Positioning}

\begin{table}[t]
\centering
\caption{Categorization of Geometric-Knowledge-Assisted Positioning Methods}
\label{tab:ai_positioning_knowledge}
\begin{tabular}{|
>{\centering\arraybackslash}p{3.1cm}|
>{\centering\arraybackslash}p{4.8cm}|}
\hline
\textbf{Category} & \textbf{Methods \& References} \\
\hline

\multirow{3}{*}{{\makecell{Knowledge from \\ geometric information}}}
& Based on distance and angle measurements \cite{wang_multipath-assisted_2024, yang_model-based_2021, liu_mlloc_2024, klus_c2r_2024}, LOS/NLOS classification \cite{yang_model-based_2021, pan2020deep}, and motion trajectory analysis \cite{zhang_context-aware_2022, li_deep_2020} \\
\hline

\multirow{2}{*}{{\makecell{Knowledge from \\ environmental priors}}}
& Based on physical maps \cite{yu_floor-plan-aided_2024, amiri_indoor_2023, zhang2025uniloc} and BS deployment \cite{GCN-2021-SUN} \\
\hline

\end{tabular}
\end{table}

Another effective approach to improving positioning accuracy is to combine AI algorithms with geometric knowledge and spatial constraints. When integrated with AI models, geometric knowledge enables the system to better adapt to variations in wireless environments and spatial dynamics. Specifically, this category leverages either geometric information or environmental priors to assist positioning. Table. \ref{tab:ai_positioning_knowledge} presents the classification and representative studies of geometric-knowledge-assisted positioning methods.

From the geometric perspective, methods utilize distance and angle measurements \cite{wang_multipath-assisted_2024, yang_model-based_2021, liu_mlloc_2024, klus_c2r_2024}, LOS/NLOS classification \cite{yang_model-based_2021, pan2020deep}, and motion trajectory analysis \cite{zhang_context-aware_2022, li_deep_2020} to infer accurate locations. For instance, in OTDOA-based positioning \cite{pan2020deep}, the authors enhance accuracy by incorporating NLOS indicator with RSTD measurements as inputs to a neural network, which effectively mitigates the NLOS problem and improves the positioning accuracy. Further extending this, the authors in \cite{yang_model-based_2021} use hybrid delay and angular measurements in a neural network-based weighted least squares (WLS) framework to enhance performance. Another study \cite{wang_multipath-assisted_2024} employs a deep variational learning method to estimate position-related parameters such as distance, TDOA, and AOA, which are subsequently used to calculate the position. To address issues with outlier positioning estimates, the authors in \cite{zhang_context-aware_2022} propose a context-aware localization technique that leverages historical trajectory information to refine the accuracy of anomalous points. In \cite{li_deep_2020}, the authors propose a DRL-based unsupervised wireless localization method, modeling localization as a Markov decision process and designing a reward-setting mechanism based on high RSS near APs for robust localization without retraining.

From the perspective of environmental and configuration priors, techniques leverage physical maps \cite{yu_floor-plan-aided_2024,amiri_indoor_2023,zhang2025uniloc} (such as floor plans) and BS deployment information \cite{GCN-2021-SUN} to enhance localization. For example, \cite{yu_floor-plan-aided_2024} introduces a zero-shot learning framework for indoor positioning using floor plan images. A graph neural network is employed to model the relationships between APs and devices for coarse localization, while floor-plan constraints refine positioning accuracy. The authors in \cite{amiri_indoor_2023} focus on multipath detection to reconstruct indoor environments. The study uses an ML model to predict dominant multipath components, identify virtual anchors, and build a generative channel model, significantly improving positioning accuracy under NLOS conditions. In \cite{GCN-2021-SUN}, the paper proposes a graph convolutional network to model spatial relationships among multiple APs, extracting features from RSSI-based fingerprints.

\color{black}

\subsection{Channel Charting based Positioning}

As mentioned above, wireless-knowledge-assisted positioning builds upon fingerprint-based positioning by leveraging wireless knowledge to further enhance localization performance and reduce the reliance on extensive data collection. Building on this foundation, Channel charting-based positioning eliminates the need for external RPs or exhaustive datasets by directly learning latent spatial relationships from CSI data, enabling relative or pseudo-positioning. This approach allows channel charting to be highly adaptable to dynamic environments while significantly reducing operational overhead, making it a promising alternative to traditional localization methods in rapidly changing wireless scenarios.

Numerous studies have investigated the potential of channel charting to enhance wireless positioning performance \cite{lei_siamese_2019, deng2021network, euchner_augmenting_2023, stahlke_velocity-based_2024, stephan_angle-delay_2024, euchner_leveraging_2024, viet_implicit_2022, stahlke_indoor_2023, aghajari_multi-point_2023, agostini_learning_2023}. The authors in \cite{lei_siamese_2019} first introduce channel charting for positioning by proposing a unified Siamese network architecture for CSI-based localization. Their framework supports supervised, semi-supervised, and unsupervised scenarios, leveraging Sammon's mapping extension and side information to achieve accurate positioning in both LOS and NLOS channels with minimal CSI measurements. Building on this, graph-based approaches are further applied to channel charting and wireless positioning. For example, the authors in \cite{deng2021network} propose a semi-supervised graph-based channel charting framework for 5G localization that utilizes distributed CSI, side-information, and constrained manifold learning to construct a 2D channel chart, achieving 5.6 m localization accuracy with minimal labeled samples. In \cite{euchner_augmenting_2023}, a comparative study between classical model-based localization methods and channel charting highlight the limitations of channel charting in achieving global geometric accuracy. To address this, augmented channel charting is proposed by integrating model-based localization information into channel charting training, thereby improving overall performance and surpassing classical methods on evaluated datasets. Leveraging temporal correlations during sampling, the authors in \cite{stahlke_velocity-based_2024} introduce a reference-free channel charting framework that incorporates velocity information and topological maps to transform relative charts into real-world coordinates. Further enhancements are made in \cite{stephan_angle-delay_2024}, where a novel dissimilarity metric is introduced, incorporating angular-domain information and a deep learning-based metric. Additionally, metric fusion is proposed to integrate temporal and CSI similarities, demonstrating superior performance in sub-6 GHz massive MIMO scenarios, even under NLOS conditions. In \cite{euchner_leveraging_2024}, a Doppler-based loss function for channel charting is introduced, requiring only frequency synchronization to enable localization with minimal assumptions. The authors in \cite{stahlke_indoor_2023} develop a geodesic distance-based metric for channel charting using synchronized CSI measurements, utilizing a Siamese network to learn global geometry for localization, which outperforms traditional fingerprinting methods in real-world 5G and UWB systems. Additionally, in \cite{aghajari_multi-point_2023}, the paper presents a multi-point channel charting approach for multi-gateway LoRa networks, applying t-SNE and k-means clustering on received power vectors to map spatial geometry and improve IoT device localization. Lastly, privacy concerns in channel charting are systematically examined in \cite{agostini_learning_2023}, focusing on user and vendor privacy risks associated with pseudo-locations and raw CSI exposure.

\subsection{Lessons Learned}

From our comprehensive analysis of direct AI/ML positioning techniques, several important observations emerge. Fingerprint-based positioning, while capable of delivering high accuracy through advanced deep learning architectures such as CNNs, LSTMs, and Transformers, is heavily reliant on dense labeled data and frequent updates, making it costly and less scalable in dynamic or large-scale scenarios. In contrast, knowledge-assisted AI positioning addresses data inefficiency and generalization issues by embedding domain knowledge (wireless channel characteristics and geometric constraints) into the learning process, thereby improving robustness and reducing dependence on exhaustive measurements. Channel charting offers a compelling alternative by learning latent spatial structures directly from CSI data without the need for labeled positions, making it highly scalable and label-efficient. However, challenges remain in achieving global geometric accuracy and ensuring interpretability. Therefore, future research must focus on how to effectively exploit both wireless and geometric knowledge in combination with cross-domain adaptation techniques, in order to enable accurate, robust, and environment-adaptive positioning. Lastly, the increasing emphasis on lightweight and interpretable models reflects the growing demand for real-time, resource-efficient localization solutions in practical deployments. These findings highlight that future advances in AI-driven positioning will likely hinge on the integration of domain knowledge, data-efficient learning, and scalable system design.

\color{black}

\section{Datasets for Wireless Positioning}\label{sec:dataset}

\begin{table*}[htbp]
\centering
\scriptsize
\caption{Comparison of Wireless Positioning Datasets}
\begin{tabular}{|p{1.5cm}|p{1.5cm}|p{1.5cm}|p{1.2cm}|p{4.5cm}|p{4cm}|}
\hline
\textbf{Dataset Name} & \textbf{Channel Model Type} & \textbf{Open Source Type} & \textbf{Mobility Support} & \textbf{Parameter Configuration} & \textbf{Limitation} \\
\hline
xG-Loc & CDL channel model (3GPP 38.901) & Simulated dataset  & No & Adjustable bandwidth (5–400 MHz), frequency (FR1/FR2/FR3), and scenario type (UMi, UMa, RMa, InF-DH, and InF-SH) & No measured data; static channels only. \\

\hline
MaMIMO & Real measurement & Measured dataset (252,004 samples)  & Yes & 2.61 GHz frequency, 20 MHz bandwidth, four antenna array topologies (URA LOS/NLOS, ULA LOS, DIS LOS) & Limited to indoor environments; no high-speed mobility support. \\

\hline
DeepMIMO & Ray tracing & Channel simulator & No & Antenna array structure, OFDM subcarriers, bandwidth, frequency, scenario selection, BS/UE range, number of paths & Fixed sampling points; cannot simulate dynamic channels. \\

\hline
5G-NR-data-generator & CDL model + ray tracing & Channel simulator & No & RBs, subcarriers, bandwidth, BS/UE antenna count, carrier frequency, delay/angle spread & Single fixed scene; limited environment diversity \\

\hline
WAIR-D & Ray tracing & Channel simulator & No & Carrier frequencies, antenna configuration, bandwidth, OFDM settings & Fixed environments; no dynamic channel or mobility modeling. \\

\hline
DataAI-6G & Ray tracing & Channel simulator & Yes & Adjustable frequency band, bandwidth, antenna config, user mobility speed/direction, sampling interval & Only one outdoor street scenario; lacks environment diversity \\
\hline
ViWi & Ray tracing + 3D visual simulation & Channel/vision simulator & No & BS/UE positions, antenna config, frequency, trajectory, camera setup, visual resolution & Synthetic data only; limited scenarios; no Doppler modeling. \\
\hline
\end{tabular}
\label{tab:channel_datasets}
\end{table*}

\subsection{Dataset Descriptions}

In recent years, several datasets have been proposed in the field of wireless positioning, including the xG-Loc \cite{10349917xGdataset}, MaMIMO \cite{MaMIMO}, DeepMIMO, \cite{alkhateeb2019deepmimo}, 5G-NR-data-generator~\cite{zhang2021generalized}, Wireless AI Research Dataset (WAIR-D) \cite{huangfu2022wair}, DataAI-6G \cite{shen2023dataai}, and the ViWi \cite{alrabeiah2020viwi}. However, the above datasets do not conform to the clustered delay line (CDL) channel model \cite{3gpp-38.901-2019channel} recommended by 3GPP. Since the CDL channel model is more suitable for link-level and system-level simulation and for the comprehensiveness of the study, we also examine datasets based on the CDL channel model. A comparison of wireless positioning datasets is shown in Table. \ref{tab:channel_datasets}.

\subsubsection{xG-Loc}
The xG-Loc dataset \cite{10349917xGdataset} is an open dataset explicitly designed for localization algorithms and services, fully compliant with 3GPP technical reports and specifications. This dataset provides a standardized and comprehensive resource for evaluating and benchmarking localization solutions across diverse scenarios, conforming to the channel models defined in 3GPP TR 38.901 \cite{3gpp-38.901-2019channel}. xG-Loc is structured into 28 compressed directories, representing 28 unique configurations of 3GPP-standardized scenarios, bandwidths, and central frequencies. These configurations span multiple frequency ranges, including FR1 (microwaves), FR2 (millimeter waves), and FR3 (upper mid-band),  as well as a variety of deployment scenarios, such as Indoor Factory Dense-High (InF-DH), Indoor Factory Sparse-High (InF-SH), Indoor Open Office (IOO), Urban Microcell (UMi), Urban Macrocell (UMa), Rural Macro (RMa). In addition, the dataset provides PRS, SRS, and other channel-related measurements. It enables a wide range of tasks, including distance estimation, angle estimation, position estimation, and wireless channel quality evaluation.

\subsubsection{MaMIMO}

MaMIMO \cite{MaMIMO} is an open-access indoor CSI dataset, providing 252,004 high-precision CSI samples collected using the 64-antenna KU Leuven Massive MIMO testbed. Measurements were conducted across four antenna topologies: Uniform Rectangular Array (URA) in both LOS and NLOS conditions, Uniform Linear Array (ULA), and Distributed ULA (DIS), each within a controlled indoor office environment. CSI was recorded over a 20 MHz bandwidth at 2.61 GHz using 100 subcarriers, with user equipment moved across a 1.25 m × 1.25 m grid via CNC-controlled XY-tables. This setup ensured sub-millimeter accuracy in positional labeling. The resulting dataset enables benchmarking of positioning algorithms, multipath component analysis, and precoder visualization.

\color{black}

\subsubsection{5G-NR-data-generator} 

The 5G-NR-data-generator~\cite{zhang2021generalized} is a generalized channel dataset generator that adopts the CDL channel model defined in the 5G NR standard~\cite{3gpp-38.901-2019channel}. This generator allows users to customize various channel parameters to meet specific needs and supports the generation of MIMO channel matrices. The configurable channel parameters include a variety of settings, such as the number of RBs, subcarriers, bandwidth, number of BS/UE antennas, carrier frequency, delay spread, and angle spread. The generator performs ray-tracing-based simulations to model the propagation environment using external tools. However, the current version only supports a single predefined scenario and does not allow flexible environment modeling or dynamic user movement.

\subsubsection{DeepMIMO}

\begin{figure}[tb]
\centering
\includegraphics[scale=0.33]{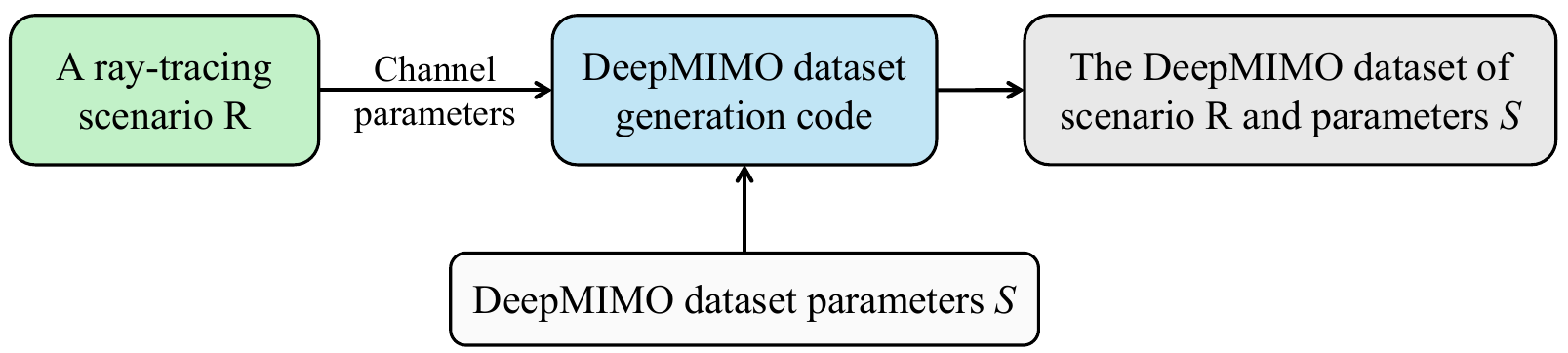}
\caption{Dataset generation process of DeepMIMO.}
\label{fig_DeepMIMO}
\end{figure}

DeepMIMO \cite{alkhateeb2019deepmimo} provides a comprehensive framework for generating channel datasets tailored to various tasks in wireless communication. The channels in the DeepMIMO dataset are constructed using accurate ray-tracing data obtained from the Wireless InSite simulator~\cite{remcom_wireless_insite}. This ray-tracing simulation captures the dependence of the wireless channels on the geometry and materials of the environment, as well as the spatial locations of the transmitter and receiver. This feature is crucial for ML applications in mmWave and massive-MIMO systems, as it ensures the dataset realistically reflects the physical characteristics of the propagation environment. DeepMIMO enables researchers to flexibly configure their simulation scenarios and wireless parameters prior to dataset generation. These parameters include antenna array structure, OFDM subcarrier configuration, bandwidth, frequency band, scenario selection, user/BS activation range, and the number of paths. Therefore, the DeepMIMO dataset is defined by two primary components: a scenario and a set of parameters. However, the channel generation in DeepMIMO is based on pre-collected ray-tracing paths at fixed sampling points, offering limited flexibility in sampling location configuration. Moreover, dynamic channels caused by user movement cannot be simulated due to the lack of channel temporal modeling. The process of using the DeepMIMO dataset generation framework is illustrated in Fig.~\ref{fig_DeepMIMO}.

\subsubsection{WAIR-D}

The WAIR-D \cite{huangfu2022wair} is developed by researchers from Huawei and Zhejiang University to provide a wireless dataset that replicates various environments closely resembling real-world conditions. The dataset is generated using a 3D ray-tracing simulator, PyLayers \cite{amiot2013pylayers}, in conjunction with the OSM \cite{haklay2008openstreetmap}. WAIR-D includes two scenarios: 1). Scenario 1 consists of 10,000 environments, each with 5 base stations and 30 sparsely distributed user equipments (UEs); 2). Scenario 2 contains 100 environments, each with 1 base station and 10,000 densely dropped UEs. WAIR-D supports five carrier frequencies, i.e., 2.6 GHz, 6 GHz, 28 GHz, 60 GHz, and 100 GHz. Users can configure antenna configuration, bandwidth, and OFDM subcarrier settings to generate channel data. Similar to DeepMIMO, WAIR-D is based on static ray-tracing results and does not support flexible sampling or dynamic channel modeling caused by user movement.

\subsubsection{DataAI-6G}


\begin{figure}[tb]
\centering
\includegraphics[scale=0.25]{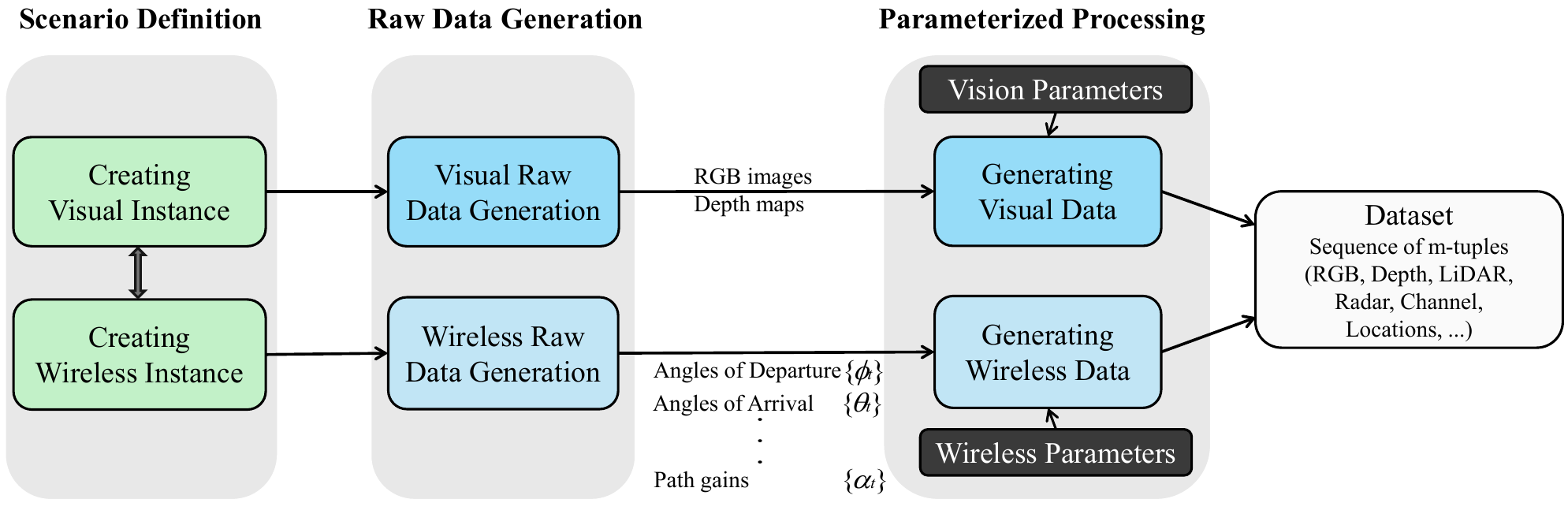}
\caption{Dataset generation process of the ViWi dataset.}
\label{fig_ViWi}
\end{figure}

The DataAI-6G dataset \cite{shen2023dataai}, developed by researchers from Beijing University of Posts and Telecommunications and the China Mobile Research Institution, is specifically designed for AI-6G research. Compared to existing datasets such as DeepMIMO and WAIR-D, DataAI-6G introduces key advancements by incorporating spatially non-stationary channel features and modeling high-mobility scenarios, including Doppler effects and time-varying multipath propagation. The dataset allows users to configure a wide range of system parameters, such as carrier frequency (uplink and downlink), bandwidth, antenna array configuration, user mobility paths and speeds, number of users, and OFDM settings. However, its current version is limited to a single outdoor street scenario, lacking environmental diversity across indoor, suburban, or industrial settings.

\subsubsection{ViWi}

The ViWi dataset \cite{alrabeiah2020viwi} is designed specifically for vision-aided wireless communications research. It serves as a parametric, systematic, and scalable data generation framework, leveraging advanced 3D modeling and ray-tracing software to produce high-fidelity synthetic wireless and vision data samples for identical scenes. By combining vision and wireless data, the dataset facilitates research at the intersection of these domains, supporting innovative approaches to wireless communication and positioning. The dataset includes wireless parameters such as AOD, path gains, and detailed user location, along with corresponding RGB images, enabling joint research in communication and localization. ViWi supports customizable scene configurations and user trajectories but is limited to synthetic data and does not model Doppler effects or continuous channel dynamics. The dataset generation process consists of three stages, as illustrated in Fig.~\ref{fig_ViWi}:
\begin{itemize}
    \item \textbf{Scenario Definition}: Defines physical layout, transmitter and receiver locations, and environmental parameters.
    \item \textbf{Raw-Data Generation}: Utilizes 3D modeling and ray-tracing tools to produce realistic wireless and visual data.
    \item \textbf{Parametrized Processing}: Processes raw data into structured formats tailored to research needs.
\end{itemize}

\subsection{Case Study in AI-Driven Positioning}

\begin{figure}[t]
    \centering
    \includegraphics[width=0.35\textwidth]{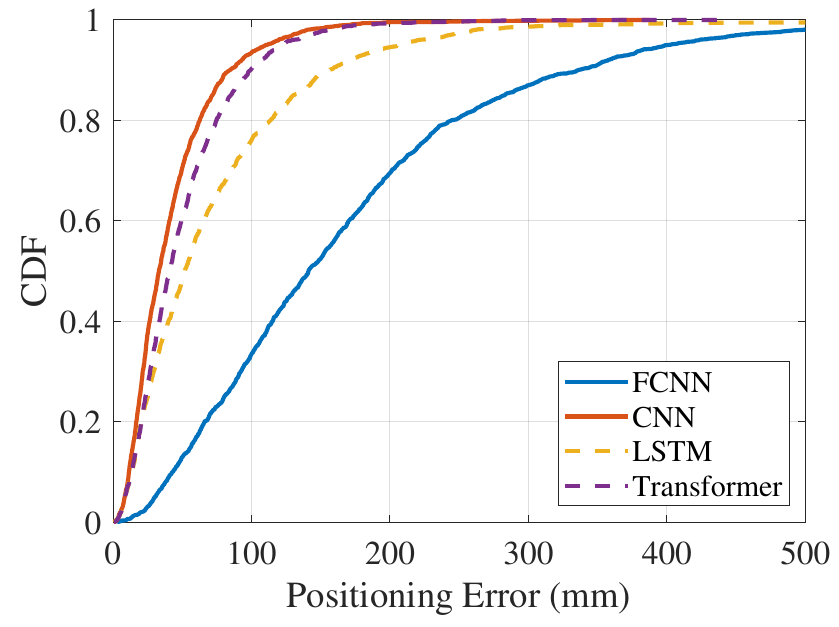}
    \caption{Localization performance comparison using the MaMIMO dataset under ULA-LOS configuration.}
    \label{fig:mamimo_results}
\end{figure}

\begin{figure}[t]
    \centering
    \includegraphics[width=0.35\textwidth]{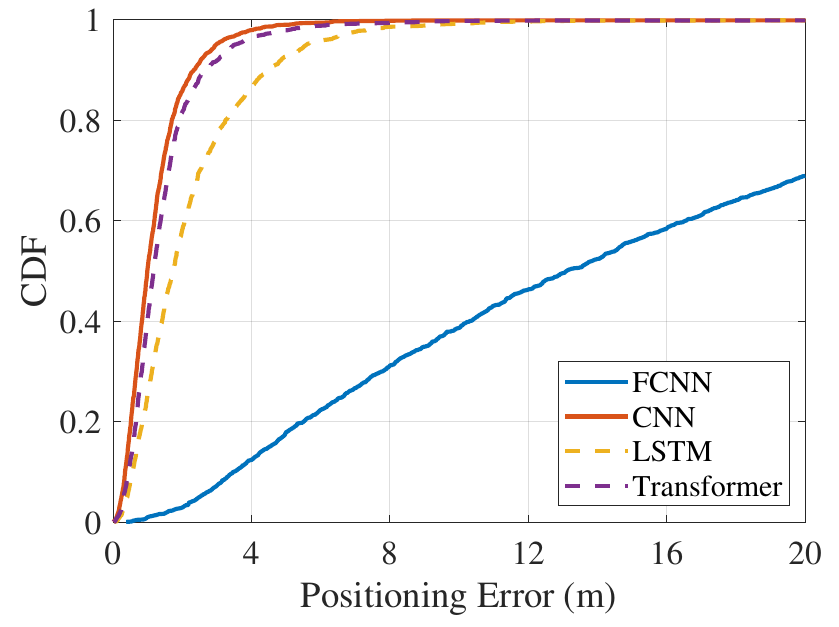}
    \caption{Localization performance comparison using the DeepMIMO ‘O1’ scenario.}
    \label{fig:deepmimo_results}
\end{figure}

To evaluate the impact of different datasets and models on positioning performance, we conduct a comparative case study using both the measured MaMIMO dataset and the simulated DeepMIMO dataset. For the MaMIMO dataset, we focus on the ULA configuration under LOS conditions. For DeepMIMO, we select the widely used urban street scenario "O1", which is a typical deployment environment. In both datasets, we randomly sample 7,000 samples for training, 1,000 for validation, and 2,000 for testing. To ensure a fair comparison, we fix key channel parameters across both scenarios: 20 MHz bandwidth, 100 OFDM subcarriers, and a 64-element ULA at the BS. We evaluate four neural network backbones: FCNN, CNN (ResNet-32), LSTM, and Transformer. Each model is trained and tested under the same data split for both datasets.

The results are summarized in Fig.~\ref{fig:mamimo_results} and Fig.~\ref{fig:deepmimo_results}. The models exhibit different performance across the two datasets. In the MaMIMO scenario, the denser and more concentrated spatial sampling allows the models to achieve higher positioning accuracy. Nevertheless, the ranking of model performance is consistent across both datasets: from best to worst, the order is CNN, Transformer, LSTM, and FCNN. Although the Transformer demonstrates stronger feature representation capabilities, it requires more data for training to fully converge. Therefore, it performs worse than CNN in situations with limited data or without pre-training. Specifically, in the MaMIMO dataset, the median positioning errors achieved by FCNN, CNN, LSTM, and Transformer are 141.2 mm, 33.1 mm, 51.8 mm, and 40.9 mm, respectively. In the DeepMIMO dataset, the corresponding median errors are 13.1 m, 1.0 m, 1.7 mm, and 1.2 m, respectively.

\subsection{Future Directions for Dataset Design}

Despite the growing availability of wireless positioning datasets, there remains no unified benchmark that supports dynamic, multi-modal, and heterogeneous conditions. Most existing datasets suffer from limitations such as lack of user mobility, restricted environmental diversity, or absence of cross-modal sensing data, which hinders the generalizability and reproducibility of AI-based localization solutions. Therefore, future dataset development should consider the following directions:
\begin{itemize}
    \item \textbf{Multi-modality integration:} Combine RF data with auxiliary sensing modalities such as IMU, LiDAR, cameras, and digital maps to enable multi-sensor fusion, learning-based SLAM, and cross-domain localization tasks.

    \item \textbf{Doppler and time-evolving channels:} Incorporate realistic time-series CSI that captures Doppler effects, user mobility, and handover dynamics to support mobility-aware learning and trajectory prediction.

    \item \textbf{Scalable and customizable simulation:} Develop open-source, extensible simulation platforms (e.g., built on Unity or Unreal Engine) that allow researchers to design custom virtual environments and generate synchronized RF and sensor data in real time.

    \item \textbf{Alignment between synthetic and real-world data:} Promote methodologies for bridging the gap between simulated datasets and real-world environments, enabling models trained in virtual domains to generalize effectively to physical deployments.

    \item \textbf{Standardized benchmarks and evaluation metrics:} Establish public benchmarks with consistent data splits, task definitions, and evaluation metrics (e.g., accuracy, latency, energy efficiency) to foster reproducibility and fair comparison across algorithms.
\end{itemize}

\subsection{Lessons Learned}

We reviewed representative datasets, including MaMIMO, DeepMIMO, WAIR-D, and xG-Loc, covering both measured and simulated data sources. Dataset design significantly influences the performance of AI-based positioning models. Measured datasets like MaMIMO offer higher positioning accuracy due to dense and precise spatial sampling, while simulated datasets such as DeepMIMO and WAIR-D provide configurability but lack temporal channel dynamics and real-world variability. Moreover, few existing datasets support user mobility, multi-band signals, or cross-modal data (e.g., vision or IMU), limiting their applicability in complex 6G scenarios. Our case studies further show that CNNs perform consistently well on DeepMIMO and MaMIMO datasets, while Transformers require larger datasets to achieve optimal performance. Overall, current datasets fall short in enabling generalizable, dynamic, and multi-modal learning, underscoring the urgent need for standardized and extensible datasets to benchmark and accelerate wireless positioning research for future networks.

\color{black}


\section{Challenges and Opportunities} \label{sec:Challenges}

Wireless AI positioning presents both significant challenges and exciting opportunities. This section explores the key challenges and potential opportunities in this field.

\subsection{Challenges in AI-driven Wireless Positioning}

Wireless AI positioning has witnessed remarkable advancements, yet several challenges hinder its widespread adoption and optimal performance. These challenges stem from the intrinsic complexities of wireless environments, computational constraints, and the integration of emerging technologies. Below, we outline the key challenges.

\subsubsection{Data Collection}

Data collection for AI-driven wireless positioning presents unique challenges, primarily due to the scenario-specific nature of positioning tasks, which necessitates customized datasets. This significantly increases the difficulty of acquiring accurate ground-truth position labels. Heterogeneous propagation characteristics across different environments (e.g., urban canyons, dense indoor layouts, rural open spaces) often lead to severe performance degradation when a model trained in one setting is transferred to another. Additionally, the dynamic nature of wireless propagation causes fingerprints to become outdated quickly, thereby increasing the demand for frequent dataset updates and incurring substantial maintenance costs.

High-quality labeled datasets are essential for supervised learning models. To overcome this challenge, one promising direction is the development of digital twin systems tailored for wireless positioning \cite{shi2025digital}. These systems leverage ray tracing techniques to generate synthetic wireless channels that can be used for model training. However, a major limitation lies in the domain gap between simulated and real-world channels, which undermines model generalization and highlights the need for further advancement in accurate channel modeling. An alternative approach involves the use of generative neural networks. GANs have already demonstrated strong performance in existing wireless positioning works \cite{zhang_rme-gan_2023}. As AI technology continues to evolve, more advanced generative models, such as diffusion models and large AI models, are expected to further bridge the gap between synthetic and real-world data. These techniques provide a promising solution for building data-efficient and scalable wireless positioning systems that work well in diverse and dynamic environments.

\color{black}

\subsubsection{Accuracy in Complex Environments}

Achieving high positioning accuracy in complex and high-interference environments remains a critical challenge for AI-driven wireless positioning systems. Real-world scenarios such as urban canyons, highways, dense indoor areas, or industrial facilities are characterized by dynamic and unpredictable signal conditions, including multipath propagation, NLOS effects, high-speed mobility, and external interference.  In such environments, wireless signals are frequently reflected, refracted, or scattered, resulting in distorted and non-stationary channel characteristics. These factors not only degrade signal quality but also lead to rapid shifts in the underlying data distribution, making it difficult for AI models. Moreover, the presence of strong interference from overlapping wireless systems or co-channel users further reduces signal fidelity, introducing noise that hampers both model training and inference. While AI models have demonstrated the capability to learn complex propagation patterns, they often require large volumes of diverse and up-to-date labeled data to generalize effectively. However, such data is costly and often infeasible to collect at scale in dynamically evolving environments.

To address this, one promising direction is to leverage foundation models pretrained on large-scale unlabeled wireless data, which can enhance the representation capability of AI models and improve localization accuracy \cite{pan2025large}. Researchers are also exploring hybrid approaches that combine data-driven learning with model-based methods, aiming to improve robustness in complex environments \cite{yang_model-based_2021}. Moreover, digital twin systems can simulate intricate propagation conditions ahead of deployment, enabling the design of data- and model-driven positioning algorithms that are more resilient and adaptable to real-world scenarios \cite{shi2025digital}. Finally, multi-modal positioning \cite{tang2024novel}, which incorporates complementary sensory data such as inertial measurements, camera input, or geomagnetic signals, can help compensate for radio signal degradation, thus improving accuracy and reliability in harsh environments.

\subsubsection{Model Generalization}
Model generalization remains one of the most critical challenges in AI-driven wireless positioning, particularly due to the high variability of deployment environments and signal propagation conditions. Wireless environments vary significantly across different geographic locations, building layouts, materials, hardware platforms, and time periods. For example, models trained in urban environments may fail to perform in rural areas or indoor industrial settings due to stark differences in channel characteristics, such as multipath density and obstruction patterns. These discrepancies result in domain shifts in both feature space (e.g., CSI or RSSI distributions) and label space (e.g., spatial layouts), making it difficult for AI models to maintain robust performance across unseen scenarios.

While advanced techniques such as domain adaptation, transfer learning, and multi-task learning can partially address this issue, their ability to fully resolve the complexity and dynamic nature of wireless positioning environments remains limited. To further enhance model generalization, one promising approach is to leverage federated learning frameworks that aggregate heterogeneous channel data from diverse deployment scenarios, enabling collaborative training \cite{10118848FL}. Additionally, large AI foundation models pretrained on large-scale unlabeled wireless datasets have the potential to learn domain-invariant and generalizable positioning representations, thereby improving adaptability to unseen environments \cite{pan2025large}. Finally, model-based approaches that incorporate digital twin systems and physical prior knowledge, such as floor plans \cite{yu_floor-plan-aided_2024} or environmental maps \cite{zhang2025uniloc}, can also help bridge the generalization gap by introducing environment information into the learning process.

\subsubsection{Resource Constraints}
Resource constraints pose a significant barrier to the deployment of AI-based positioning systems, particularly on edge devices such as smartphones, IoT sensors, and UAVs. AI model inference and training typically demand substantial computational power, memory, and energy, which often exceed the capabilities of these devices. Additionally, processing wireless positioning data (e.g., high-dimensional CSI matrices from massive MIMO or raw CIR signals from mmWave systems) typically involves deep neural networks with millions of parameters, resulting in high inference latency and power consumption. This challenge is particularly severe in latency-critical applications like autonomous vehicles or AR/VR. Moreover, model updates in dynamic environments demand continual learning or on-device adaptation, further straining computational budgets.

To address these issues, model compression techniques such as weight pruning \cite{zhu2017prune}, quantization \cite{zhou2018adaptive}, and knowledge distillation \cite{wang2019private} can significantly reduce model size and computation cost while maintaining accuracy. Beyond compression, Edge-AI paradigms have demonstrated great potential. In particular, federated learning \cite{chen2024towards} enables collaborative model training across distributed devices without transmitting raw data, thereby reducing communication costs and preserving user privacy. Meanwhile, distributed inference techniques \cite{pan2023joint, sun2025energy} leverage the computing power of nearby edge servers to offload heavy computation, alleviating the burden on individual devices. These approaches together offer a promising direction for enabling scalable and efficient AI-driven wireless positioning in constrained environments.

\subsubsection{Scalability Across Network Configurations}

Scalability across different wireless network configurations presents another key challenge for AI-driven positioning systems. On one hand, scalability across BS configurations refers to the ability of AI models to operate effectively under diverse network settings. In practice, BSs may employ different configurations in terms of carrier frequency (e.g., sub-6 GHz vs. mmWave), bandwidth, MIMO setup, antenna layout, and pilot signaling schemes to meet varying deployment requirements. These differences lead to significant variability in the format and characteristics of the CSI, such as input dimensionality, spectral properties, and temporal structure. As a result, models trained under one configuration often struggle to generalize across others, reducing their transferability and deployment flexibility. On the other hand, scalability at the system level involves handling large-scale, complex network scenarios, such as multi-BS, multi-user interference, and ultra-dense infrastructure deployments. Emerging technologies like massive MIMO, RIS, and NTN networks introduce more intricate propagation behaviors and generate vast amounts of high-dimensional data. These challenges highlight the necessity for flexibly scalable AI algorithms that can adapt to heterogeneous network parameters and system conditions.

To address the scalability challenge, future solutions may leverage large AI foundation models tailored for positioning tasks. By pretraining on large-scale unlabeled CSI data from diverse network configurations, these models can learn domain-invariant representations that align features across different BS settings and infrastructures (e.g., RIS, NTN). This enables the model to generalize across varying input formats and supports multi-source CSI fusion, improving adaptability and robustness in complex deployment scenarios.

\subsubsection{Security and Privacy Concerns}

Security and privacy are also very important for positioning applications in future networks \cite{XAI_Security}. AI-based wireless positioning systems face three primary categories of security and privacy risks: (1) Vulnerabilities during signal acquisition: Raw positioning signals, such as CSI and RSS, can be intercepted, spoofed, jammed, or perturbed by adversaries. These threats degrade signal integrity and can significantly undermine localization accuracy, especially in dynamic or hostile environments \cite{gao2024surgical, huang2024attacking}. (2) Privacy leakage during model training: The training phase often involves large volumes of sensitive user data containing precise location traces. Such data is vulnerable to privacy attacks, including data reconstruction, membership inference, and gradient leakage, which can reveal personal information or re-identify individuals from aggregated datasets. (3) Attacks during model inference: The AI models themselves may be exploited as attack vectors. Techniques such as model inversion, membership inference, or backdoor injection can be used to extract private information or manipulate the model’s output. In addition, attackers may introduce imperceptible perturbations to input signals, tricking the model into outputting incorrect or misleading positions.

To mitigate the security and privacy risks in wireless positioning, several countermeasures can be applied at different stages of the system \cite{XAI_Security}. During the signal acquisition phase, physical-layer security techniques can be employed to detect and exclude adversarial sources, thereby mitigating spoofing and jamming attacks. In addition, signal preprocessing and robust feature extraction methods can reduce the susceptibility of raw signals to malicious perturbations \cite{elsisi2024robust}. During the model training phase, federated learning offers a promising solution by enabling decentralized model updates without sharing raw location data, thus preserving user privacy while also improving model generalization \cite{chen2024towards}. For threats occurring during inference, techniques such as differential privacy can increase the difficulty of model inversion, membership inference, or backdoor injection attacks. Furthermore, incorporating adversarial training during the learning process can enhance model robustness against input perturbations and adversarial manipulation \cite{huang2024attacking}.

\subsubsection{Explainability and Trustworthiness}

AI-driven wireless positioning systems often rely on complex deep learning models that function as black boxes with limited interpretability. This lack of transparency presents a major obstacle to practical deployment, particularly in safety-critical applications such as autonomous navigation, emergency response, and industrial automation, where black-box AI algorithms cannot provide sufficient guarantees of reliability. Without clear explanations, it becomes difficult to trace, debug, or correct erroneous outputs, thereby undermining user confidence and accountability \cite{guo2020explainable}. Therefore, understanding the underlying rationale behind model predictions is essential for building trustworthy systems.

To address this challenge, one direction is to pursue rigorous theoretical analysis of AI model reliability, although this remains one of the most difficult open problems in the research community. On the other hand, a more practical approach is to develop data-model hybrid AI algorithms that combine data-driven learning with physics-based models. By integrating domain knowledge and physical priors \cite{Chaccour_Knowledge}, these hybrid methods can enhance interpretability and improve the trustworthiness of AI-driven positioning systems.

\color{black}

\subsection{Opportunities in AI-Driven Wireless Positioning}

While wireless AI positioning faces numerous challenges, it also presents unprecedented opportunities for innovation and application across various domains. These opportunities arise from advancements in AI technologies, the evolution of wireless networks, and the growing demand for high-precision positioning solutions. Building on the aforementioned challenges and their potential solutions, we outline below several key opportunities emerging in this field:

\subsubsection{Enhanced Accuracy Through Advanced AI Technologies}

Recent progress in AI has opened new possibilities for improving the accuracy of wireless positioning systems. Advanced models such as Transformers, self-supervised learning, and large-scale foundation models have shown strong capabilities in capturing the nonlinear and complex nature of wireless signal propagation. Especially, inspired by the success of foundation models in vision and language, similar paradigms are now emerging in wireless communications. By pretraining on large-scale unlabeled datasets from diverse environments, these models can learn transferable and generalizable channel semantics, significantly reducing reliance on labeled data and improving robustness across deployment scenarios \cite{pan2025large}. Moreover, their scalability enables them to handle heterogeneous inputs, including varying frequency bands, antenna configurations, signaling protocols, and even emerging technologies such as RIS and NTN. They can also integrate multi-modal data (e.g., vision and LiDAR), making them highly adaptable to diverse and evolving wireless infrastructures.
\color{black}

\subsubsection{Multi-Source and Cross-Modal Data Fusion}

AI-driven wireless positioning is increasingly benefiting from the fusion of data collected across multiple network infrastructures and sensing modalities. Emerging network architectures, such as cell-free massive MIMO, NTN, and UAV-assisted networks, offer complementary spatial and temporal coverage, enriching the available wireless signal space for localization tasks. Simultaneously, multi-modal sensor data, including vision, LiDAR, and inertial measurements, provides additional context that can compensate for the limitations of radio-based methods under challenging conditions such as NLOS or deep indoor environments.  By leveraging AI models capable of cross-modal representation learning and feature alignment, positioning systems can integrate heterogeneous data sources to achieve more robust, accurate, and environment-aware localization.
Moreover, large AI foundation models provide a unified framework for learning consistent representations across diverse modalities and input formats. Their capacity for multi-source feature abstraction and semantic alignment enables deeper fusion at both the feature and decision levels, laying the foundation for scalable and generalizable positioning systems in complex, multi-modal environments.
\color{black}

\subsubsection{Data-Model Co-Design through Digital Twins}

The emergence of digital twin technologies introduces a promising paradigm for both dataset generation and the co-design of data- and model-driven wireless positioning algorithms \cite{shi2025digital,pan2025rate}. By simulating realistic radio environments through ray tracing and physical channel modeling, digital twins enable the creation of large-scale, scenario-specific synthetic datasets, significantly reducing the cost and effort of data collection. These virtual environments also support controlled model pretraining, rigorous evaluation under diverse conditions, and reproducible benchmarking. Beyond data generation, digital twins serve as a bridge between physics-based modeling and AI learning frameworks. By embedding physical priors (such as geometry, material properties, and propagation constraints) into AI models \cite{zhang2025uniloc}, they enable hybrid learning strategies that fuse model knowledge with data-driven optimization. This co-design approach enhances model interpretability, generalizability, and robustness.
\color{black}

\subsubsection{Resource-Efficient AI Models and Edge Deployment}
The development of lightweight and resource-efficient AI models offers a promising pathway for the widespread deployment of wireless positioning systems on resource-constrained devices. Techniques such as model pruning, quantization, and knowledge distillation enable the creation of compact AI models that reduce computational and energy requirements while maintaining high accuracy. In addition, offloading positioning tasks to edge servers or adopting distributed inference frameworks allows for efficient utilization of edge computing resources, alleviating the processing burden on local devices. As a result, resource-efficient AI and edge deployment together enable the practical integration of AI-driven positioning into a broader range of applications, even on devices with limited hardware capabilities such as IoT nodes, wearable devices, and mobile platforms.

\subsubsection{ISAC-Enhanced AI Positioning}
Integrated sensing and communication (ISAC) systems \cite{ISAC1, zhang2025joint} enhance positioning by utilizing echo signals to sense the surrounding environment and extract key parameters such as AOA, TOA, and Doppler shift. This capability provides a deeper understanding of the network environment, such as NLOS detection \cite{ISAC2}. ISAC systems can infer the position and motion of surrounding objects even without explicit transmissions from them. In complex and dynamic scenarios, AI can be used to jointly extract and fuse spatial features from both positioning and sensing modalities, enabling high-precision localization through the synergy of active and passive information. This opens up new possibilities for robust, environment-aware, and context-adaptive positioning in future wireless networks.

\subsubsection{Privacy- and Security-Aware AI Positioning}

As AI-driven wireless positioning systems are increasingly deployed in user-centric and safety-critical scenarios, ensuring data privacy and system security has become a fundamental design requirement. This opens new opportunities for building AI frameworks that are both privacy-preserving and resilient to attacks. It is essential to address not only the inherent privacy and security challenges of the positioning task itself, but also the vulnerabilities introduced by AI models during training and inference. Techniques like federated learning \cite{gao2022federated} and differential privacy enable collaborative model training without transmitting raw location data, significantly reducing the risk of data leakage. Meanwhile, adversarial training and robust feature extraction methods can enhance model resistance to spoofing, jamming, and model inversion attacks. By jointly considering the threats from both wireless data and AI components, future systems can achieve secure and trustworthy AI-based positioning.

\subsubsection{Robust and Trustworthy AI for Mission-Critical Applications}
AI-based wireless positioning is becoming increasingly relevant in safety-critical applications such as emergency response, industrial automation, and autonomous driving. This trend underscores both the need and the opportunity to develop AI models that are not only accurate but also trustworthy and interpretable. Research on robust AI, adversarial defenses, and explainable machine learning can help build reliable systems capable of operating under uncertain or adversarial conditions. Furthermore, the integration of data- and model-driven algorithm design offers a promising path toward enhancing both the transparency and reliability of AI-based localization

\section{Conclusion} \label{sec:conclusion}
In this work, we primarily explore the potential of AI-driven wireless positioning technologies from the perspective of integrating AI with wireless positioning. While the focus is on cellular positioning scenarios, the study also incorporates insights from WiFi, Bluetooth, and UWB positioning to enhance the comprehensiveness of the algorithmic understanding. Specifically, we introduce the foundational knowledge of AI technologies and wireless positioning techniques, followed by a summary of 3GPP standards related to positioning and AI advancements. Subsequently, we also reviews the SOTA research in both AI/ML-assisted positioning and direct AI/ML positioning, as well as datasets commonly used for wireless positioning. Finally, we summarize the challenges and opportunities in AI-driven wireless positioning. With the continued advancement of AI technologies, the integration of AI and positioning is expected to deepen. This review aims to inspire researchers in both industry and academia, contributing to the advancement of this promising field.

\bibliographystyle{IEEEtran}
\bibliography{reference}

\end{document}